\documentclass[10pt,journal,compsoc]{IEEEtran}

\ifCLASSOPTIONcompsoc
    \usepackage[nocompress]{cite}
    \usepackage{textcomp}
    \usepackage{stfloats}
    \usepackage{url}
    \usepackage[caption=false,font=normalsize,labelfont=sf,textfont=sf]{subfig}
    \usepackage{caption}
    \usepackage{verbatim}
    \usepackage{graphicx}
    \usepackage{booktabs}
    \usepackage{amsmath,amsfonts}
    \usepackage{amssymb}
    \usepackage{mathtools}
    \usepackage{amsthm}
    \usepackage{hyperref}
    \usepackage{multirow,tabularx}
    \usepackage{makecell}
    \usepackage{color}
    \usepackage{utfsym}
    \usepackage{microtype}
    \usepackage{booktabs} % for professional tables
    \usepackage{amsmath,amsfonts}
    \usepackage{algorithmic}
    \usepackage{algorithm}
    \usepackage{array}
    \usepackage{tikz,xcolor,hyperref}% Make Orcid icon
    \definecolor{lime}{HTML}{A6CE39}
    \DeclareRobustCommand{\orcidicon}{%
        \begin{tikzpicture}
        \draw[lime, fill=lime] (0,0) 
        circle [radius=0.16] 
        node[white] {{\fontfamily{qag}\selectfont \tiny ID}};    \draw[white, fill=white] (-0.0625,0.095) 
        circle [radius=0.007];    \end{tikzpicture}
        \hspace{-2mm}}
    \foreach \x in {A, ..., Z}{%
        \expandafter\xdef\csname orcid\x\endcsname{\noexpand\href{https://orcid.org/\csname orcidauthor\x\endcsname}{\noexpand\orcidicon}}
        }
\else
    \usepackage{cite}
\fi

\ifCLASSINFOpdf

\else

\fi

\begin{document}

\title{
NineRec: A Benchmark Dataset Suite for Evaluating Transferable Recommendation
}

\author{Jiaqi Zhang$^{1}$, Yu Cheng$^{1}$, Yongxin Ni$^{1}$, Yunzhu Pan$^{1}$, Zheng Yuan$^{1}$ \\
        Junchen Fu$^{1}$, Youhua Li$^{1}$, Jie Wang$^{1}$, Fajie Yuan$^{1\dagger}$\thanks{The paper has been accepted by IEEE Transactions on Pattern Analysis and Machine Intelligence. \\ $^{1}$ Westlake University. $\dagger$ Corresponding author: Fajie Yuan (e-mail: yuanfajie@westlake.edu.cn).}
        }

% The paper headers
\markboth{IEEE TRANSACTIONS ON PATTERN ANALYSIS AND MACHINE INTELLIGENCE, 2024}%
{Shell \MakeLowercase{\textit{et al.}}: Bare Advanced Demo of IEEEtran.cls for IEEE Computer Society Journals}

\IEEEtitleabstractindextext{%
\begin{abstract}
Large foundational models, through upstream pre-training and downstream fine-tuning, have achieved immense success in the broad AI community due to improved model performance and significant reductions in repetitive engineering. By contrast, the \underline{trans}ferable one-for-all models in the \underline{rec}ommender system  field, referred to as   TransRec, have made limited progress.  The development of TransRec has encountered multiple  challenges, among which the lack of large-scale, high-quality transfer learning recommendation dataset and benchmark suites is one of the biggest obstacles. To this end, we introduce NineRec, a TransRec dataset suite  that comprises a large-scale source domain recommendation dataset and \underline{nine} diverse target domain recommendation datasets. Each item in NineRec is accompanied by a descriptive text and a high-resolution cover image. Leveraging NineRec, we enable the implementation of TransRec models by learning from raw multimodal features instead of relying solely on pre-extracted off-the-shelf features. 
% Through an extensive and rigorous evaluation, 
Finally, we present robust TransRec benchmark results with several classical network architectures, providing valuable insights into the field. To facilitate further research, we will release our code, datasets, benchmarks, and leaderboards at \textcolor{blue}{\url{https://github.com/westlake-repl/NineRec}}.

\end{abstract}

}

\maketitle

\IEEEdisplaynontitleabstractindextext

\IEEEpeerreviewmaketitle

\ifCLASSOPTIONcompsoc
\IEEEraisesectionheading{\section{Introduction}\label{sec:introduction}}
\else
\section{Introduction}
\label{submission}
\fi

\IEEEPARstart{R}{ecommender} system (RS)  models play a crucial role in predicting user preferences for unseen items based on their previous interactions. These highly successful models have found wide-ranging applications, such as in advertising systems, e-commerce websites, search engines, and streaming services. In the past few decades, extensive research has been conducted on both content-based~\cite{pazzani2007content} and collaborative filtering~\cite{koren2009matrix,rendle2012bpr,he2017neural} recommendation models. Among these approaches, ID-based collaborative filtering models  (known as IDRec), which leverage unique IDs to represent users and items have dominated the RS field for over 10 years.

Meanwhile, the IDRec paradigm encounters several key bottlenecks due to its inherent characteristics. \textcolor{black}{Firstly, IDRec struggles to handle cold-start scenarios because new userIDs and itemIDs cannot be effectively trained before being deployed in live environments. Secondly, the design philosophy of IDRec diverges from the fundamental principle of modern "foundation" models~\cite{bommasani2021opportunities} in the deep learning community, which emphasizes the adaptability of pre-trained parameters to multiple downstream tasks.}
% The first well-known shortcoming is that IDRec cannot handle cold-start settings because new userIDs and itemIDs cannot be well-trained before serving them online. Second, the design philosophy of IDRec in essence deviates from the core principle of modern \textit{foundation} models~\cite{bommasani2021opportunities} in the deep learning community --- allowing the pre-trained parameters to be adapted to multiple downstream tasks. 
This is because IDRec usually requires either shared data or overlapped IDs (i.e., userIDs and itemIDs) 
to realize cross-domain recommendation~\cite{yuan2020parameter,yuan2021one,li2009transfer,hu2018conet,fu2023exploring, zhu2021personalized}. \textcolor{black}{However, achieving such cross-domain recommendation often proves impractical due to concerns related to data privacy and overlap rates between different systems. For instance, platforms like TikTok may not share their userIDs or videoIDs with platforms like YouTube.}

\textcolor{black}{To overcome these limitations, an intuitive approach is to abandon the use  of userID and itemID features, particularly itemID.\footnote{This is because the userID can be represented by the itemIDs that the user has interacted with, as seen in most sequential recommendation models.} Instead, we can leverage the multimodal content of items to represent them~\cite{yuan2023go,fu2023exploring,hou2022towards,ramesh2021zero,wang2022transrec,li2023exploring}. We refer to this approach as MoRec~\cite{yuan2023go}.  For example, if the item is a news article or text, we can utilize BERT~\cite{devlin2018bert} or RoBERTa~\cite{liu2019roberta} to represent it. If the item is an image, we can employ ResNet~\cite{he2016deep} or Vision Transformer (ViT)~\cite{dosovitskiy2020image} to represent it. By representing items with modality features, recommendation models can naturally possess  transfer learning capabilities  across domains and systems. This paradigm, called TransRec, shares similarities with universal models in natural language processing (NLP) \& computer vision (CV).}

\textcolor{black}{However, TransRec models have received less attention and success than NLP and CV. So far, the RS community does not have a downloadable TransRec model, such as on platforms like HuggingFace, whose pre-trained parameters can be directly applied to other recommendation datasets,  akin to the usage of BERT in NLP. There are several  challenges for the successful deployment of TransRec models in practical applications. One major challenge  is the strong establishment and dominance of the IDRec paradigm, which has represented state-of-the-art baselines for over a decade, especially in non-cold start scenarios. TransRec or MoRec that rely solely on multimodal features often struggle to outperform these IDRec models,\footnote{A typical example is that to date, except for the recent M6-Rec~\cite{cui2022m6}, almost no real-world  online recommender systems have explicitly claimed that they have completely abandoned the itemID feature.} particularly in past years when highly expressive modality encoders, such as large BERT or top-performing ViT, were not yet available. This situation has seen some progress in recent months, as evidenced by recent literature~\cite{yuan2023go,li2023exploring}, which confirms that even in non-cold start and warm item scenarios, itemID features can be replaced  with an advanced multimodal encoder.}

Another challenge for the TransRec paradigm is the scarcity of large-scale multimodal pre-trained recommendation datasets and diverse downstream  datasets. While Microsoft provides MIND ~\cite{wu2020mind}, a high-quality news recommendation dataset, it lacks diverse downstream datasets for evaluation and does not include raw image features. Several e-commerce datasets, such as Amazon\footnote{http://snap.stanford.edu/data/web-Amazon.html}, Yelp\footnote{https://www.yelp.com/dataset/download}, and GEST~\cite{yan2022personalized}, can provide raw image features, \textcolor{black}{but the items in these datasets often revolve around simple objects (as depicted in Figure~\ref{fig:ninerecdataexample}) or have limited visual diversity,\footnote{For example, item images in Yelp and Gest are mostly about food and restaurants.} making them less suitable as pre-training datasets for general or semantically richer images. More importantly,  these datasets are intuitively less optimal  for studying pure modality (visual or textual) recommendations as user intent in e-commerce datasets is heavily  influenced by other factors, such as price, sales, brand, location, and most importantly, the user's actual purchase needs.} 
% That is also the reason why visual features are mostly used as auxiliary information rather then the main features for in most previous RS literature.

\textcolor{black}{In this paper, our primary goal is to solve the dataset challenges for the community, and subsequently provide reliable benchmarks.} Specifically, we introduce NineRec, a TransRec dataset collection consisting of a very large source domain dataset (with 2 million users, 144 thousand items, and 24 million user-item interactions) and nine diverse target domain datasets (including five from the same platform with different scenarios and four from different platforms). Each item is represented by an original descriptive text and a high-resolution cover image.
% and an itemID, and an itemID for baseline or other research purposes.
\textcolor{black}{To the best of our knowledge, NineRec is the first large-scale and highly diverse datasets for  streaming content recommendation, encompassing various types of raw content, including  short videos, news, and images. One distinctive characteristic  of our NineRec datasets is that user watching intent in streaming media can be primarily  inferred from the visual
appearance of items, with minimal influence from non-visual factors like price in e-commerce recommendation datasets or distance in location recommendation datasets. From this perspective, NineRec is a more ideal dataset for studying multimodal  content-focused recommendation.
% \footnote{Existing image recommendation datasets, such as Amazon, Yelp, and GEST, usually treat visual features as auxiliary information, since user intent is influenced by  many too important features.}
We then report several representative TransRec baselines for visual and text recommendation tasks on the source dataset and nine target datasets by replacing ID embeddings with advanced
modality encoders.}
% We emphasize that learning item representation from raw modality features using fine-tuned modality encoders is computationally  very expensive but more effective than using pre-extracted features.
% \footnote{Please note that collecting such a large-scale dataset and running computationally intensive baselines is expensive in terms of computation, time, and financial resources. Conservatively, it can be 100 times more expensive than typical IDRec research. In total, we have spent approximately \$300,000  on GPU services (using NVIDIA A100 80G memory) and over 10 months collecting data, as well as over a year running all related experiments  (see Appendix Table 5). }
Our rigorous empirical studies on NineRec have uncovered several interesting findings.
% Last, we scale TransRec models and data, both up and down, and provide useful yet unknown findings for the community.
To facilitate future research, we release our code, datasets, benchmarks, and leaderboard.
Beyond this, we envision NineRec as a  useful dataset for the NLP \& CV  researchers, who can use recommendation as a  downstream task to evaluate the generality of new image/text encoders. Given this, NineRec helps unify the fields of RS, NLP \& CV.

\section{NineRec Dataset Suite}

\subsection{Dataset Summary}
\label{datasetsummary}

To facilitate the TransRec research, we curate a suite of benchmark datasets which comprise a large-scale source domain dataset from Bili and nine different downstream target domain datasets, namely, Bili\_Food, Bili\_Dance, Bili\_Movie, Bili\_Cartoon, Bili\_Music, QB, TN, KU and DY.\footnote{Bili: https://www.bilibili.com/; QB: https://browser.qq.com/; TN: https://news.qq.com/; KU: https://www.kuaishou.com/new-reco; DY: https://www.douyin.com/. } Bili, KU and DY are three most famous short-video RS platforms in China, where each item is a short-video\footnote{Videos with play time longer than 10 minutes are not collected.}, while TN and QB are two  large streaming recommendation platforms where an item can be either a news article, short-video or an advertisement. Each item in all of the above datasets contains a textual description and an image cover. Each positive user-item interaction is either a thumb-up or a comment, which is a  strong signal for user preference.   Note that we do not retain the contents of comments as we consider the textual  description (i.e. title) of an item to be more representative than comments or reviews. 
% However, throughout this paper, each item is represented by an original descriptive text and high-resolution cover image (also with an itemID --- not our focus).

We provide two source datasets: Bili\_500K and Bili\_2M, where 500K and 2M stand for 500 thousand and 2 million users, respectively.  Bili\_500K is a subset of  Bili\_2M.
Their collection strategies are similar and will be given in the following subsection.
The user-item interactions for the source datasets were collected from both the main channel and 20 vertical channels, resulting in a highly diverse range of item categories. In contrast, the Bili\_* datasets in the target domain were collected from five vertical channels (excluding those from the source datasets) of the Bili website, where the items on each channel page are mostly from the same category. For example, items in Bili\_Food are mainly about food and cooking, while  items in Bili\_Music are  music videos.
Bili\_2M and Bili\_* have no overlapped  items or user-item interactions.
There might be a small number of users who visited both the main and these vertical channels. But we do not consider the overlapping users as it is not our focus. 

\begin{figure}[htbp]  
    \centering
    \includegraphics[width=0.9\columnwidth]{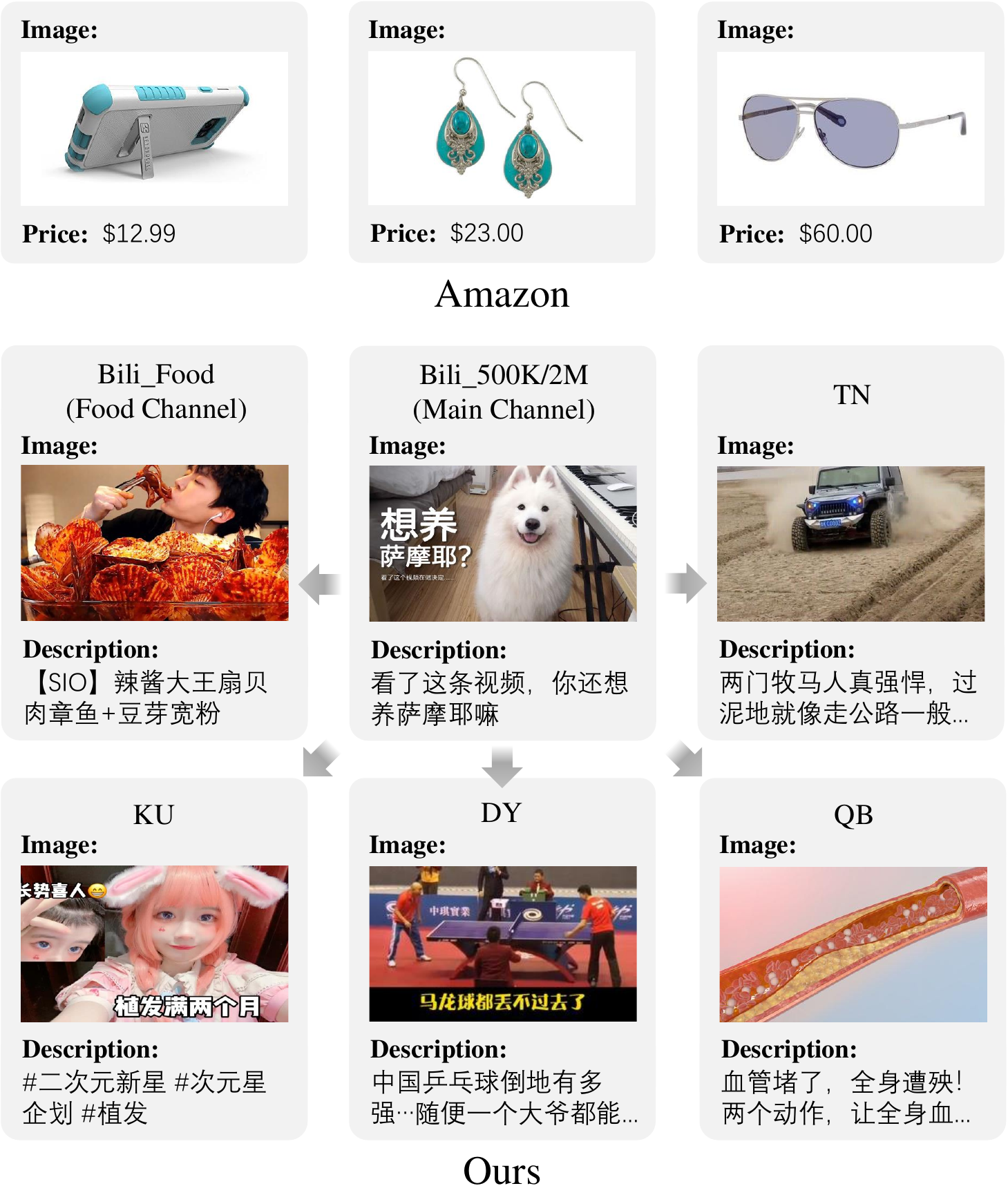}\\
      \caption{\textbf{Image cases of NineRec vs. Amazon.
      } (a) Compared to Amazon, the images in NineRec are more abstract and semantically enriched; (b) NineRec supports cross-platform recommendation;  Bili, TN, KU, DY, QB are different recommender systems;
      % arrows are the directions of transfer learning.
      (c) User intent in Amazon is largely influenced by item price, which cannot be achieved by learning only appearance or visual features.} 
    \label{fig:ninerecdataexample}
    % \vspace{-0.7cm}
\end{figure}
\subsection{Dataset Construction \& Analysis}
% During the data collecting process, we crawl 
The data collection process lasted approximately 10 months from September 2021 to July 2022. Taking Bili source as an example, we collected short videos from more than 20 channels (including main channels with various categories). By frequently requesting the webpage, we could collect about 1000-2000 videos per channel. We then went to the pages of all these videos, which often contained many links to other videos. In each page, we randomly selected 3-5 videos. We did this many times. Then, we merged all videos and removed duplicates.  As for user feedback, we went through all pages of collected videos
and collected the public comments (including the bullet-screen comment) for each video, and ensured that each video has at most 3,500 user comments. We crawled the comment data page by page, and the more comments we collect, the longer it will take. We only recorded the latest pair of user-video interactions even though the user might commented on a video multiple times. 

% Again, for the source dataset, we randomly collect videos from the main channel of Bili with various categories, while for downstream Bili\_*, we collect videos  from only the * channel. 
For the source dataset, we first got Bili\_500K after a few months of data collection, then aggregated all the data and removed users with less than 10 behaviors. We then proceeded to crawl more data over several months and aggregated all existing data, but only removed users with less than 5 behaviors, resulting in Blil\_2M.
Due to  the slow collection time of Bili\_2M, we conducted the major experiments on Bili\_500K.
% In fact, we have collected around 200K videos with over 20M users in total.
% But after removing cold users, only a few 
% We then simply filter out cold users as mentioned above.
% Finally, we have Bili 500K and Bili\_2M.
Similarly, we collected data from 5 vertical channels of Bili and
 four other platforms, i.e. QB, TN, KU and DY. 
For these downstream datasets, we maintain the same data collection procedure.
% users that have at least 5 interactions. 
In this paper, we only keep the cover image and title description to represent an article or a short-video instead of the news contents or original videos. After basic processing, we recruited five students who manually checked the quality of images and text and removed about 1\% of items that were poor quality (e.g. images with just black background, image-text mismatches, overly sensational text descriptions, etc.). 
We preserve main properties of these datasets without more pre-processing since they might be important for other research.
The statistics of the final datasets are in Table~\ref{Datasets Statistics}. 
\textcolor{black}{The source dataset Bili\_2M contains 144,146 raw images with an average resolution of 1920x1080. All downstream Bili\_* datasets have the same resolution, and the four cross-platform datasets have at least 300x400 resolution, meeting the basic requirements of popular vision encoders. The average word length of all these datasets falls within the range of 16-34.}
% Note our downstream datasets are smaller than the source dataset, but not small compared to datasets used in many well-known recommendation literature~\cite{rendle2012bpr,he2017neural,wei2021model}.

\begin{figure*}
\centering
\subfloat[Bili\_2M]{
    \begin{minipage}[htbp]{0.48\columnwidth}
        \includegraphics[width=1\columnwidth]{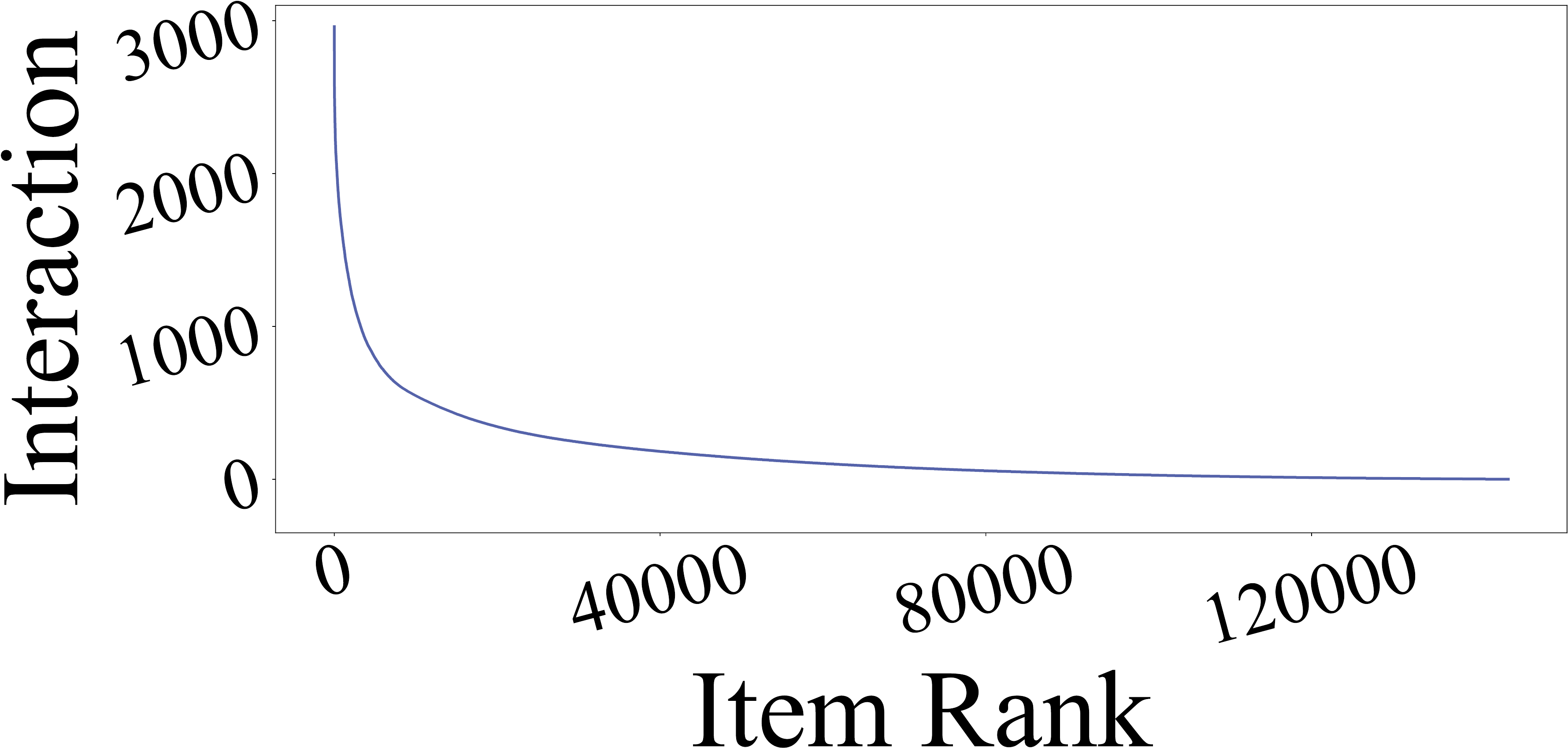}\vspace{1mm} \\
            \includegraphics[width=1\columnwidth]{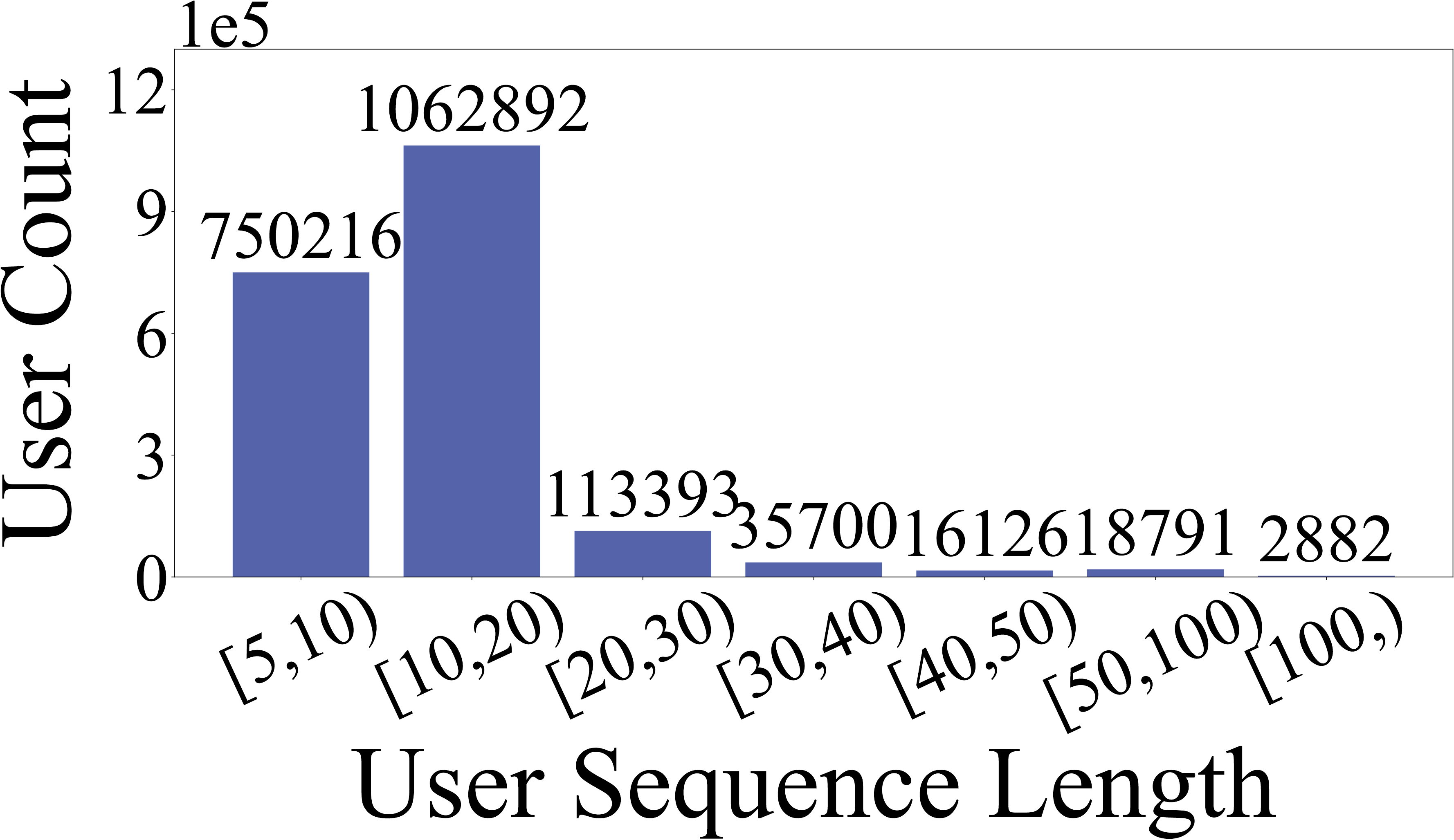}\vspace{1mm} \\ 
            \includegraphics[width=1\columnwidth]{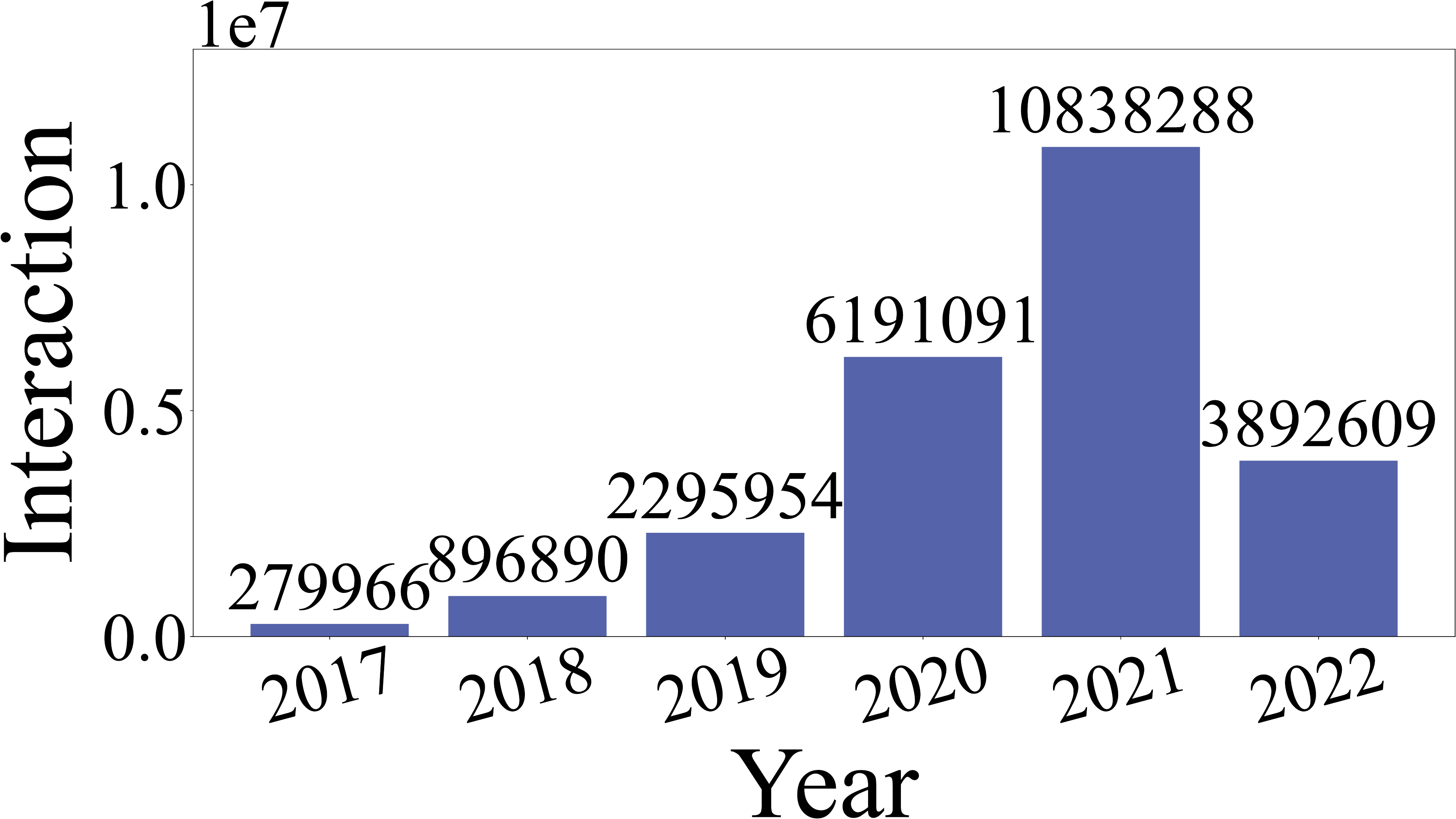}\vspace{1mm}
    \end{minipage}
    \label{fig:Data distribution a}}
\subfloat[DY]{
    \begin{minipage}[htbp]{0.48\columnwidth}
        \includegraphics[width=1\columnwidth]{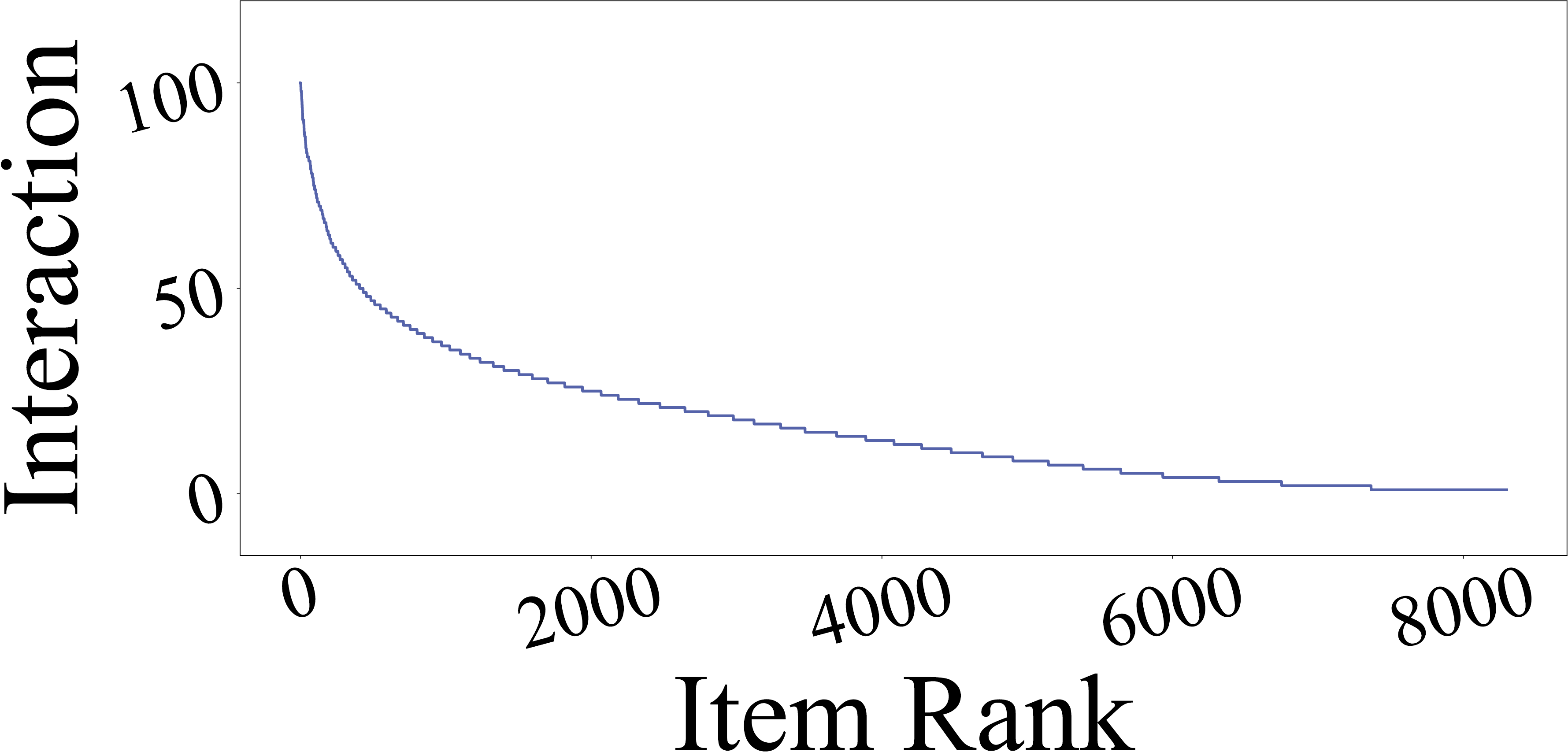}\vspace{1mm} \\
            \includegraphics[width=1\columnwidth]{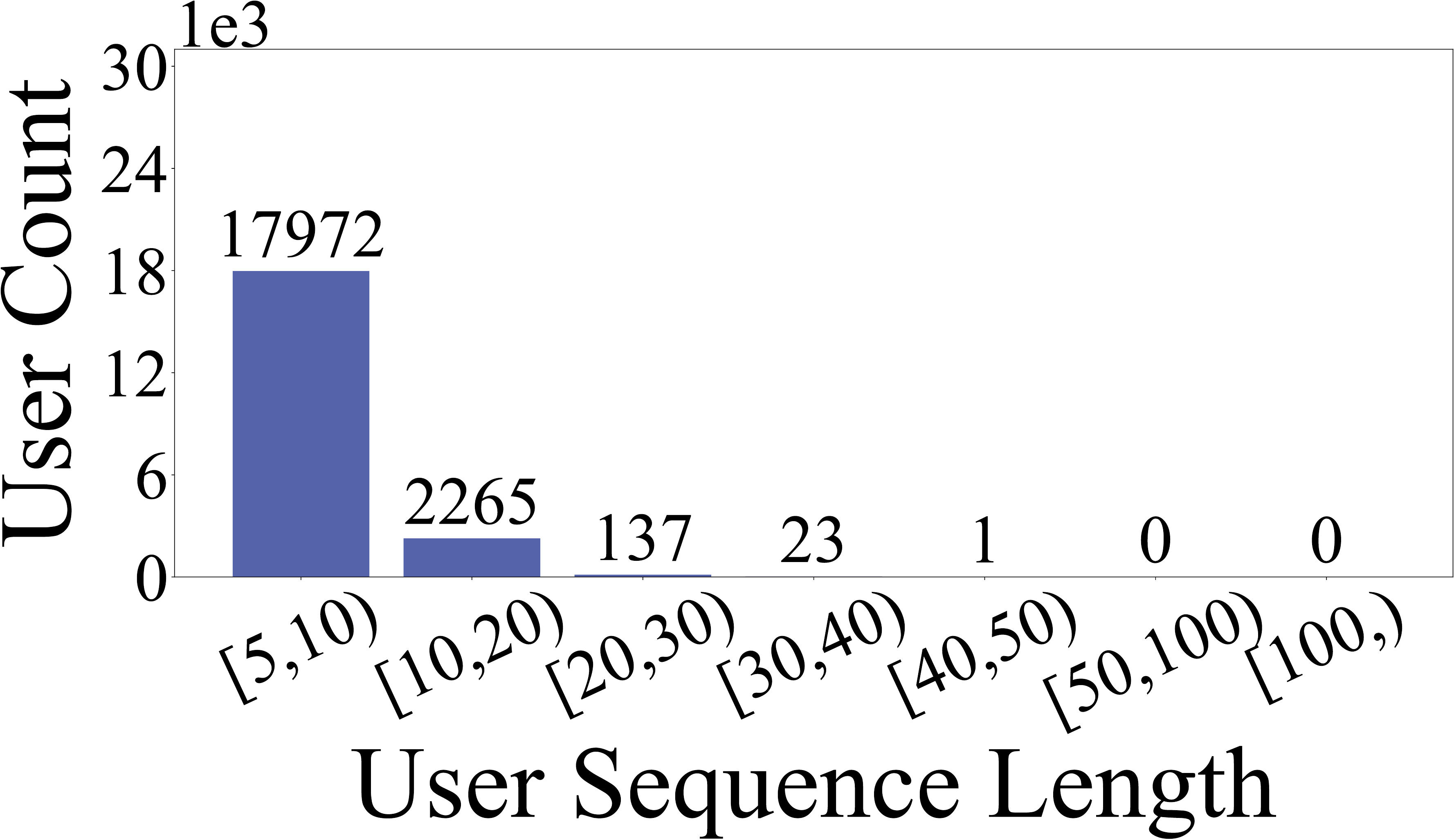}\vspace{1mm} \\
            \includegraphics[width=1\columnwidth]{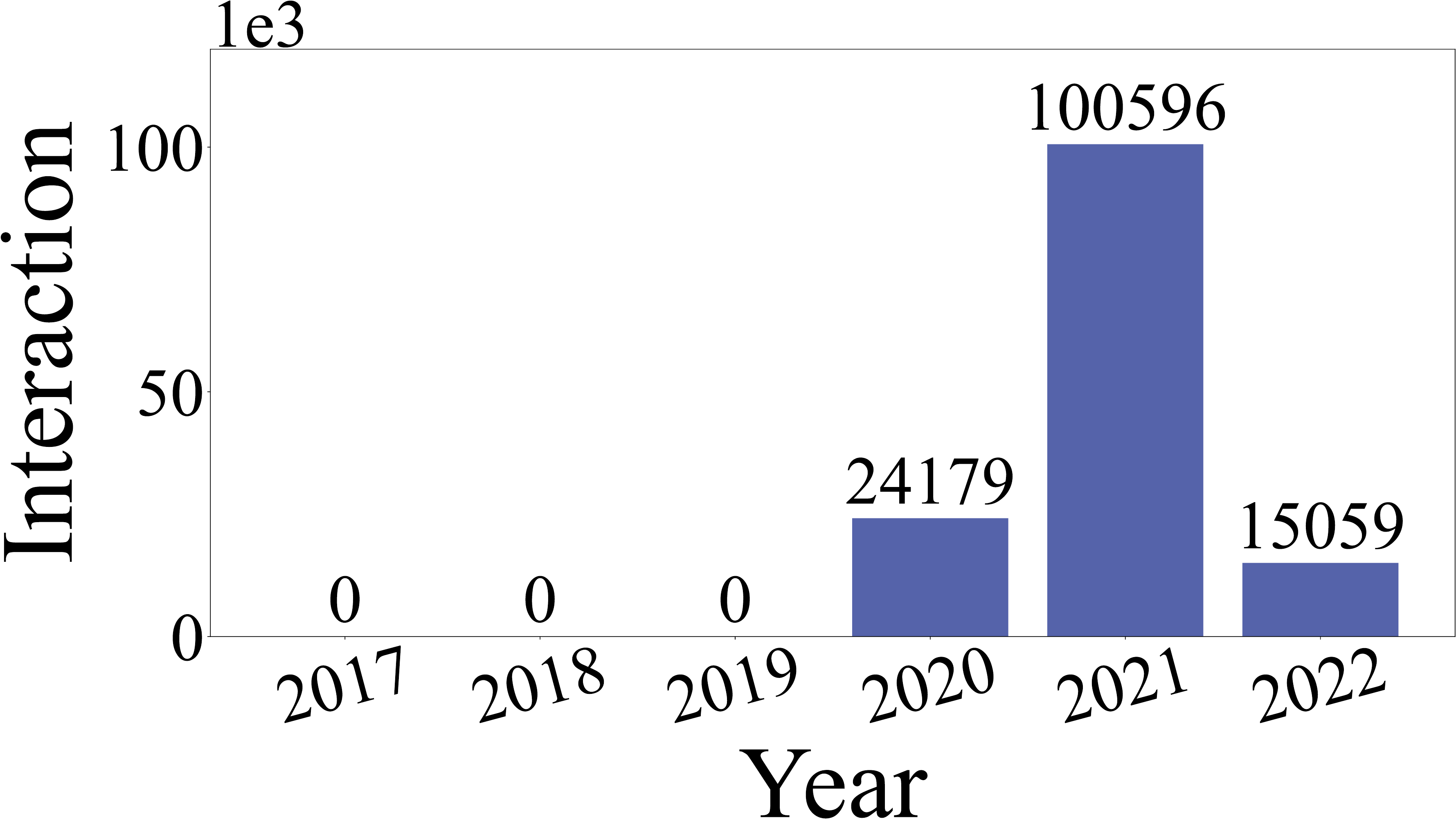}\vspace{1mm}
    \end{minipage}
    \label{fig:Data distribution b}}
\subfloat[KU]{
    \begin{minipage}[htbp]{0.48\columnwidth}
        \includegraphics[width=1\columnwidth]{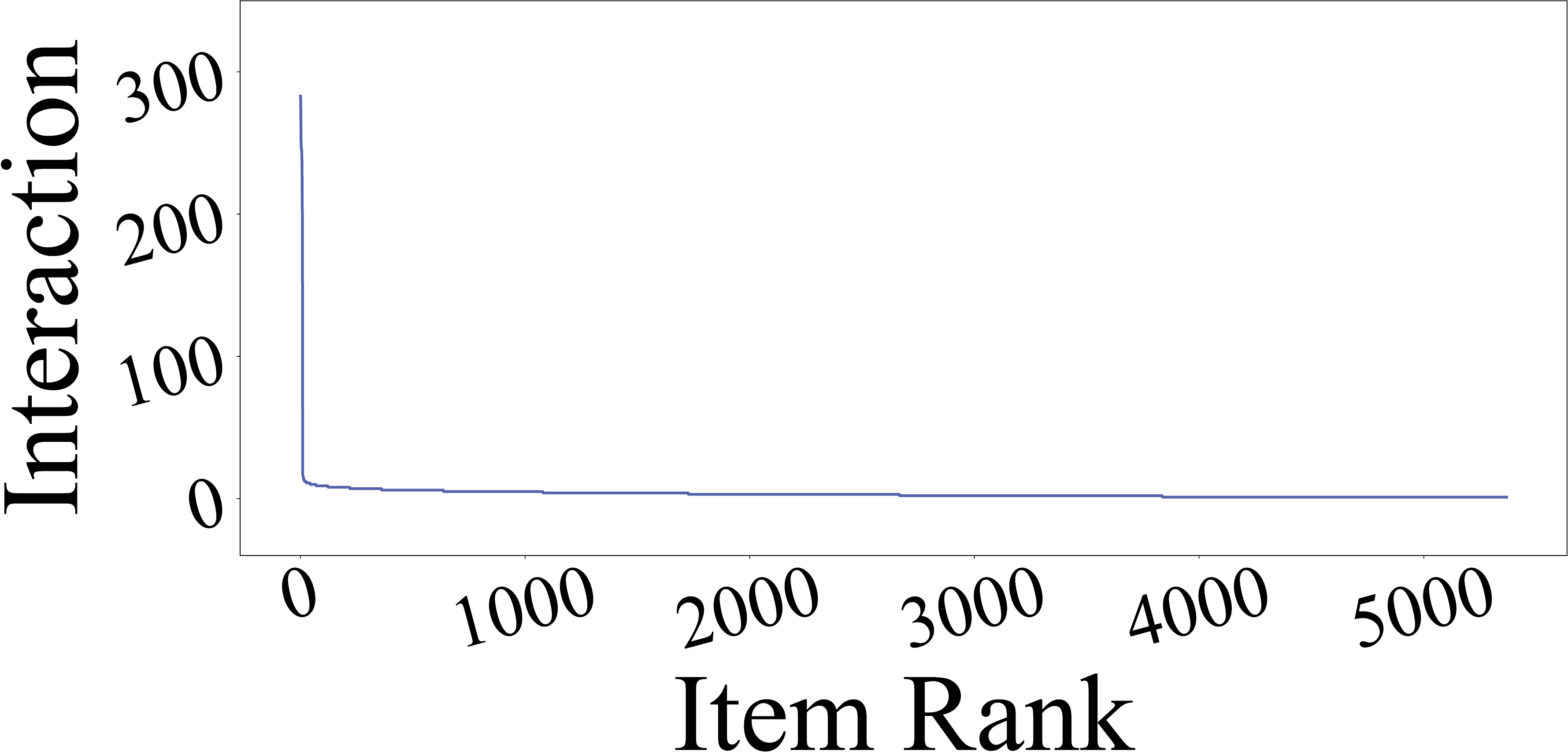}\vspace{1mm} \\
            \includegraphics[width=1\columnwidth]{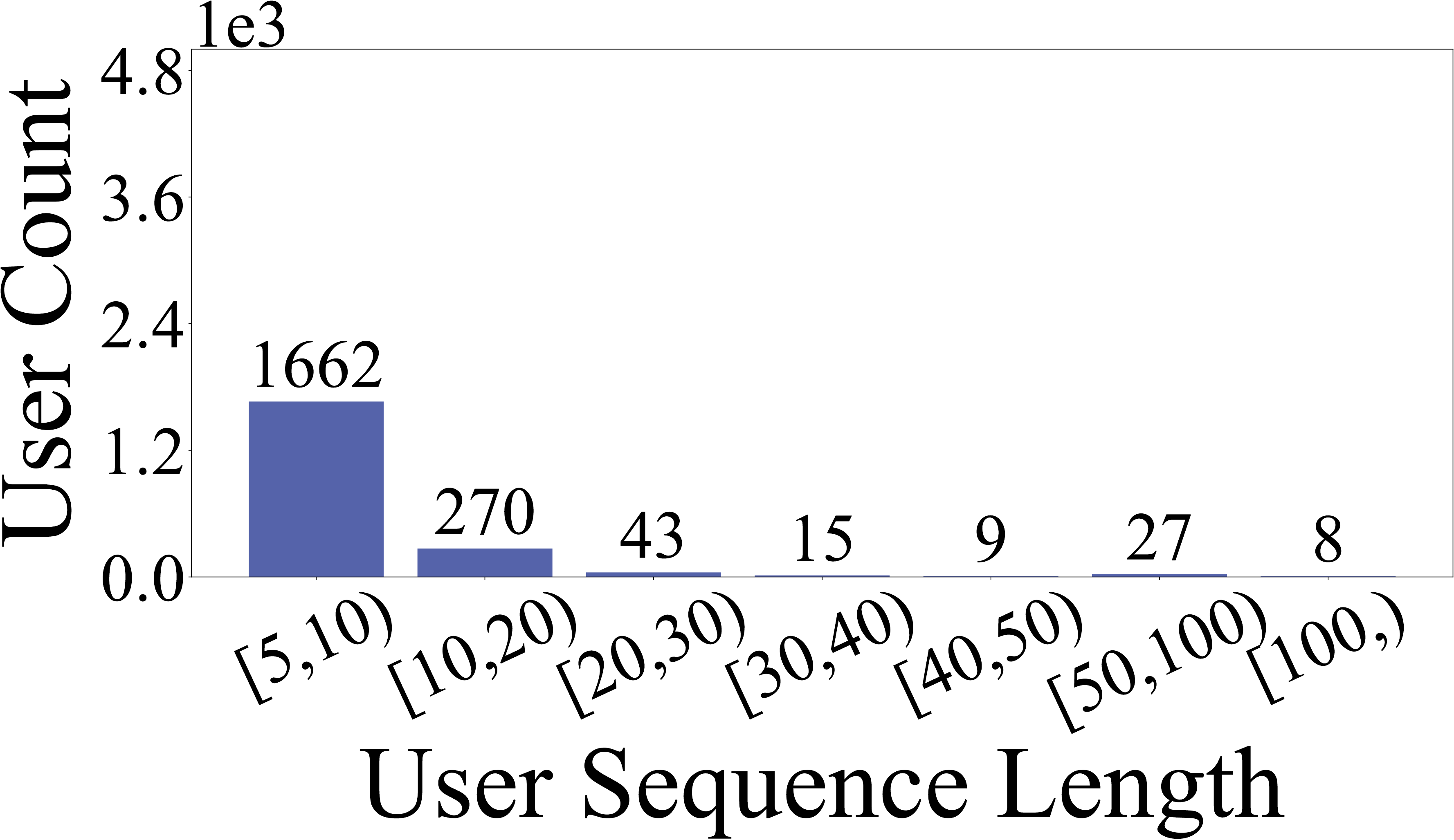}\vspace{1mm} \\
            \includegraphics[width=1\columnwidth]{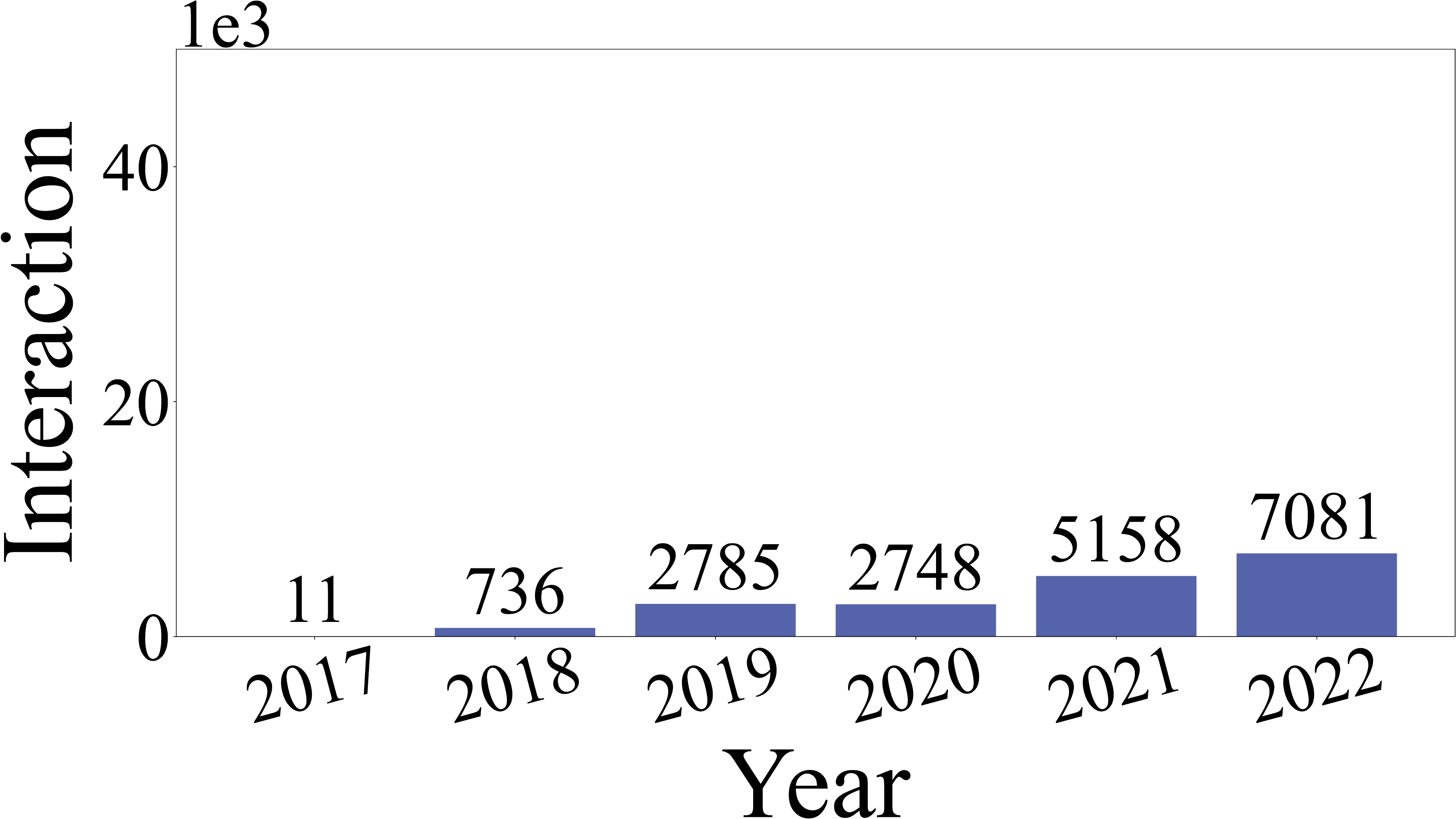}\vspace{1mm}
    \end{minipage}
    \label{fig:Data distribution c}}
\subfloat[QB\&TN]{
    \begin{minipage}[htbp]{0.47\columnwidth}
        \includegraphics[width=1\columnwidth]{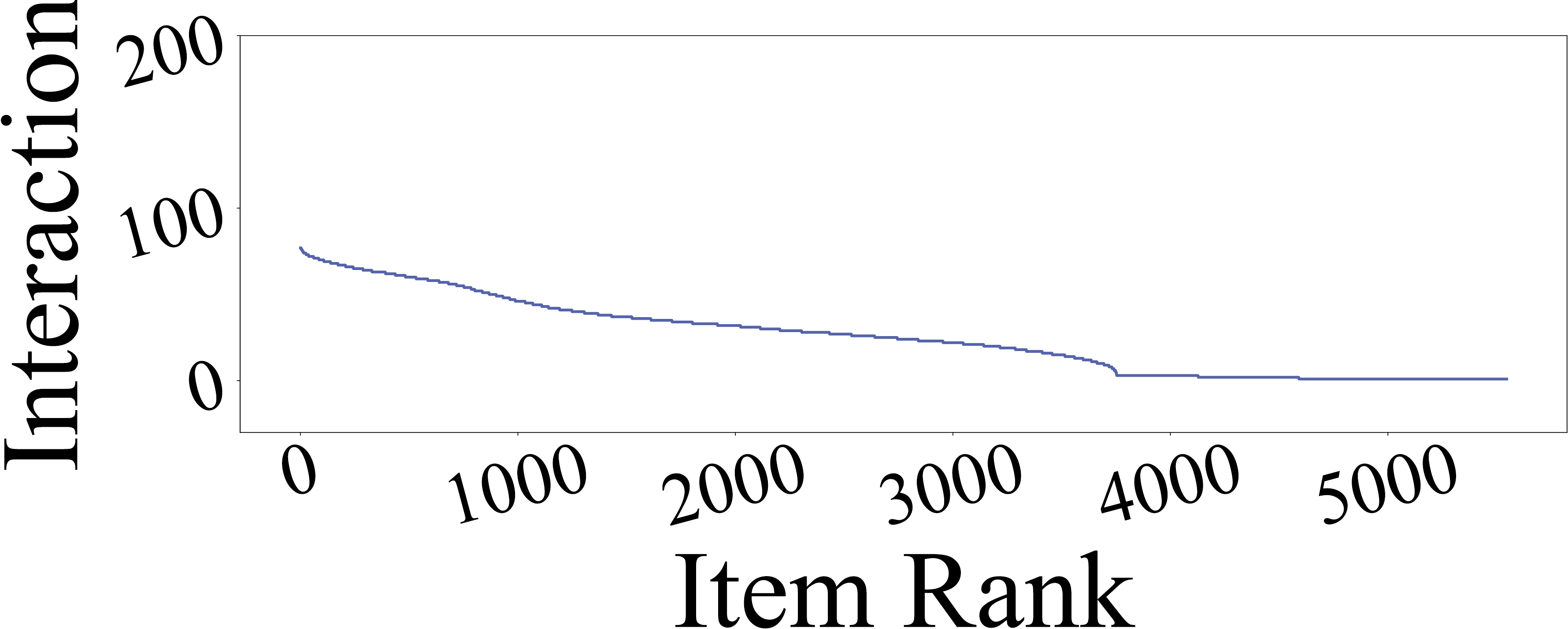}\vspace{0.5mm} \\
            \includegraphics[width=1\columnwidth]{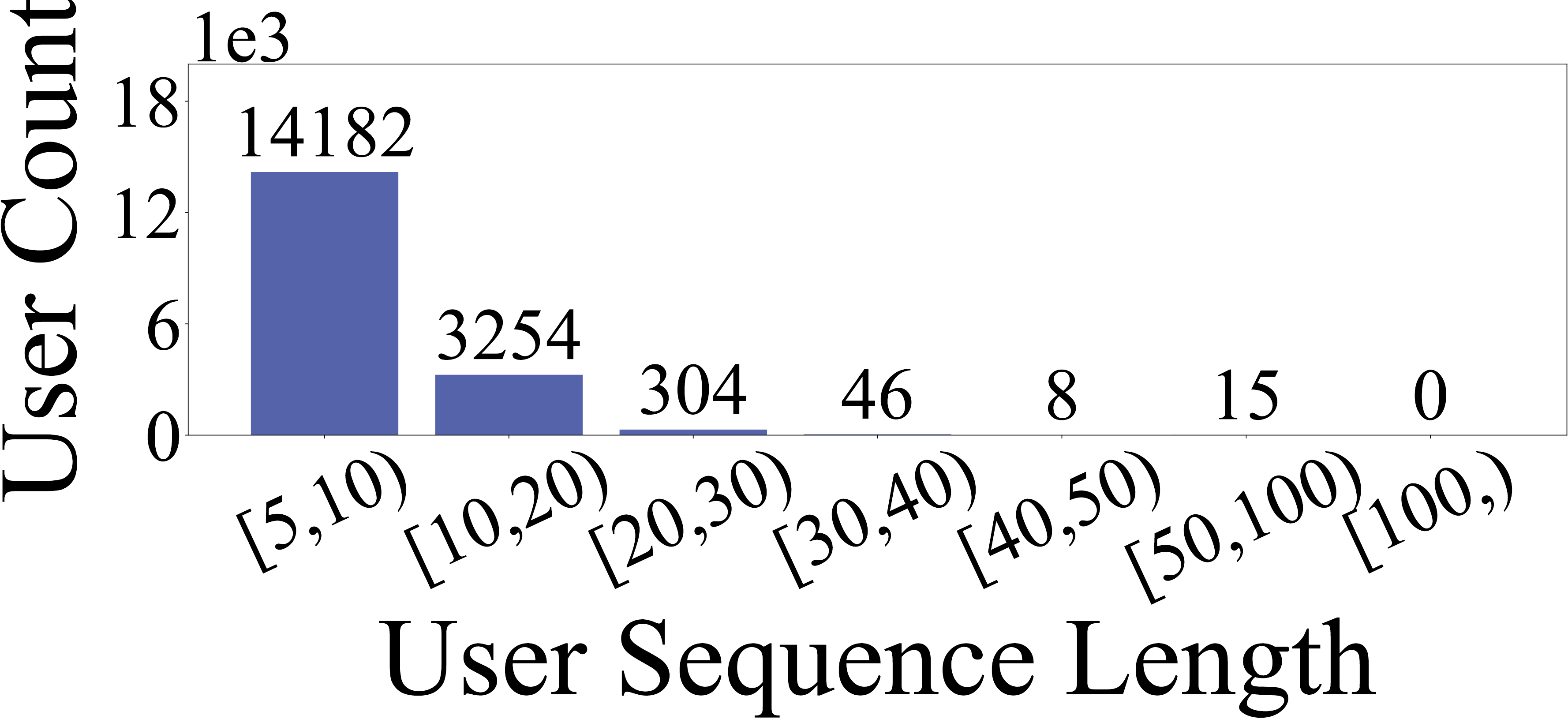}\vspace{0.5mm} \\
        \includegraphics[width=1\columnwidth]{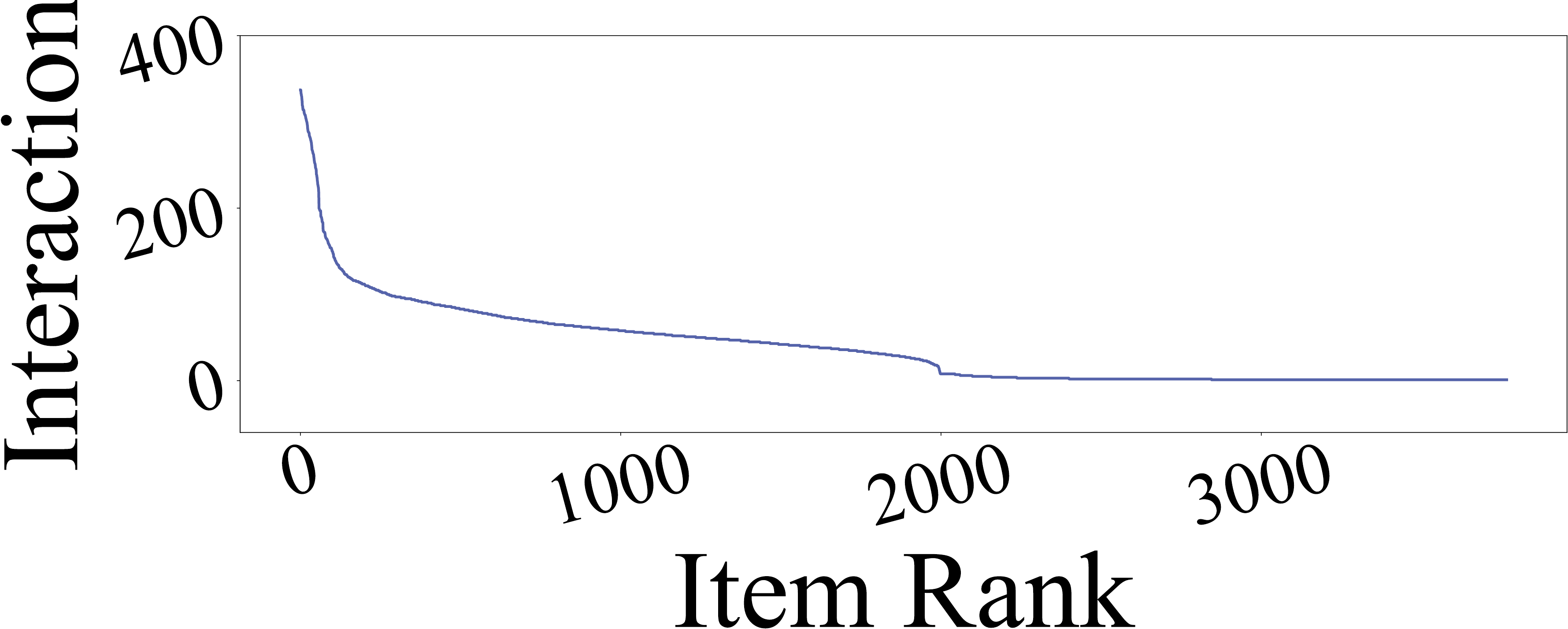}\vspace{0.5mm} \\
            \includegraphics[width=1\columnwidth]{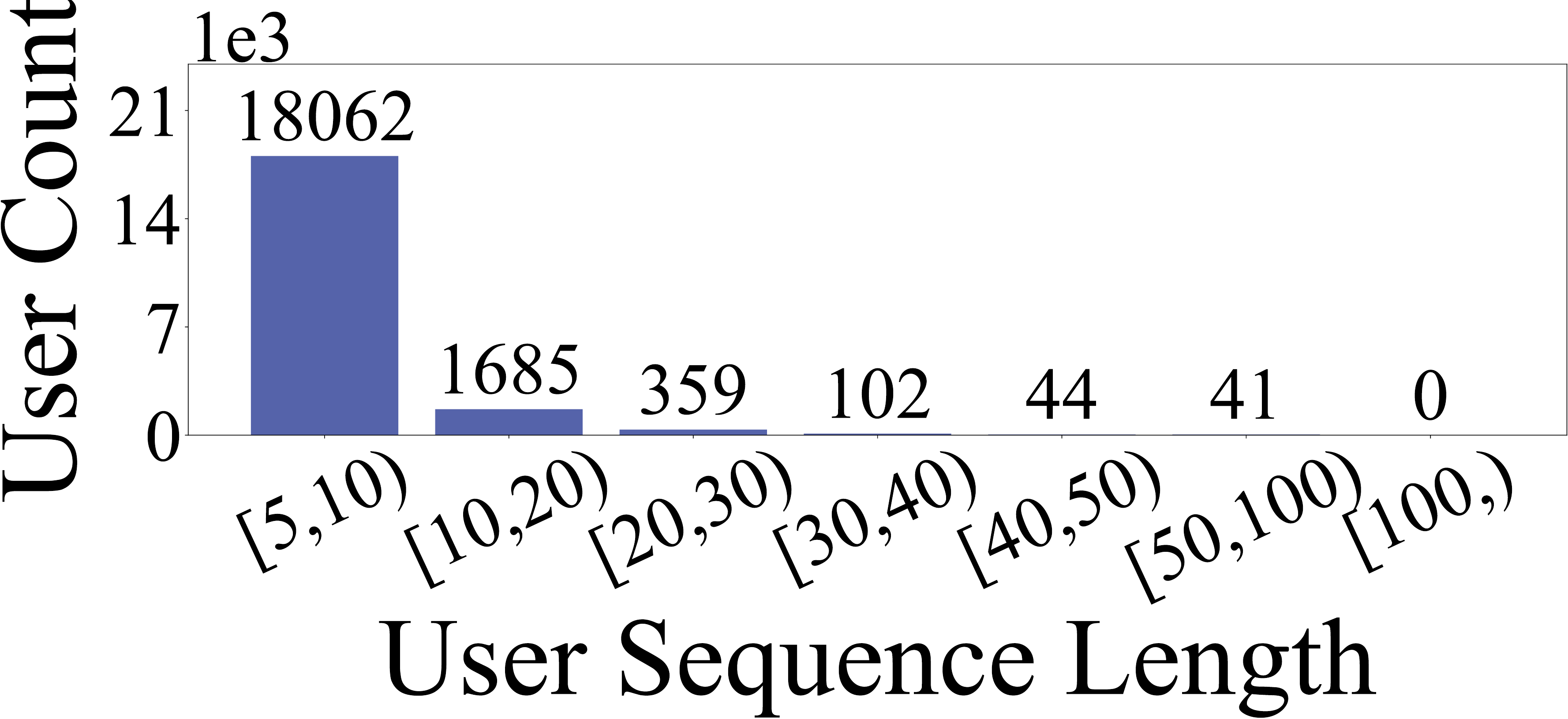}\vspace{0.5mm}
    \end{minipage}
    \label{fig:Data distribution d}}
\caption{Dataset details. Top:  item popularity distribution; Middle:  user interaction length distribution;  Bottom: the occurring time of user-item interactions. See more in Appendix Figure 2.}
\label{tab:dataDetails}
\vspace{-0.3cm}
\end{figure*}

Other statistics of the NineRec dataset are given in  Figure~\ref{tab:dataDetails}.  
First, we can see the item distributions of all datasets typically follow the long-tail distribution, which is widely observed in much prior literature~\cite{yuan2022tenrec}. 
% The most popular item has 2961 comments, while the coldest item has only 1 comment.
Second, we can see that the number of user interactions are mainly in the range of [5,100], where [5,20) is the majority. We therefore run TransRec experiments by setting the maximum user sequence length to 23 and padding with zeros when user interaction is insufficient.
% (the last two interactions are used for validation and testing, the remaining 21 is used in the training set).
Third, the interactions of Bili\_2M occurred mainly between 2017 and 2022.

\begin{table*}[ht]
\caption{Datasets. The sequential order of actions are retained. \#Description (Chinese) and \#Description (English) is the average word length of text description in Chinese and English, respectively.
% A positive action is either a like or a comment on an item. 
% \textcolor{red}{Image Resolution denotes the averaged image resolution. 
% }
}
\label{Datasets Statistics}
% \vskip 0.15in
\begin{center}
\begin{small}
% \begin{sc}
\begin{tabular}{l r r r r r r p{1.8cm}<{\centering} p{1.8cm}<{\centering} p{1.8cm}<{\centering} p{1.8cm}<{\centering} }
\toprule
\multirow{2}{*}{Dataset}  &&\multirow{2}{*}{\#Users}  &&\multirow{2}{*}{\#Items}  &&\multirow{2}{*}{\#Actions.}  &\multirow{2}{*}{Sparsity}  &\multirow{2}{*}{\makecell[c]{Image \\ Resolution}}  &\multirow{2}{*}{\makecell[c]{\#Description \\ (Chinese)}}  &\multirow{2}{*}{\makecell[c]{\#Description \\ (English)}}  \\ % &\multirow{2}{*}{\makecell[c]{\#Text\_char \\ (en)}}
&&&&& \\
\midrule
Bili\_500K    &&500,000   &&138,033 &&7,845,805  &99.99\%  &1920x1080  &21.91  &19.28  \\ %&69.93  \\ 
Bili\_2M      &&2,000,000 &&144,146 &&24,497,157 &99.99\%  &1920x1080  &21.90  &19.27  \\ %&69.93  \\
\midrule
Bili\_Food    &&6,549     &&1,579   &&39,740     &99.62\%  &1920x1080  &24.91  &22.15  \\ %&79.68  \\
Bili\_Dance   &&10,715    &&2,307   &&83,392     &99.66\%  &1920x1080  &19.62  &18.00  \\ %&65.86  \\
Bili\_Movie   &&16,525    &&3,509   &&115,576    &99.80\%  &1920x1080  &24.93  &21.99  \\ %&78.70  \\
Bili\_Cartoon &&30,300    &&4,724   &&215,443    &99.85\%  &1920x1080  &19.26  &16.20  \\ %&58.53  \\
Bili\_Music   &&50,664    &&6,038   &&360,177    &99.88\%  &1920x1080  &21.60  &19.70  \\ %&68.94  \\
KU            &&2,034     &&5,370   &&18,519     &99.83\%  &360x640    &25.00  &19.13  \\ %&74.16  \\
QB            &&17,722    &&6,121   &&133,664    &99.88\%  &496x280    &25.26  &20.74  \\ %&82.62  \\
TN            &&20,211    &&3,334   &&122,576    &99.82\%  &496x280    &26.76  &22.12  \\ %&88.36  \\
DY            &&20,398    &&8,299   &&139,834    &99.92\%  &300x400    &34.46  &26.92  \\ %&101.06  \\
\bottomrule
\end{tabular}
% \end{sc}
\end{small}
\end{center}
% \vskip -0.1in
\end{table*}

\subsection{Copyrights and Privacy}

\textcolor{black}{
In this paper, we strictly adhere to privacy protection measures by collecting only public user behaviors. We have not collected any private user behaviors such as clicks or watching time. Furthermore, the item content we collected, including thumbnails and descriptive texts, is itself freely accessible on the platform's webpages without any limitations. User account IDs and item IDs are also publicly displayed on these platforms. Despite that, we have taken precautions to anonymize them to mitigate potential attacks. These anonymous item IDs can be used to construct item URLs using our mapping algorithm. We have implemented the mapping algorithm and URL construction within our download software, ensuring that researchers can only access the data through our downloader.} 

\textcolor{black}{
Regarding copyright concerns, we do not directly provide item cover images. Instead, we offer the downloader tool that allows data users to directly download content from the respective platforms by parsing the provided URLs (the ULRs are embedded in the downloader and are not exposed to the public). This approach ensures that copyright issues are not involved and is a widely adopted practice in academic literature~\cite{zeng2022tencent,nielsen2022mumin}.}
\textcolor{black}{Additionally, for videos that may be expired or unavailable, our downloader automatically locates them in a backup directory to ensure permanent access and downloadability.}

\subsection{Comparison to Existing Datasets}
The datasets used for the TransRec research can be categorized into three types:  
datasets with overlapped categorical IDs~\cite{yuan2022tenrec,gao2022kuairec,yuan2021one}, datasets with pre-extracted features by multimodal encoder~\cite{he2016vbpr, He_2016,wu2019hierarchical,feige2019invariantequivariant,wei2023multi}, and datasets with raw modality features. 
While there are numerous public datasets available for the former two types, there are very few for the latter type.
MIND (for text RS)
, Amazon (for product RS), Pinterest\footnote{The final released version of Pinterest has only 46,000 users in total and 37,000 items (more than 10  clicks) and no timestamps. }~\cite{geng2015learning} (for image RS), WikiMedia~\cite{moskalenko2020scalable} (for image RS),  GEST \& Yelp (for food recommendation), have raw modality features.
Among them, MIND, Amazon, Yelp and GEST (a.k.a. Google Restaurants) have a  large scale. 
However, MIND does not have downstream datasets. Though items in Amazon have category information, they are more like cross-category recommendation rather than a strict cross-domain recommendation as there is not a clear concept of domain~\cite{zhu2021cross}.\footnote{\textcolor{black}{In the Amazon dataset, recommendations for different categories of products are likely to be based on a unified recommendation algorithm. There are some differences compared to true cross-domain or cross-platform recommendation in the strict sense.}} 
By contrast, NineRec enables both cross-domain and cross-platform recommendation as the target data of NineRec is collected from either different recommendation channels or different systems. \textcolor{black}{Detailed comparison with related  datasets is shown in Appendix Table 10.} 
  
% , including both News and short-video images.
% Amazon is suited for the cross-domain recommendation, but does not support the cross-system/platform recommendation (since all data is from the same platform).
Another drawback of Amazon  is that its images are mainly about single products (e.g. shoes, books, food, electronic products),
% \footnote{Note that Amazon’s user purchase behaviors are largely influenced by item price, not just by the item appearance.}, 
so models trained on them cannot reflect the real performance in other more complex and practical image scenarios (see Figure~\ref{fig:ninerecdataexample}).  Similarly, Yelp and GEST also suffer from the image diversity issue since most item images are about food and restaurants. 
% A clear  comparison of them is in Appendix Figure 1.
% A preprint paper~\cite{yan2022personalized} just released a large dataset called GEST for food recommendation. 

\textbf{Novelty and Limitations.} \textit{First}, while there are several large-scale public datasets available with raw modality features, their visual or semantic diversity is limited, making them unsuitable as pre-training datasets. For an ideal pre-training model, it is crucial to learn from data with good diversity. In contrast, the source dataset of NineRec contains items from over 20 different video channels, providing a much wider range of visual diversity.
In addition, NineRec provides nine target tasks which support both cross-domain and cross-platform recommendation tasks.
\textit{Second}, user behavior observed in existing datasets, such as Amazon and GEST, is not primarily driven by item appearance or modality features. \textcolor{black}{Instead, it is influenced by a myriad of other significant factors, including price, sales, brand, location, and the user's actual purchase needs. 
That is, user preference cannot be mostly learned from visual features. For instance, when a user purchases baby milk powder on Amazon, it is more likely due to the quality and brand of the product rather than its image characteristics.
In contrast, the appearance features in our NineRec dataset which are all collected from content-sharing platforms are \textit{reasonably more important}   signals to attract user viewing or clicking actions. Specifically, in the context of short videos and information streams, users tend to passively accept recommendations from the platform, rather than having a specific intent as seen in e-commerce scenarios. Furthermore, a user's decision of whether to click or watch a video is intuitively influenced by the attractiveness of the thumbnail and title.}
% \footnote{The underlying assumption here is that if a user leaves a comment or assigns a like, it can be reasonably inferred that they have viewed the item. }
Therefore, from this persepctive, NineRec is a more ideal dataset for conducting pure modality feature based recommendation research. \textcolor{black}{Besides, we believe that the recommender systems community requires datasets not only from e-commerce scenarios but also from short videos and information stream contexts. These are highly distinct application domains and it is crucial to take them into consideration when developing recommendation algorithms.}

\textcolor{black}{It is worth noting that NineRec also has some limitations: (1) certain user interactions may be influenced by clickbait video thumbnails and titles; (2) the NineRec dataset is sourced from real-world recommendation platforms, resulting in a data distribution that may contain exposure and popularity biases. These factors can potentially impact the fairness of recommender systems. However, we have retained the original data distribution of NineRec to foster research on diversity to the greatest extent possible.}

\section{Related work of TransRec}
\label{relatedworkonmodels}
\textit{Foundation} models~\cite{bommasani2021opportunities}, trained on broad data at scale and adaptable to a diversity of downstream tasks, have  shifted the research paradigm of the AI community from task-specific models  to general-purpose models. A broad spectrum of foundation models have been developed in recent years. Amongst them, BERT, RoBERTa, GPT~\cite{radford2018improving,radford2019language,brown2020language}, and  ChatGPT\footnote{https://openai.com/blog/chatgpt/} are renowned  for
encoding and generating textual data,  ResNet, ViT, Swin Transformer~\cite{liu2021swin} and various diffusion  models~\cite{yang2022diffusion} are known for   encoding and generating   visual data, while CLIP~\cite{radford2021learning} and DALL.E~\cite{ramesh2021zero} are known for the  multimodal research. 

Unlike NLP and CV, so far, there are no highly recognized pioneering work on foundation models in the RS community. 
Recent work such as PeterRec~\cite{yuan2020parameter}, DUPN~\cite{ni2018perceive}, STAR~\cite{sheng2021one} and Conure~\cite{yuan2021one} have made some meaningful explorations in learning universal (user or item) representation. However, they all belong to the IDRec category, which has limited transfer learning capabilities when the downstream dataset lacks overlapping userIDs or itemIDs~\cite{gao2019cross}. More recently, researchers started to learn the RS models directly from the raw modality features~\cite{yuan2023go,li2023text}. ZESREC~\cite{ding2021zero} is the first paper that achieved the zero-shot transfer learning ability for text RS without using user or item overlapping information. Similar work includes ShopperBERT~\cite{shin2021one4all}, 
PTUM~\cite{wu2020ptum}, UniSRec~\cite{hou2022towards}, IDA-SR~\cite{mu2022id}, VQ-Rec~\cite{hou2022learning},  LLM4Rec~\cite{li2023exploring}.
All these work focused only on text modality and mainly based on pre-extracted textual features from a frozen text encoder.
Three preprints, i.e. TransRec~\cite{wang2022transrec}, AdapterRec~\cite{fu2023exploring},  LLM-REC~\cite{tang2023model} and concurrent works Recformer~\cite{li2023text} and LMRec~\cite{shin2023pivotal} performed joint or end-to-end (E2E) training of modality encoder,
but most of them only investigated one type of UE and ME, whereas it remains unknown for other more advanced UE and ME, and training manners.  
% Conure~\cite{yuan2021one} and 
P5~\cite{geng2022recommendation}, M6-Rec~\cite{cui2022m6} and Conure~\cite{yuan2021one} proposed a unified model to serve multiple tasks, such as review summary, rating prediction, user profile prediction, and item recommendation. 

In this paper, we present benchmark results on E2E-learned TransRec, which is computationally expensive but performs much better than pre-extracted features.

\begin{figure*}
    \centering
    \includegraphics[width=1\textwidth]{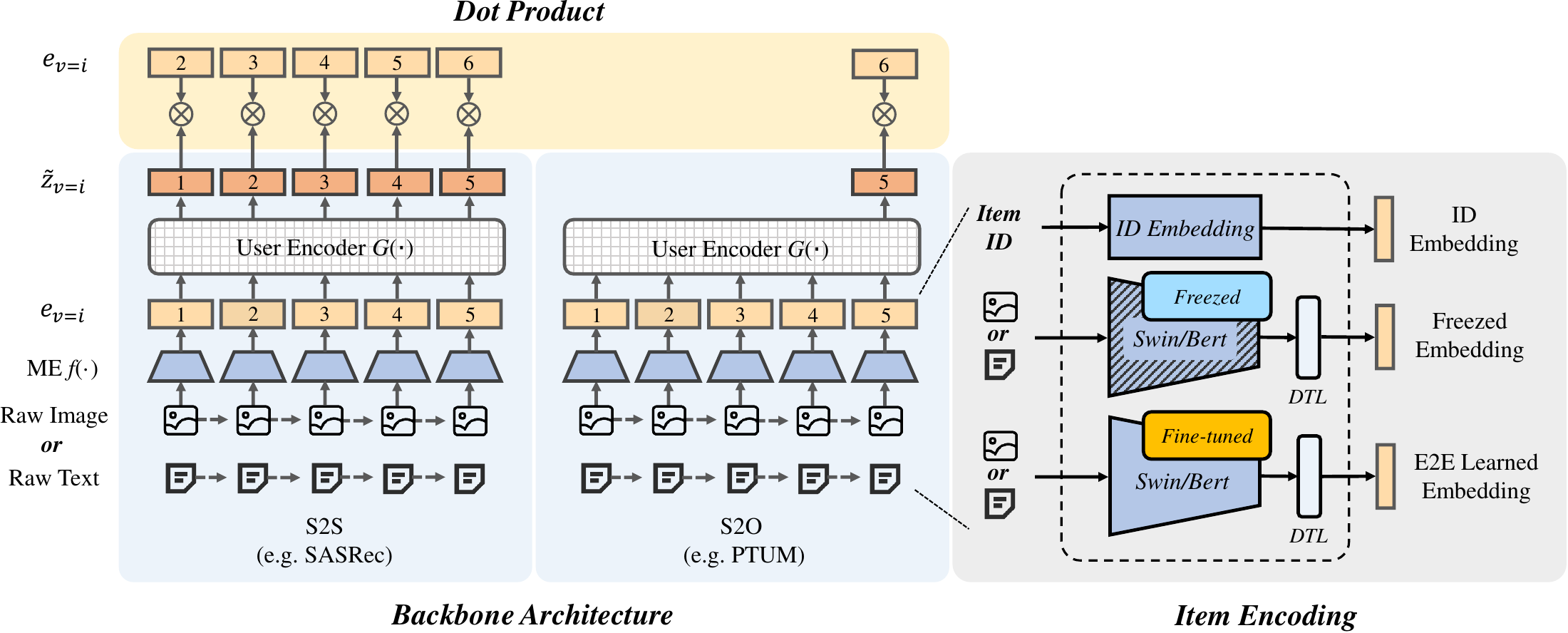}\\
    % \hfill
    \caption{TransRec architectures (S2S \& S2O). BERT and Swin-B (Swin Transformer) are used as  ME. 
    DTL is the DNN layers for dimension transformation. 
    UE can be a stack of DNN, RNN, CNN, or MHSA layers. 
    $\tilde{Z}_{v=1},...,\tilde{Z}_{v=n}$ are vector generated by UE, $e_{v=1},...,e_{v={n+1}}$ are vectors generated by ME.
    }
\label{fig:Illustration}
\vspace{-0.3cm}
\end{figure*}

\section{Baselines Overview}

\textbf{Modality-based recommendation (MoRec).} 
Let  $\mathcal {U}, \mathcal {I}$ be the set of users and items respectively. The goal of RS is to predict the potential interaction of user $u \in \mathcal {U}$ by exploiting her past behaviors $\mathcal {I}_u$ = $\{i_1,...,i_n\}$. In a classical IDRec setup, users and items are usually represented by their unique IDs. Accordingly, the userIDs and itemIDs can be embedded into a series of dense vectors, denoted as $\beta_u \in R^d$ and $\beta_i \in  R^d$, where $d$ is the embedding size, and each of them is the representation of a user or item. MoRec instead applies an modality encoder (ME), denoted as $f(x_i)$,  to encode xmodality features $x_i$ of an item $i$. 
MoRec can basically inherit other modules in IDRec, such as the user encoder or recommendation backbone. In theory, various MoRec models can be constructed by simply replacing the  $\beta_i$ of IDRec with $f(x_i)$. In this paper, we limit the scope of MoRec to learn recommendation models only from pure modality features instead of treating them as auxiliary features of ID features.
% \footnote{However, the datasets we provided can be used for more studies, e.g., combining both modality \& ID features for recommendation.} 
This distinguishes from a majority of previous works~\cite{he2016vbpr,he2016ups} using ID as the main feature and modality as the side feature. \textcolor{black}{However, such paradigms are not well-suited to achieve the goal of transferable recommendation due to the practical challenges associated with sharing or transferring ID features~\cite{hou2022towards,wang2022transrec,yuan2023go}.}

\textbf{TransRec.} 
A RS model is usually composed of a user encoder (UE) $g(x_u)$, item encoder $f(x_i)$ and their dot product $\hat{y}= g(x_u) \otimes f(x_i)$.\footnote{On top of $f(x_i)$ there are usually one or more DNN layers for dimensionality transformation. For simplicity, we omit related formulas.} To realize a foundation TransRec model, both UE and ME should be transferable. That is, the widely used userID should not exist in TransRec either. The common approach is to replace userID by a sequence of her interacted items $\mathcal {I}_u$, which are again encoded by ME~\cite{ding2021zero}, that is, $g(x_u)= G(f(x_{i_1}),...,f(x_{i_n})$ where $G(\cdot )$ can usually be a sequential encoder. \textcolor{black}{In view of this, existing TransRec models are mainly sequential recommendation models or sequential MoRec}, e.g. PTUM~\cite{wu2020ptum}, CLUE~\cite{shin2021scaling}, TransRec~\cite{wang2022transrec}, UniSRec~\cite{hou2022towards}, VQ-Rec~\cite{hou2022learning} and AdapterRec~\cite{fu2023exploring}. 
In this paper, we benchmark TransRec using the most well-known $G(\cdot )$, including $RNN$-based GRU4Rec~\cite{hidasi2015session}, $CNN$-based NextItNet, multi-head self-attention ($MHSA$) based SASRec,  BERT4Rec, and a standard $DNN$-based  encoder. While there are some new SOTA sequential models in literature, we find that most of them can be seen as variants of the above classic models  (especially a variant of Transformer~\cite{vaswani2017attention}).

\textbf{Training Details.} 
The TransRec model will first be pre-trained with sufficient data in the source domain, and then fine-tuned to serve various target domains with relatively less data.
The training process of TransRec has no big difference from IDRec models. It involves computing the dot product(s) of user embeddings  and item embeddings  for both a positive user-item pair and a randomly selected negative pair. Subsequently, the typical binary cross-entropy loss is calculated based on these dot products. Recent literature~\cite{yuan2023go,li2023text,yang2022gram}
 clearly demonstrates that end-to-end (E2E) learning is significantly more effective compared to using pre-extracted modality features from a frozen multimodal encoder. Therefore, in our baselines, we adopt E2E learning to report baseline results.

Second, we evaluate two popular training modes:
 sequence-to-sequence (S2S) and sequence-to-one (S2O), see Figure~\ref{fig:Illustration}. 
They both encode a sequence of items as input, S2O  predicts only the last item while S2S predicts a sequence of items. That is, the (input $\to $ output) format of S2S is $i_1$, $i_2$, ..., $i_{n-1}$ $\to$ $i_2$, $i_3$, ..., $i_{n}$, and format of S2O is $i_1$, $i_2$, ..., $i_{n-1}$ $\to $ $i_{n}$.
Clearly, the S2O training architecture is essentially a variant of two-tower DSSM~\cite{huang2013learning} model, where one tower represents user sequence and the other represents target item.
In this paper, we optimize all parameters  on both the source and target datasets.
In practice,  comparable results may be obtained by tuning a few top layers for some datasets. 

\begin{table*}[t]
\caption{TransRec results (\%) on downstream datasets.  NoPT means that it is directly trained on the dataset without pre-training (PT) on the
source or other RS data (Note ME in NoPT was pre-trained on NLP or CV data.). HasPT means it has been pre-trained on the source
dataset and then fine-tuned on the target dataset.  The NDCG@10 results are given in Appendix Table 1. \textcolor{black}{Results in italics} denote model collapse. Results in bold indicate the maximum between NoPT \& HasPT.  Underlined results are the maximum value among all.
}
\label{tab:Comparative_StudyHR}
% \vskip 0.15in
\begin{center}
\begin{small}
% \begin{sc}
\begin{tabular}{p{1.3cm}<{\centering}  p{1.2cm}<{\centering}  p{0.7cm}<{\centering}  p{0.8cm}<{\centering}  p{0.9cm}<{\centering}  p{0.7cm}<{\centering}  p{0.8cm}<{\centering}  p{0.9cm}<{\centering}  p{0.7cm}<{\centering}  p{0.8cm}<{\centering}  p{0.9cm}<{\centering}  p{0.7cm}<{\centering}  p{0.8cm}<{\centering}  p{0.9cm}<{\centering}}
\toprule
\multirow{2}{*}{Dataset} & \multirow{2}{*}{Metric} &\multicolumn{3}{c}{SASRec} &\multicolumn{3}{c}{BERT4Rec}  &\multicolumn{3}{c}{NextItNet}  &\multicolumn{3}{c}{GRU4Rec} \\
\cmidrule(r){3-5}\cmidrule(r){6-8}\cmidrule(r){9-11}\cmidrule(r){12-14}
    &&IDRec &NoPT &HasPT &IDRec &NoPT &HasPT &IDRec &NoPT &HasPT &IDRec &NoPT &HasPT \\
\midrule 
\multicolumn{14}{c}{BERT (base version) for text recommendation} \\
\midrule     % &            SASRec                               &           BERT4Rec                       &           NextItNet                      &            GRU4Rec
Bili\_Food     &H@10  &18.09 &18.03 &\underline{\textbf{19.59}}  &17.07 &16.70 &\underline{\textbf{20.00}}  &11.75 &15.80 &\underline{\textbf{17.26}}  &11.49 &16.12 &\underline{\textbf{17.21}} \\ 
Bili\_Dance    &H@10  &23.46 &23.49 &\underline{\textbf{25.41}}  &21.92 &22.14 &\underline{\textbf{26.84}}  &17.52 &21.11 &\underline{\textbf{23.33}}  &17.38 &21.25 &\underline{\textbf{22.47}} \\
Bili\_Movie    &H@10  &11.40 &11.64 &\underline{\textbf{12.63}}  &10.75 &11.12 &\underline{\textbf{13.47}}  &7.95  &10.34 &\underline{\textbf{11.57}}  &6.85  &9.67  &\underline{\textbf{11.70}} \\ 
Bili\_Cartoon  &H@10  &11.63 &11.94 &\underline{\textbf{13.75}}  &11.54 &12.14 &\underline{\textbf{14.01}}  &8.17  &11.07 &\underline{\textbf{11.85}}  &8.73  &11.04 &\underline{\textbf{12.69}} \\
Bili\_Music    &H@10  &19.16 &19.42 &\underline{\textbf{20.65}}  &17.73 &17.44 &\underline{\textbf{20.95}}  &14.69 &17.84 &\underline{\textbf{19.29}}  &15.52 &16.19 &\underline{\textbf{17.79}} \\ 
\midrule
KU             &H@10  &28.36 &30.77 &\underline{\textbf{31.36}}  &24.18 &24.13 &\underline{\textbf{29.30}}  &22.22 &27.38 &\underline{\textbf{27.92}}  &20.40 &28.61 &\underline{\textbf{29.60}} \\
QB             &H@10  &33.92 &34.27 &\underline{\textbf{34.60}}  &32.80 &32.28 &\underline{\textbf{33.40}}  &\underline{31.77} &\textbf{30.28} &29.98  &30.75 &32.82 &\underline{\textbf{33.24}} \\
TN             &H@10  &15.74 &15.11 &\underline{\textbf{16.85}}  &15.77 &15.14 &\underline{\textbf{16.86}}  &\underline{12.86} &\textbf{12.79} &11.96  &14.16 &\underline{\textbf{15.73}} &14.01 \\
DY             &H@10  &\underline{15.92} &14.35 &\textbf{14.49}  &13.43 &9.72  &\underline{\textbf{13.60}}  &\underline{12.06} &\textbf{10.78} &9.51   &10.45 &16.24 &\underline{\textbf{16.34}} \\ 
\midrule

\multicolumn{14}{c}{Swin Transformer (base version) for image recommendation} \\
\midrule     % &                SASRec                             &               BERT4Rec                          &            NextItNet                       &            GRU4Rec
Bili\_Food     &H@10  &18.09 &17.20 &\underline{\textbf{18.72}}    &17.07 &\textcolor{black}{\textit{2.15}} &\underline{\textbf{19.01}}    &11.75 &14.52 &\underline{\textbf{17.89}}    &11.49 &16.33 &\underline{\textbf{17.74}} \\
Bili\_Dance    &H@10  &\underline{23.46} &21.63 &\textbf{22.16}    &\underline{21.92} &16.65                  &\textbf{20.90}    &17.52 &17.12 &\underline{\textbf{21.92}}    &17.38 &19.56 &\underline{\textbf{21.84}} \\
Bili\_Movie    &H@10  &11.40 &10.30 &\underline{\textbf{11.50}}    &\underline{10.75} &\textcolor{black}{\textit{1.64}} &\textbf{10.11}    &7.95  &8.21  &\underline{\textbf{10.73}}    &6.85  &8.99  &\underline{\textbf{10.44}} \\
Bili\_Cartoon  &H@10  &11.63 &11.09 &\underline{\textbf{11.78}}    &11.54             &10.54      &\underline{\textbf{11.82}}    &8.17  &9.00  &\underline{\textbf{10.92}}    &8.73  &8.95  &\underline{\textbf{10.48}} \\
Bili\_Music    &H@10  &\underline{19.16} &17.17 &\textbf{17.56}    &\underline{17.73} &11.92                  &\textbf{15.42}    &14.69 &15.58 &\underline{\textbf{16.76}}    &15.52 &15.20 &\underline{\textbf{16.04}} \\
\midrule
KU             &H@10  &28.36 &\underline{\textbf{33.08}} &33.03    &24.18             &23.75      &\underline{\textbf{25.42}}    &22.22 &27.23 &\underline{\textbf{32.64}}    &20.40 &\underline{\textbf{31.36}} &30.18 \\  
QB             &H@10  &\underline{33.92} &32.39 &\textbf{33.57}    &\underline{32.80} &\textbf{25.96}         &22.33             &\underline{31.77} &29.91 &\textbf{31.75}    &30.75 &32.53 &\underline{\textbf{33.04}} \\ 
TN             &H@10  &\underline{15.74} &14.12 &\textbf{14.44}    &\underline{15.77} &\textbf{13.59}         &12.98             &\underline{12.86} &11.40 &\textbf{12.15}    &14.16 &\underline{\textbf{14.27}} &13.81 \\ 
DY             &H@10  &\underline{15.92} &14.08 &\textbf{14.68}    &\underline{13.43} &\textbf{10.83}         &9.40              &12.06 &11.60 &\underline{\textbf{12.49}}    &10.45 &13.26 &\underline{\textbf{13.93}} \\
\bottomrule
\end{tabular}
% \end{sc}
\end{small}
\end{center}
\vskip -0.1in
\end{table*}

\section{TransRec Benchmark}

\subsection{Evaluation}
We adopt the leave-one-out strategy to split each dataset, namely, the last interaction per user is used for testing, the second to last is used for validation, and the rest are used for training. 
The popular H@10 (Hit Ratio @10) and N@10 (Normalized Discounted Cumulative Gain @10) are used as the evaluation metrics~\cite{yuan2022tenrec}. To save space, we report results of N@10 in Appendix.
We rank the predicted item among all items in the  pool instead of drawing 100 random items~\cite{krichene2022sampled}.

\subsection{Experimental Setting}
\textcolor{black}{Considering the early stage of TransRec\footnote{\textcolor{black}{In fact, the community is still unable to provide a definitive answer regarding the possibility of developing a one-for-all foundation model for recommender systems.}}, it is crucial to conduct a fair comparison between TransRec and the well-established and dominant IDRec models. To ensure fairness, we make sure that both IDRec and TransRec employ the same network backbone and training approach. This includes using identical loss functions and samplers, with the only difference being the item encoder, which is replaced with a state-of-the-art modality encoder in TransRec. This setup enables a fair and direct comparison between the two models. Some literature  utilized relatively smaller ID embedding sizes for IDRec, making their MoRec or TransRec easier to achieve improvements in performance. Additionally, there are also studies that compare TransRec and IDRec using different network backbones and samplers. However, we believe that conducting a fair comparison between the two models becomes challenging when multiple factors differ between them.\footnote{\textcolor{black}{Repeatability is a growing concern in the recommender system community. The absence of a public and community-assessed benchmark and leaderboard creates difficulties in accurately assessing the real progress made in the field, see~\cite{shehzad2023everyone,ferrari2019we,rendle2019difficulty,krichene2020sampled}.}}}

Regarding the hyper-parameter setting, our first principle is to ensure that IDRec on both upstream and downstream datasets are \textit{extremely} tuned, 
including learning rate $\gamma$, embedding/hidden size $d$, layer number $l$, dropout $\rho$, batch size $b$, etc. For example, we tune $\gamma$ by searching from 
[5e-6, 1e-5, 5e-5, 1e-4, 5e-4, 1e-3], $d$ from [64, 128, 256, 512, 1024, 2048]. Similarly, we find optimal values for other hyperparameters.
While for TransRec, we first use the same set of hyper-parameters as IDRec and then perform the search around the best choices (the search range and step size are kept exactly the  same as IDRec). 
This is a faster and fair way to find better hyper-parameters for TransRec on both source and target datasets. It is worth noting that iterating over all hyper-parameter combinations for TransRec is infeasible since training it usually takes 100x larger compute and time than IDRec by the E2E manner (see details in Appendix Table 5). 

\textcolor{black}{All images in NineRec are resized to the shape of 224 × 224 pixels. The text descriptions are limited to a maximum of 30 Chinese/English words.}

\subsection{Benchmarking User Encoders}
In Table~\ref{tab:Comparative_StudyHR}, our benchmark  covers several most classical recommendation backbones (RNN-based GRU4Rec, CNN-based NextItNet, MHSA-based SASRec and BERT4Rec, and two DNN-based models in 
Appendix Table 6),
trained end-to-end on \textit{two} single modalities in \textit{nine} target tasks,
by replacing their original itemID with item ME. \textcolor{black}{We also report two additional peer-reviewed baseline UniSRec~\cite{hou2022towards} and VQ-Rec~\cite{hou2022learning} in Appendex Table 11.  Our results indicate that these models do not outperform the classical methods under the fair comparison setting.} 

Regarding the modality encoder,  we use BERT\footnote{In this paper, we use the Chinese BERT as ME, while due to lack of other Chinese text ME, we use English ME by translating the text into English for RoBERTa, OPT, CLIP and ViLT. Details are provided in Appendix Table 4.} for text recommendation, and use Swin Transformer  for image recommendation.  Without special mention, all models here are trained using the S2S mode (see Figure~\ref{fig:Illustration}).
 In addition, 
 (1) we report the results of multimodal recommendation  with the SASRec backbone in Table~\ref{tab:MMRec-target} and Appendix Table 9; 
(2) we report two DNN backbone baselines in Appendix Table 6;
(3) we report the S2O training baseline  in Table~\ref{tab:CPCresults}; (4) we report the results of the source  Bili\_500K dataset in Appendix Table 7; 
(5) We report results on the source and target datasets using the larger Bili\_2M dataset in Appendix Table 2 and 8. 

Beyond the benchmark results, we show some insightful findings as below. Note in this paper, we mainly use Bili\_500K as the source dataset unless otherwise stated, and report some key results using Bili\_2M  given the ultra-high training cost.
\begin{itemize}
\item  Table~\ref{tab:Comparative_StudyHR}, ~\ref{tab:MMRec-target}, ~\ref{tab:CPCresults}, Appendix Table 6 show that TransRec pre-trained on the source dataset (i.e. HasPT) mostly performs better than its NoPT  version.
\textcolor{black}{These results highlight the effectiveness of pre-training and indicate that the NineRec dataset is well-suited for research on transfer learning.}
\item Table~\ref{tab:Comparative_StudyHR}, ~\ref{tab:CPCresults} show TransRec pre-trained on  text modality  in general obviously outperforms its IDRec counterpart on these downstream datasets, meanwhile, it  sometimes  performs worse than  IDRec if  pre-trained on image modality. 
Similar results can be even observed on the source datasets and two very warm datasets in Appendix Table 7 and 8. \textcolor{black}{Previous works have primarily focused on TransRec beating IDRec in cold-start scenarios. However, defeating IDRec in non-cold-start scenarios  signifies a significant advancement and potentially heralds a paradigm shift in the future of recommender systems. This is particularly noteworthy considering that IDRec has remained the state-of-the-art approach for over 10 years.
}

\item Table~\ref{tab:MMRec-target} shows that TransRec models trained on multimodal (text and image) features do not consistently outperform those trained on a single modality (i.e. Table~\ref{tab:Comparative_StudyHR}). This is a reasonable observation \textcolor{black}{(also observed in literature~\cite{zhou2023comprehensive})}, as effectively fusing text and image modalities in recommendation models poses a non-trivial challenge that remains largely unexplored within the E2E learning paradigm.

\item Table~\ref{tab:Comparative_StudyHR} and Appendix Table 7 indicate that a recommendation network with higher accuracy on IDRec, such as SASRec compared to BERT4Rec, may not necessarily result in higher accuracy on TransRec or MoRec, even when utilizing the same ME.

\end{itemize}
A surprising result is that model collapse may happen during learning MoRec/TransRec, as shown in Table~\ref{tab:Comparative_StudyHR} \textcolor{black}{marked with italics}. We find that it is quite difficult to jointly learn BERT4Rec with Swin Transformer sometimes even many hyper-parameter searches are performed.
This is unknown to the community.
More interesting findings can be made according to such extensive results (see Appendix). 

\begin{figure*}[t]
\centering
\subfloat[Bili\_Food]{
    \begin{minipage}[htbp]{0.52\columnwidth}
        \includegraphics[width=1\columnwidth]{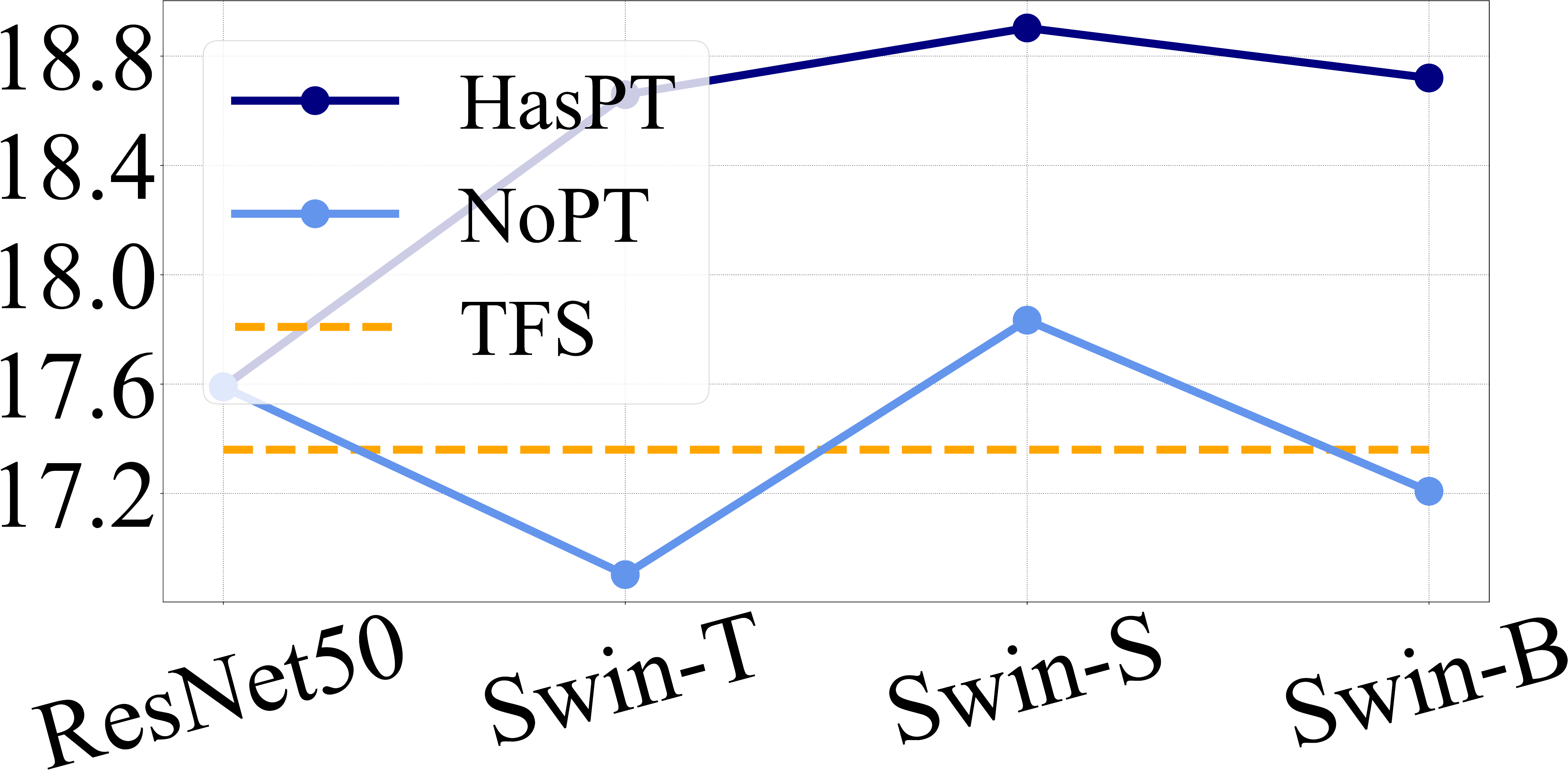}
    \end{minipage}}
\subfloat[Bili\_Dance]{
    \begin{minipage}[htbp]{0.52\columnwidth}
        \includegraphics[width=1\columnwidth]{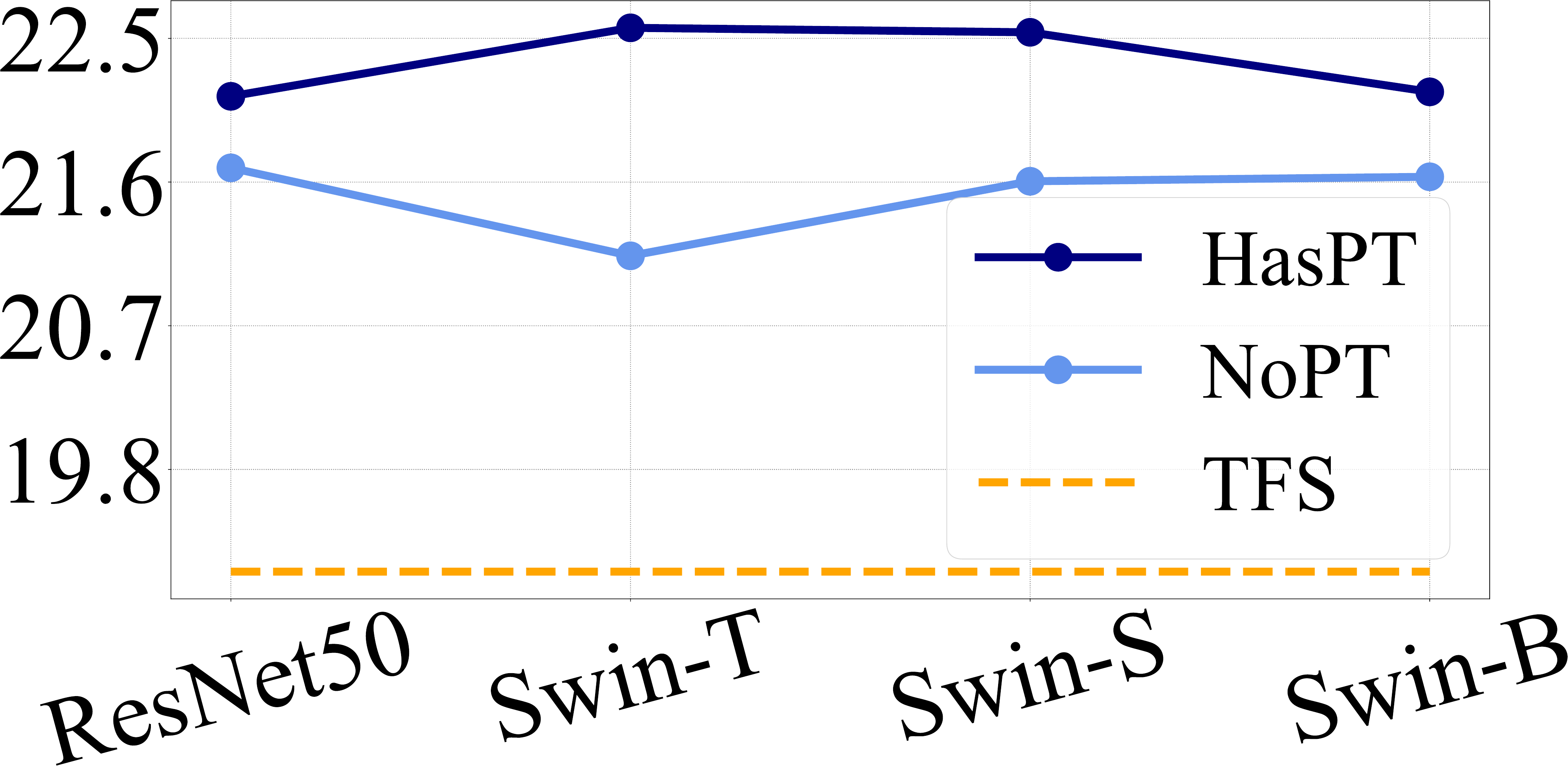}
    \end{minipage}}
\subfloat[Bili\_Movie]{
    \begin{minipage}[htbp]{0.52\columnwidth}
        \includegraphics[width=1\columnwidth]{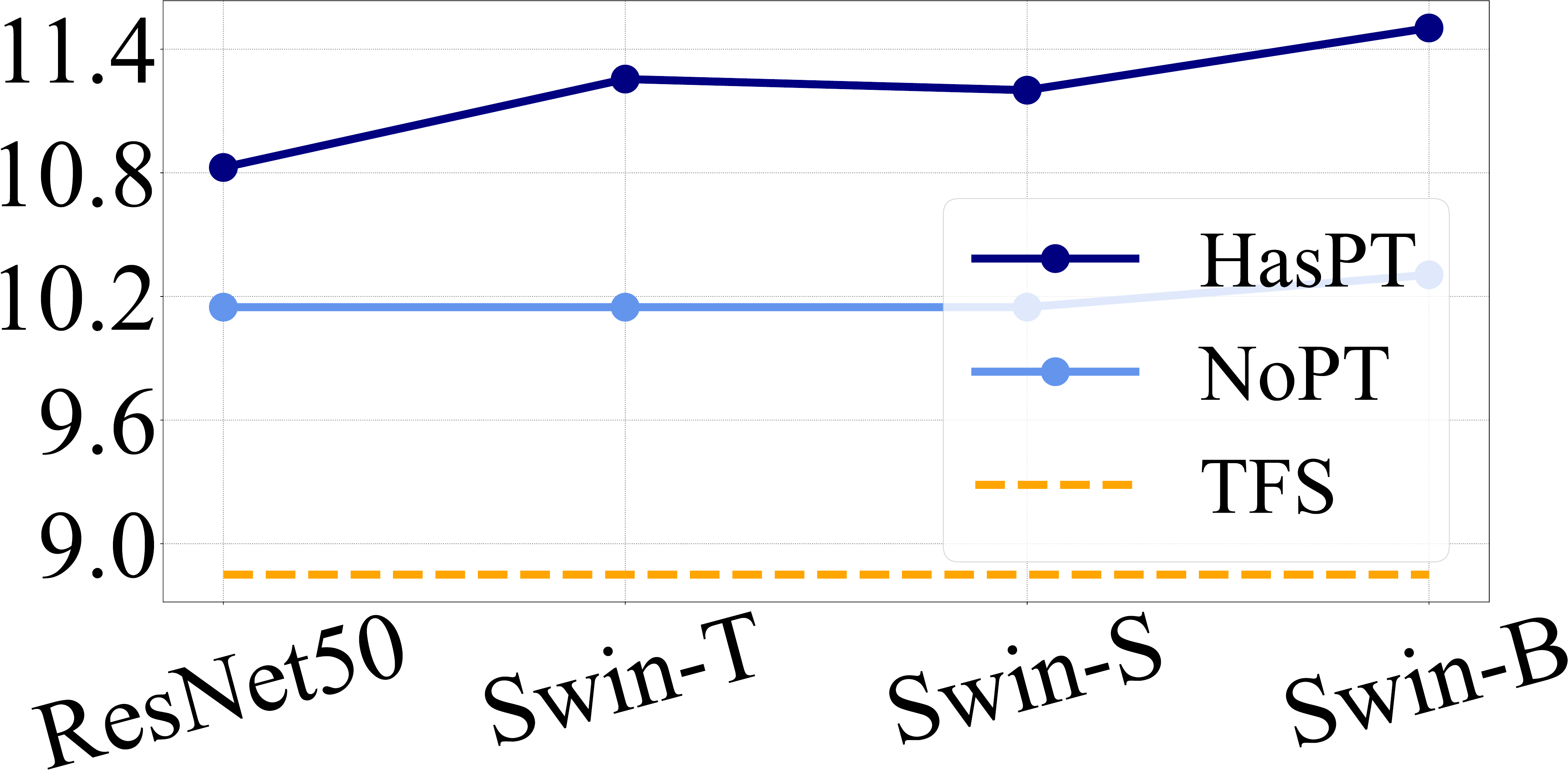}
    \end{minipage}}
\qquad
\subfloat[Bili\_Cartoon]{
    \begin{minipage}[htbp]{0.52\columnwidth}
        \includegraphics[width=1\columnwidth]{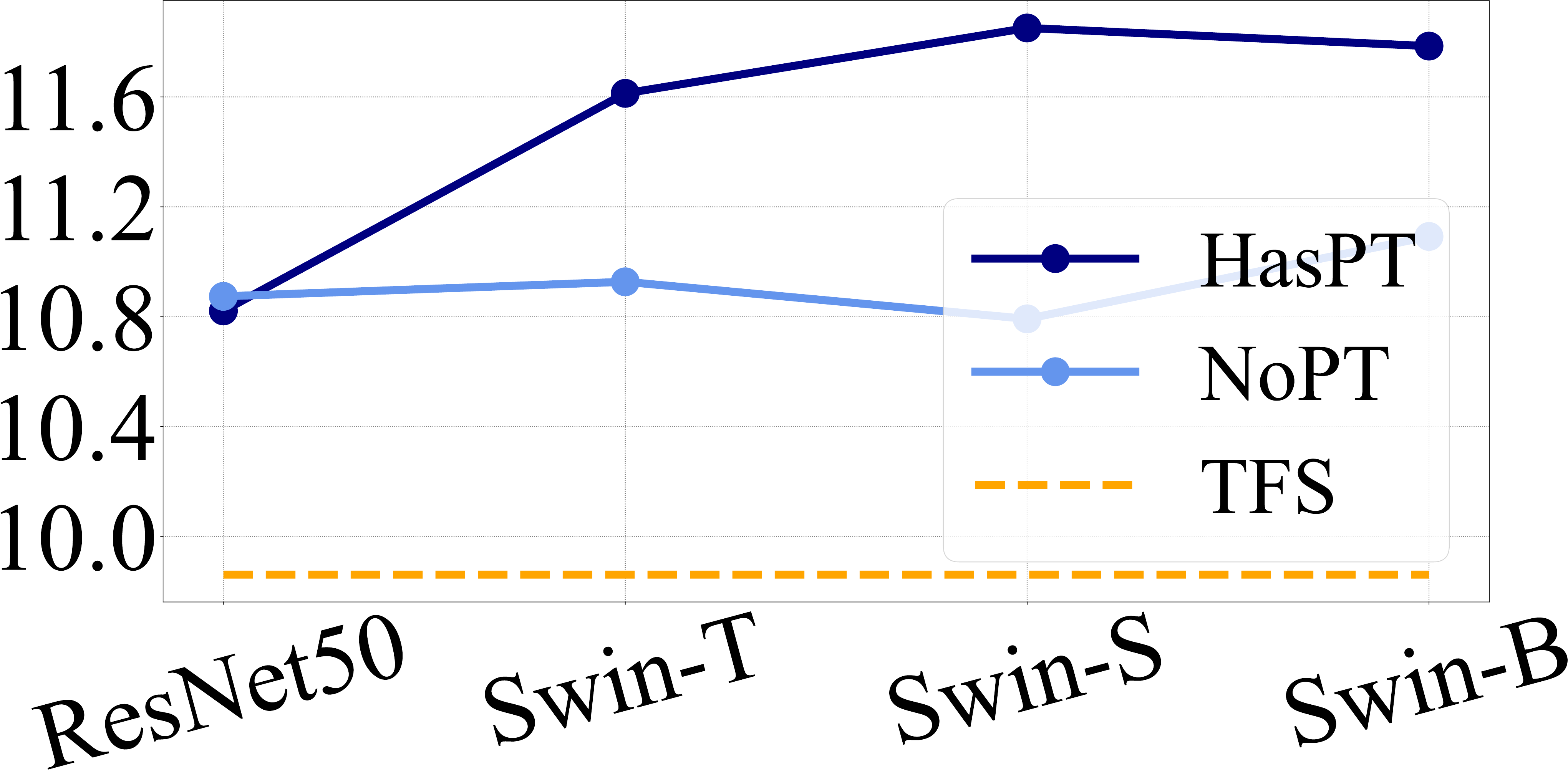}
    \end{minipage}}
\subfloat[Bili\_Music]{
    \begin{minipage}[htbp]{0.52\columnwidth}
        \includegraphics[width=1\columnwidth]{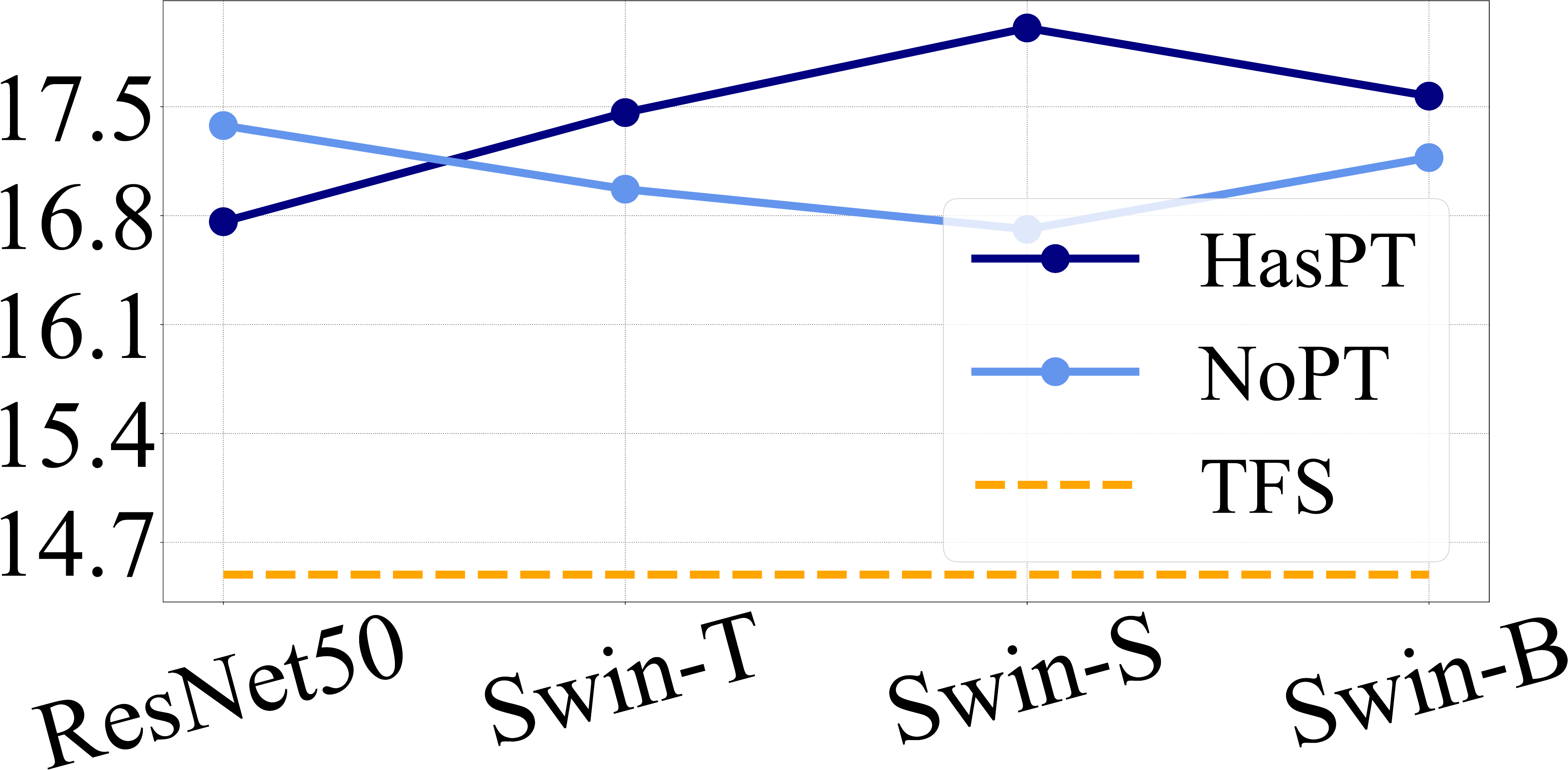}
    \end{minipage}}
\subfloat[KU]{
    \begin{minipage}[htbp]{0.52\columnwidth}
        \includegraphics[width=1\columnwidth]{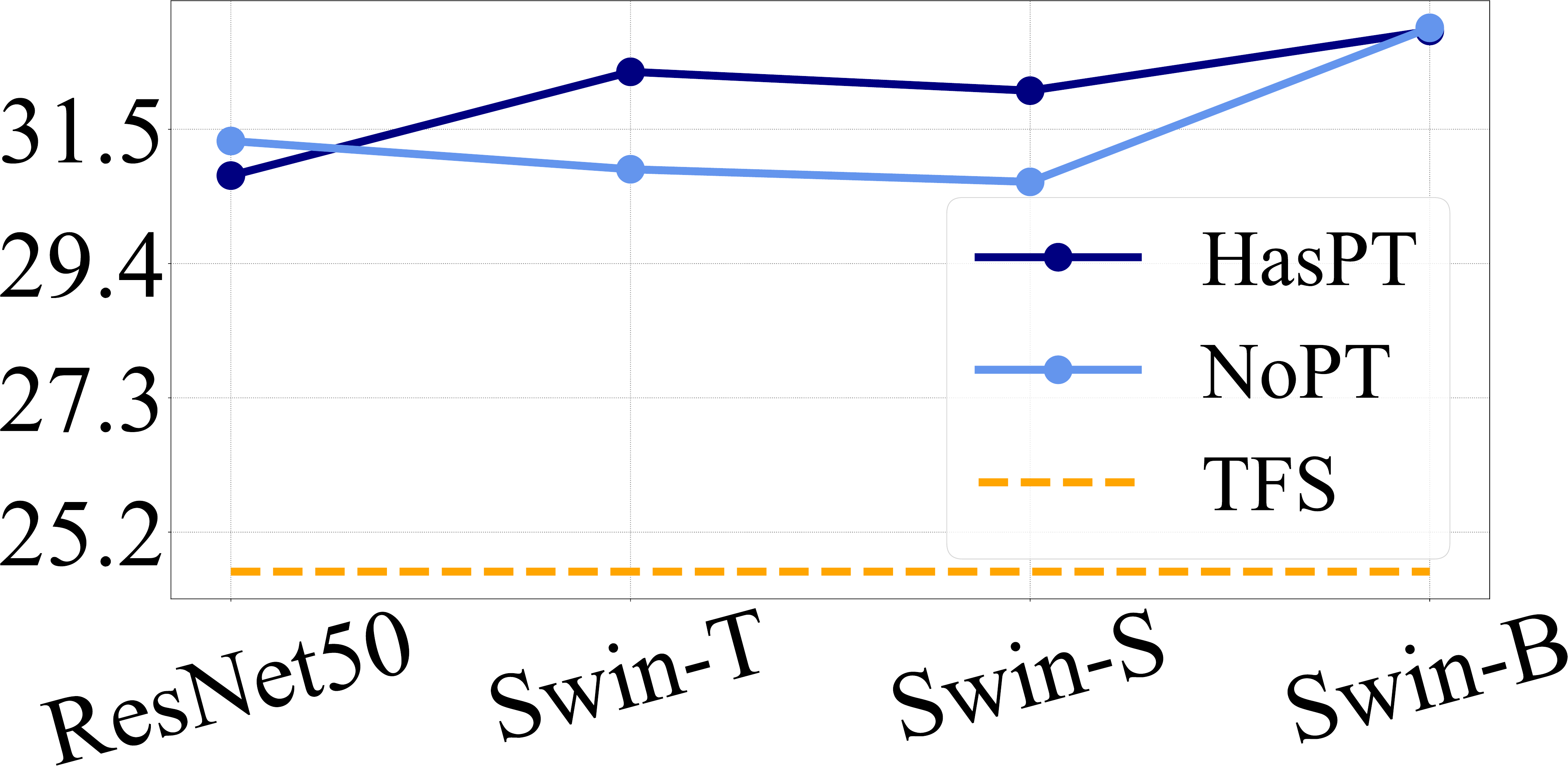}
    \end{minipage}}
\qquad
\subfloat[QB]{
    \begin{minipage}[htbp]{0.52\columnwidth}
        \includegraphics[width=1\columnwidth]{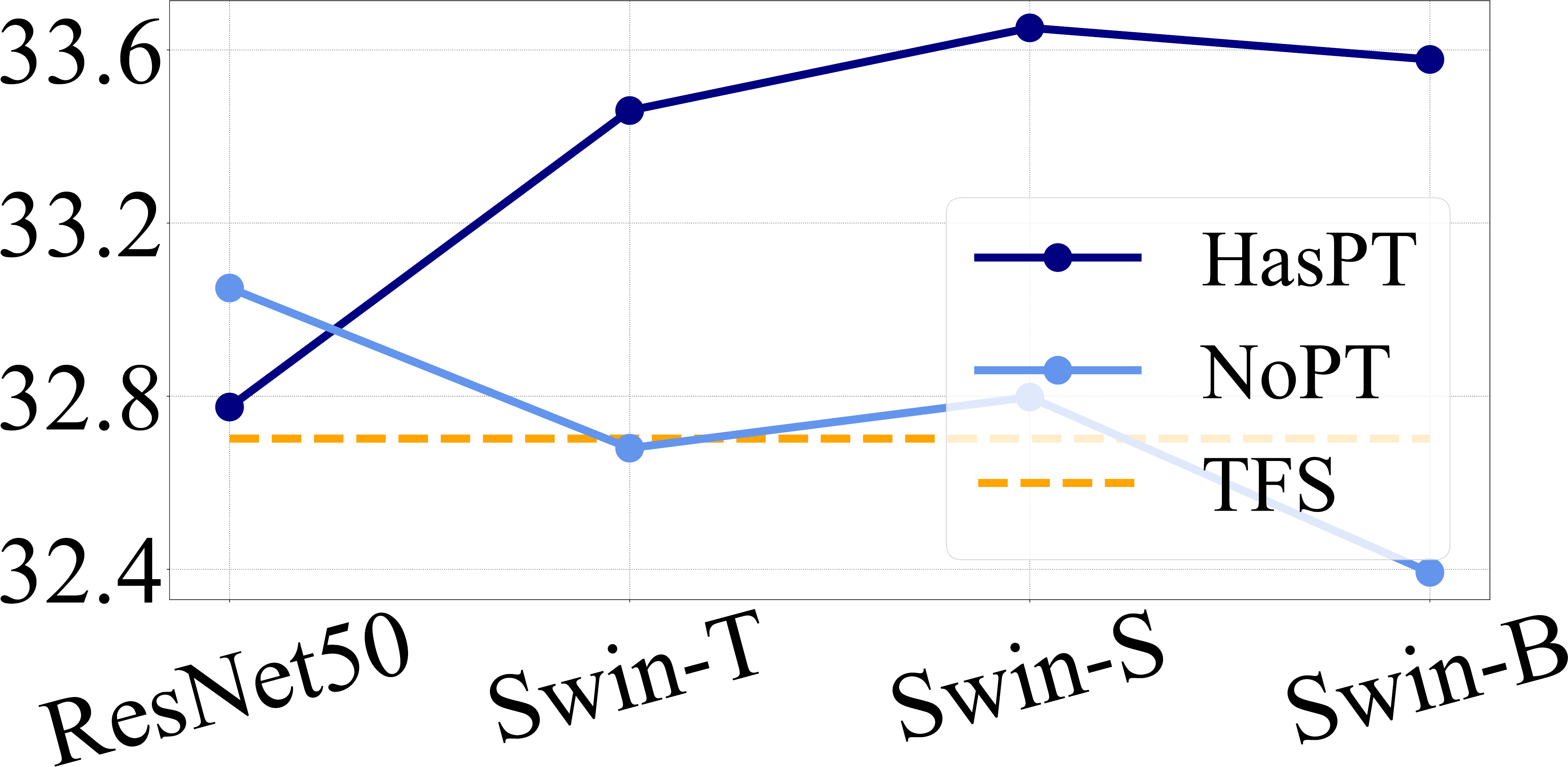}
    \end{minipage}}
\subfloat[TN]{
    \begin{minipage}[htbp]{0.52\columnwidth}
        \includegraphics[width=1\columnwidth]{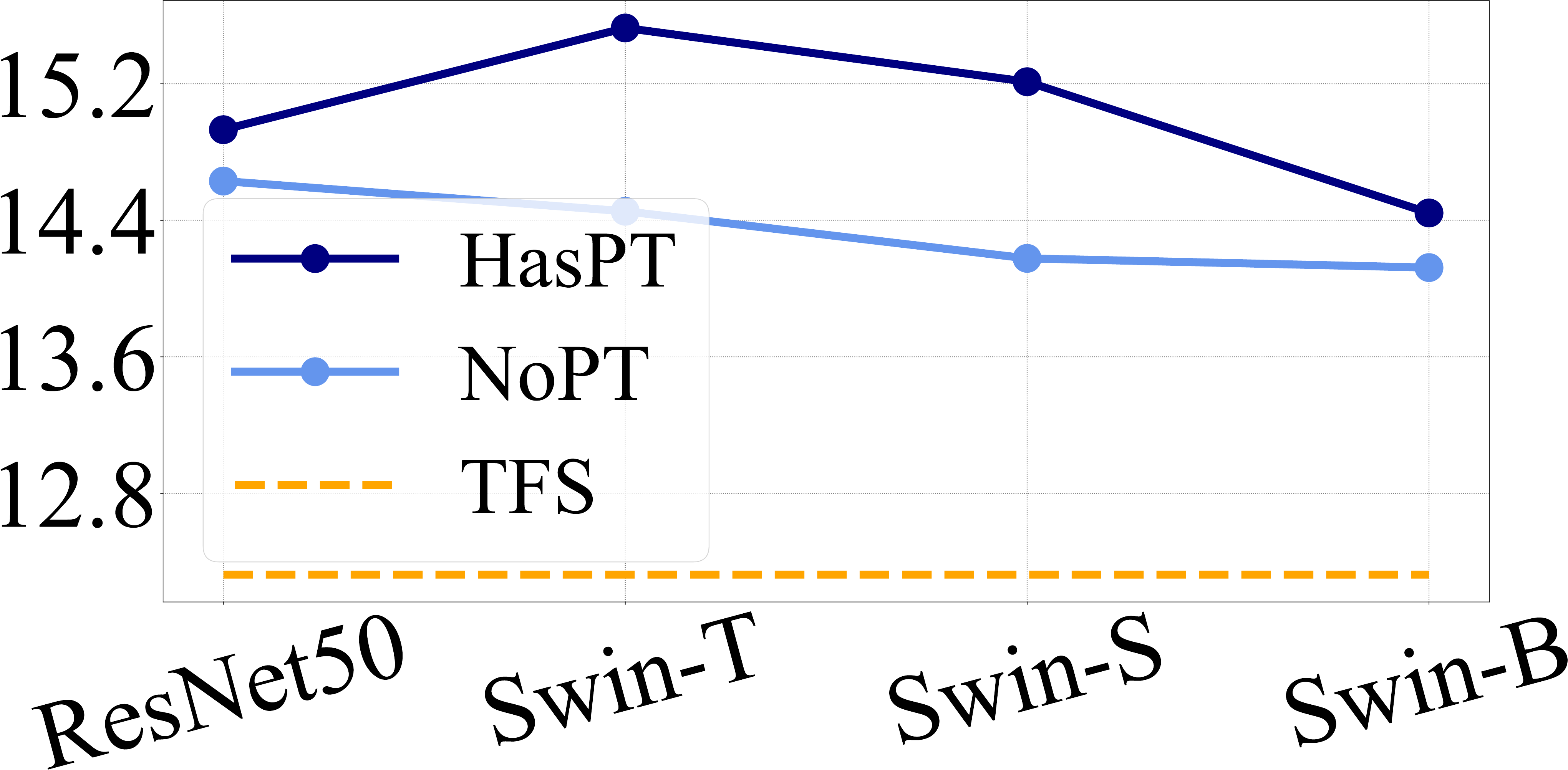}
    \end{minipage}}
\subfloat[DY]{
    \begin{minipage}[htbp]{0.52\columnwidth}
        \includegraphics[width=1\columnwidth]{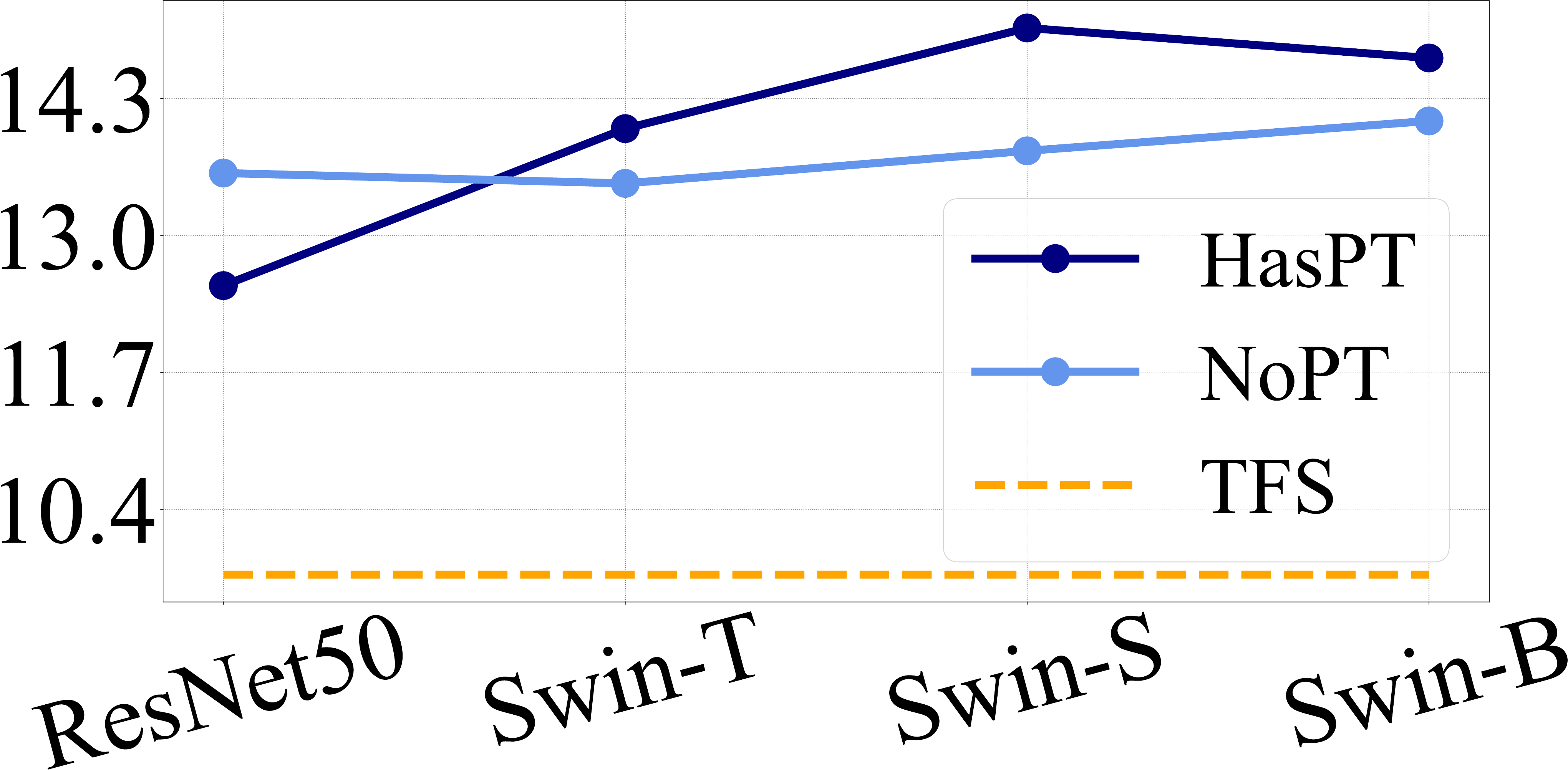}
    \end{minipage}}

\caption{Benchmark results (y-axis:\%) of item ME (with SASRec as UE). The details of ResNet50, Swin-T, Swin-S and Swin-B are provided in Appendix Table 4. 
All hyper-parameters are kept the same for NoPT, HasPT, and TFS. 
TFS  means TransRec is not pre-trained on the source dataset, and its ME  is not pre-trained on ImageNet. The dashed yellow line only shows ResNet50.
}
\label{fig:imageME}
\vspace{-0.3cm}
\end{figure*}

\subsection{Benchmarking Item Encoders}
Figure~\ref{fig:imageME} and Table~\ref{tab:textME} present the evaluation of several well-known item ME models, such as ResNet and Swin Transformer with different model sizes, for image recommendation, and RoBERTa and OPT~\cite{zhang2022opt} for text recommendation.
Unless otherwise specified, the SASRec backbone and S2S training mode are utilized by TransRec in the subsequent sections. The majority of observations align with the above findings. Interestingly, we  discovers that pre-training TransRec on the source Bili\_500K dataset (HasPT), using ResNet50 as the ME, does not consistently yield superior outcomes compared to its NoPT version. This outcome is somewhat unexpected, as it suggests that the parameters of ResNet50 become degraded after pre-training on Bili\_500K.
This  is unexpected but is  not impossible.
In fact, Table~\ref{tab:Comparative_StudyHR} also showed several similar results.
To see why, we show the results of ResNet50 without pre-training on ImageNet (i.e. TFS). It can be clearly seen that TFS largely underperforms
both NoPT and HasPT. This indicates that ResNet50 parameters pre-trained on ImageNet are highly beneficial as  an initialization step, and as a result, additional  training on Bili\_500K may not always bring significant  benefit for other downstream tasks.

\subsection{End-to-End (E2E) vs. Two-Stage (TS) Benchmark}
By surveying the literature, we found that most previous MoRec/TransRec studies adopt a two-stage (TS) training approach~\cite{mcauley2015image,Zhou2023BM3,Liang2023MMMLP}: first pre-extract offline modality features via ME, and then incorporate them into the recommendation model as regular features. In recent two years, the E2E training approach has attracted attention, but mainly for text recommendation~\cite{li2023text,yang2022gram}. We report the results of E2E vs. TS in Table~\ref{tab:E2E_vs_TS}. Clearly,  TransRec by E2E training of ME  outperforms the TS method substantially on both text and image modalities. For some text recommendation tasks (e.g. Bili\_Movie, Bili\_C·artoon, TN, and DY), E2E could achieve about more than 200\% higher accuracy. 
The results indicate  that the off-the-shelf representation features  extracted directly from the pre-trained modality encoders have a considerable  gap between the NLP, CV and recommendation tasks, i.e. these features are not universal or at least not specific enough to  the recommendation task.  
Parameter retraining on the target datasets is a key way to obtain desired results. Thereby, NineRec with raw text and image features will serve as an important dataset for studying E2E-learning based MoRec and TransRec, although the computational cost is high (see Appendix~Table 5).

\begin{table}[t]
\caption{E2E vs TS on HR@10 (\%). 
% TransRec uses the SASRec backbone as UE.
TS means the parameters of ME (pre-trained on NLP or CV data) are not allowed to be optimized during training on both the source and target datasets.
}
\label{tab:E2E_vs_TS}
% \vskip 0.15in
\begin{center}
\begin{small}
% \begin{sc}
% \begin{tabular}{p{1.9cm}<{\centering} p{1cm}<{\centering}  p{1cm}<{\centering} p{1cm}<{\centering}  p{1cm}<{\centering}}
\begin{tabular}{p{1.2cm}<{\centering} r p{0.5cm}<{\centering}  p{0.6cm}<{\centering} p{1cm}<{\centering}  p{0.5cm}<{\centering}  p{0.6cm}<{\centering}  p{1cm}<{\centering}}
\toprule
\multirow{2}{*}{Dataset} &&\multicolumn{3}{c}{BERT}  &\multicolumn{3}{c}{Swin-B}\\
\cmidrule(r){3-5}\cmidrule(r){6-8}
              &&TS    &E2E            &Improv. &TS    &E2E            &Improv. \\
\midrule
Bili\_500K    &&0.57  &\textbf{4.69}  &8.22x   &0.41  &\textbf{3.57}  &8.71x \\
\midrule
Bili\_Food    &&11.52 &\textbf{19.59} &1.7x    &14.20 &\textbf{18.72} &1.32x \\
Bili\_Dance   &&14.32 &\textbf{25.41} &1.77x   &17.57 &\textbf{22.16} &1.26x \\
Bili\_Movie   &&5.73  &\textbf{12.63} &2.2x    &7.01  &\textbf{11.50} &1.64x \\
Bili\_Cartoon &&6.28  &\textbf{13.75} &2.19x   &6.70  &\textbf{11.78} &1.76x \\
Bili\_Music   &&11.01 &\textbf{20.65} &1.88x   &12.83 &\textbf{17.56} &1.37x \\
\midrule
KU            &&27.00  &\textbf{31.36} &1.16x   &30.33 &\textbf{33.03} &1.09x \\
QB            &&28.14 &\textbf{34.60} &1.23x   &29.75 &\textbf{33.57} &1.29x \\
TN            &&8.28  &\textbf{16.85} &2.04x   &10.37 &\textbf{14.44} &1.39x \\
DY            &&6.90  &\textbf{14.49} &2.1x    &10.27 &\textbf{14.68} &1.43x \\
\bottomrule
\end{tabular}
% \end{sc}
\end{small}
\end{center}
\vspace{-0.5cm}
\end{table}

\begin{table}[ht]
\caption{Results on HR@10 (\%)  for zero-shot text recommendation (Appendix Table 3 for image recommendation).  ZeroRec is TransRec that is pre-trained on the source dataset, and then directly predicts on the target datasets. Random baseline is equivalent to the accuracy of IDRec in the new item setting.
}
\label{tab:zero-shot-0}
% \vskip 0.15in
\begin{center}
\begin{small}
% \begin{sc}
\begin{tabular}{p{1.3cm}<{\centering}  p{1.3cm}<{\centering}  p{1.3cm}<{\centering} p{0.7cm}<{\centering}  p{0.7cm}<{\centering}  p{0.7cm}<{\centering}}
\toprule
Type       &Bili\_Food  &Bili\_Music &KU    &QB   &DY \\
\midrule
Random     &0.63        &0.16        &0.49  &0.18 &0.12 \\
ZeroRec    &\textbf{4.76}       &\textbf{11.30}       &\textbf{4.96} &\textbf{7.58}&\textbf{0.88}\\
% Improv.     &\textbf{7.55x} &\textbf{70.75x} &\textbf{10.12x} &\textbf{42.11x} &\textbf{7.33x} \\
\midrule
           &Bili\_Dance &Bili\_Movie &\multicolumn{2}{c}{Bili\_Cartoon} &TN \\
\midrule
Random     &0.43        &0.28        &\multicolumn{2}{c}{0.21}          &0.26 \\
ZeroRec    &\textbf{8.30}       &\textbf{5.16}       &\multicolumn{2}{c}{\textbf{5.95}}          &\textbf{1.21}\\

\bottomrule
\end{tabular}
% \end{sc}
\end{small}
\end{center}
% \vskip -0.1in
\end{table}

\section{Zero-shot Recommendation}
Zero-shot learning is a very challenging task in NLP and CV. Although pre-trained TransRec achieves competitive results via fine-tuning, we ideally want it to achieve satisfactory results 
without parameter fine-tuning on the downstream dataset. This is also an important goal of foundation models. We refer to such recommendation setting as zero-shot recommendation in agreement with~\cite{ramesh2021zero}. 

We report results in Table~\ref{tab:zero-shot-0}. First, we can see that the  TransRec model  after pre-training (but without fine-tuning) can achieve 7x-70x (e.g. 0.16 vs. 11.32) better results than the random baseline. This clearly shows that the pre-trained representations in the source domain have some generality.
Second, we also find that TransRec's zero-shot prediction performance is far behind its fine-tuning method (see Table~\ref{tab:Comparative_StudyHR}). This suggests that the pre-trained representations in the source domain are far from perfect. 
We speculate it might have improved performance by pre-training on a significantly  larger source dataset (e.g. 100x-1000x larger)  or multiple diverse source datasets with a larger model size. This phenomenon is known as the emergent abilities of foundation models~\cite{wei2022emergent}. In other words, like NLP and CV, recommendation models also face great challenges on zero-shot tasks. We are unsure whether NineRec could be used to address this issue, but we believe NineRec will inspire new work and new datasets.

\begin{table}[t]
\caption{Results of more text ME (with SASRec as UE). The upper results denote HR@10, the lower denotes NDCG@10. Note that since RoBERTa
and OPT have no Chinese version, we translated the Chinese text into English (by  DeepL: https://www.deepl.com/translatorand) then
performed the evaluation. English translation will be provided with the datasets.  Again, NoPT means that it is directly trained on the dataset without pre-training (PT) on the
source or other RS dataset (Note ME in NoPT was pre-trained on NLP or CV data.). HasPT means it has been pre-trained on the source dataset and then fine-tuned on the target. 
% Unlike encoding Chinese text by Chinese version of BERT (see Table~\ref{apx:tab:MEDetails}), we benchmark other text ME, $\text{RoBERTa}_{\text{base}}$ and $\text{OPT}_{\text{125M}}$, on English text translated by DeepL (https://www.deepl.com/translator).
}
\label{tab:textME}
% \vskip 0.15in
\begin{center}
\begin{small}
% \begin{sc}
\begin{tabular}{p{1cm}<{\centering} r p{0.6cm}<{\centering}  p{0.6cm}<{\centering} p{1cm}<{\centering}  p{0.6cm}<{\centering}  p{0.6cm}<{\centering}  p{1cm}<{\centering}}
\toprule
\multirow{2}{*}{Dataset}  &&\multicolumn{3}{c}{$\text{RoBERTa}_{\text{base}}$}  &\multicolumn{3}{c}{$\text{OPT}_{\text{125M}}$}\\
\cmidrule(r){3-5}\cmidrule(r){6-8}
&& NoPT & HasPT & Improv. & NoPT & HasPT & Improv.\\
\midrule
\multirow{2}{*}{Bili\_500K}      &&3.82 &- &- &3.48 &- &- \\
                                 &&2.00 &- &- &1.82 &- &- \\
\midrule
\multirow{2}{*}{Bili\_Food}      &&17.94 &18.08 &+0.77\% &17.29 &18.67 &+7.39\% \\
                                 &&9.52  &10.18 &+6.48\% &9.76  &10.69 &+8.70\% \\  % 2- Food-T1-2
\multirow{2}{*}{Bili\_Dance}     &&22.97 &23.76 &+3.32\% &22.56 &23.80 &+5.21\% \\
                                 &&13.00 &13.56 &+4.13\% &12.99 &13.38 &+2.91\% \\  % 7- Dance-T5-7
\multirow{2}{*}{Bili\_Movie}     &&11.18 &12.43 &+10.06\%&11.24 &12.25 &+8.24\% \\
                                 &&5.95  &6.58  &+9.57\% &6.12  &6.53  &+6.28\% \\  % 5- Movie-T3-5
\multirow{2}{*}{Bili\_Cartoon}   &&11.77 &12.44 &+5.39\% &11.75 &12.51 &+6.08\% \\
                                 &&6.48  &6.74  &+3.86\% &6.53  &6.92  &+5.64\% \\  % 4- Cartoon-T2-4
\multirow{2}{*}{Bili\_Music}     &&18.63 &19.47 &+4.31\% &18.59 &19.12 &+2.77\% \\
                                 &&10.72 &11.07 &+3.16\% &10.73 &10.85 &+1.11\% \\  % 6- Music-T4-6
\midrule
\multirow{2}{*}{KU}              &&29.11 &30.92 &+5.85\% &28.71 &30.73 &+6.57\% \\
                                 &&24.34 &25.88 &+5.95\% &24.81 &25.25 &+1.74\% \\
\multirow{2}{*}{QB}              &&34.11 &33.62 &-1.46\% &33.02 &33.28 &+0.78\% \\
                                 &&26.26 &26.38 &+0.45\% &25.59 &26.49 &+3.40\% \\
\multirow{2}{*}{TN}              &&15.68 &14.89 &-5.31\% &14.40 &14.40 &+0.00\% \\
                                 &&8.76  &8.64  &-1.39\% &8.00  &8.19  &+2.32\% \\
\multirow{2}{*}{DY}              &&13.97 &14.29 &+2.24\% &13.28 &13.65 &+2.71\% \\
                                 &&7.50  &8.14  &+7.86\% &7.39  &7.82  &+5.50\% \\
\bottomrule
\end{tabular}
% \end{sc}
\end{small}
\end{center}
% \vskip -0.1in
\end{table}

\section{Conclusion, Limitations,  Broader Impacts}
Developing a research direction in the field of recommender systems (RS) is challenging without access to large-scale and real-world datasets; similarly, measuring genuine progress without a public benchmark is equally difficult. In this paper, we present the NineRec dataset suite and benchmarks, which are designed to advance transfer learning and foundation models in the RS field by leveraging raw and pure modality features. Through empirical studies, we also report several noteworthy findings. Given the rapid advancements in the field and the high computational demands, it is not feasible to evaluate all existing RS architectures, variants, and settings (such as various samplers and loss functions). However, we can establish public leaderboards to facilitate tracking of the latest state-of-the-art models by the community.

There are  many limitations and challenges not addressed in this paper. First, while we have developed TransRec by assembling popular user encoder (UE) from IDRec and popular item modality  encoder (ME) from NLP and CV, we acknowledge that this may not be the optimal approach. It is possible that only specifically designed UE and ME can fully realize the transfer learning potential of TransRec.
Second, we need to consider whether the optimization and hyper-parameter search techniques developed for IDRec over the past decade are also applicable to MoRec and TransRec.
% Third, what are the performance of TransRec in other recommendation settings, e.g. using different sampler and loss, different data di
Third, we need to investigate the proper alignment and fusion of multimodal features within the end-to-end learning paradigm. In addition, we need
to address the significant computational costs associated with end-to-end training TransRec in practical systems. This is particularly crucial when dealing with datasets that are 100x or 1000x larger than the ones used in this study.
In reality, TransRec or \textit{foundation} models are still in the early stages of development for recommendation problems. To date, there is no widely recognized TransRec paradigm. However, we believe that NineRec can help advance the field by inspiring new questions, new ideas, and new research.
% Again, these challenges may require the efforts of the entire community.

In this paper, we primarily study NineRec for transferable recommendation research. However, there are several other potential applications of NineRec in the RS field.
For instance, many widely-used RS datasets only provide itemID information, which limits researchers' ability to fully understand what their recommender systems are recommending, beyond an accuracy score. 
% \vskip 0.1in
\begin{table}[htbp]
\caption{Results of multimodal recommender system on the target datasets. Given much worse results of CLIP and ViLT on the source dataset (see Appendix Table 9), we only evaluate BERT+Swin-T as ME here. SASRec is used as UE.}
\label{tab:MMRec-target}
% \vskip 0.1in
\begin{center}
\begin{small}
% \begin{sc}
\begin{tabular}{p{1.2cm}<{\centering}  p{1.5cm}<{\centering}  p{1.2cm}<{\centering}  p{1.2cm}<{\centering}   p{1.5cm}<{\centering}}
\toprule
\multirow{2}{*}{Dataset} & \multirow{2}{*}{Metric} &\multicolumn{3}{c}{BERT+Swin-T}\\
\cmidrule(r){3-5}
&& NoPT & HasPT & Improv. \\
\midrule
\multirow{2}{*}{Bili\_Food}  &H@10 &16.68 &17.36 & +3.92\%\\
                             &N@10 &9.05  &9.51  & +4.84\%\\
\multirow{2}{*}{Bili\_Dance} &H@10 &20.74 &24.03 & +13.69\%\\
                             &N@10 &11.60 &13.62 & +14.83\%\\
\multirow{2}{*}{Bili\_Movie} &H@10 &10.27 &12.54 & +18.10\%\\
                             &N@10 &5.46  &6.47  & +15.61\%\\
\multirow{2}{*}{Bili\_Cartoon}&H@10 &10.32 &12.35 & +16.44\%\\
                             &N@10 &5.19  &6.72  & +22.77\%\\
\multirow{2}{*}{Bili\_Music} &H@10 &16.93 &18.84 & +10.14\%\\
                             &N@10 &9.52 & 10.58 & +10.02\%\\
\midrule
\multirow{2}{*}{KU}         &H@10 &28.96 &29.60	& +2.16\%\\
                            &N@10 &23.80 &24.67 & +3.53\%\\
\multirow{2}{*}{QB}         &H@10 &31.52 &33.15	& +4.91\%\\
                            &N@10 &22.53 &24.15 & +6.71\%\\
\multirow{2}{*}{TN}         &H@10 &14.88 &15.53 & +4.19\%\\
                            &N@10 &8.14  &9.00  & +9.56\%\\
\multirow{2}{*}{DY}         &H@10 &11.29 &12.90	& +12.48\%\\
                            &N@10 &5.87  &6.92  & +22.77\%\\
\bottomrule
\end{tabular}
% \end{sc}
\end{small}
\end{center}
\vskip -0.1in
\end{table} 
By utilizing NineRec, researchers can gain a better understanding of their RS models, particularly for interpretable RS~\cite{zhang2020explainable} and visually-aware RS evaluation~\cite{tsai2019evaluating} problems. 
This can ultimately lead to more effective and explainable RS models.
Furthermore, many researchers in the  NLP and CV fields are currently working on developing modality encoders with universal representations~\cite{wu2022nuwa}. 
However, these models are often only evaluated on standard NLP and CV tasks, such as image classification. We contend that the recommendation task, which involves predicting user preferences, is more challenging than these basic downstream tasks. Thus, NineRec could be crucial for NLP and CV researchers and may even facilitate the integration of RS with the NLP and CV fields.

\vskip 0.1in
\begin{table}[ht]
% \vskip 0.15in
\caption{Results of S2O framework (see Figure~\ref{fig:Illustration}) with the MHSA or SASRec backbone. The upper results denote HR@10, while the lower denotes NDCG@10.
The only difference of S2O training and DSSM-variant (see Appendix Figure 3) is that S2O uses the MHSA layers as the backbone but  DSSM-variant uses the DNN layers as the backbone.  We can see that the S2O training mode is very worse than the S2S model (Table~\ref{tab:Comparative_StudyHR}) even they both use  MHSA layers as the backbone.
‘-’ means that there is no pre-training stage on the source dataset.}
\label{tab:CPCresults}
% \vskip 0.15in
\begin{center}
\begin{small}
% \begin{sc}
\scalebox{0.88}{
\begin{tabular}{p{1.2cm}<{\centering}  p{1cm}<{\centering} p{0.5cm}<{\centering} p{0.5cm}<{\centering} p{1.2cm}<{\centering} p{0.5cm}<{\centering}  p{0.5cm}<{\centering}  p{1.2cm}<{\centering}}
\toprule
\multirow{2}{*}{Dataset} &\multirow{2}{*}{IDRec} &\multicolumn{3}{c}{BERT} &\multicolumn{3}{c}{Swin-B}\\
\cmidrule(r){3-5}\cmidrule(r){6-8}
&&NoPT &HasPT &Improv. &NoPT &HasPT &Improv. \\
\midrule
\multirow{2}{*}{Bili\_500K}   & 0.87  & 2.48  &-      &-         &1.75   &-      &-        \\
                              & 0.45  & 1.31  &-      &-         &0.87   &-      &-        \\
\midrule                      %     &ID     &TFS    &PE     &Impro.    &TFS    &PE     &Impro.
\multirow{2}{*}{Bili\_Food}   & 11.91 & 15.36 & 16.58 & +7.37\%  & 14.44 & 14.97 & +3.57\% \\
                              & 7.14  & 8.18  & 9.10  & +10.06\% & 7.95  & 8.15  & +2.47\% \\
\multirow{2}{*}{Bili\_Dance}  & 18.15 & 20.02 & 22.00 & +8.99\%  & 17.49 & 19.59 & +10.71\% \\
                              & 11.09 & 11.56 & 12.65 & +8.65\%  & 9.49  & 10.88 & +12.83\% \\
\multirow{2}{*}{Bili\_Movie}  & 7.68  & 9.36  & 10.05 & +6.86\%  & 8.15  & 8.65  & +5.80\% \\
                              & 4.51  & 5.21  & 5.32  & +2.06\%  & 4.41  & 4.43  & +0.41\% \\
\multirow{2}{*}{Bili\_Cartoon}& 8.51  & 9.89  & 11.18 & +11.57\% & 6.11  & 9.31  & +1.91\% \\
                              & 4.95  & 5.32  & 6.11  & +12.97\% & 5.13  & 4.98  & -2.97\% \\
\multirow{2}{*}{Bili\_Music}  & 16.68 & 17.12 & 17.62 & +2.83\%  & 15.26 & 15.27 & +0.12\% \\
                              & 10.16 & 9.51  & 10.06 & +5.56\%  & 8.44  & 8.61  & +1.92\% \\
\midrule
\multirow{2}{*}{KU}           & 20.74 & 26.59 & 26.84 & +0.92\%  & 28.26 & 26.74 & -5.39\% \\
                              & 19.40 & 23.38 & 23.38 & +0.00\%  & 23.77 & 23.31 & -1.93\% \\
\multirow{2}{*}{QB}           & 27.03 & 28.78 & 29.13 & +1.21\%  & 28.09 & 28.51 & +1.46\% \\
                              & 23.94 & 22.01 & 22.47 & +2.03\%  & 21.28 & 23.16 & +8.09\% \\
\multirow{2}{*}{TN}           & 11.62 & 12.74 & 12.73 & -0.08\%  & 11.79 & 12.38 & +4.74\% \\
                              & 7.16  & 7.41  & 6.92  & -6.57\%  & 6.69  & 7.11  & +5.89\% \\
\multirow{2}{*}{DY}           & 13.84 & 12.11 & 10.95 & -9.59\%  & 11.97 & 12.19 & +1.81\% \\
                              & 8.99  & 7.25  & 6.09  & -15.95\% & 6.97  & 7.13  & +2.25\% \\
\bottomrule
\end{tabular}}
% \end{sc}
\end{small}
\end{center}
\vskip -0.1in
\end{table}

\section{Author Contributions}
Fajie is the corresponding author who designed, supervised and funded this research; Jiaqi performed this research, in charge of technical parts; Fajie, Jiaqi wrote the paper and answered the reviewers' questions; Chengyu, Zheng, Youhua assisted some key experiments; Yunzhu, Yongxin, Jie together collected the 10 datasets.

%%%%%%%%% Main Paper REFERENCES
\bibliography{reference}
\bibliographystyle{IEEEtran}

%%%%%%%%% Appendix
% \newpage
\clearpage
\appendix

\begin{figure*}[hb]
    \centering
    \includegraphics[width=1\textwidth]{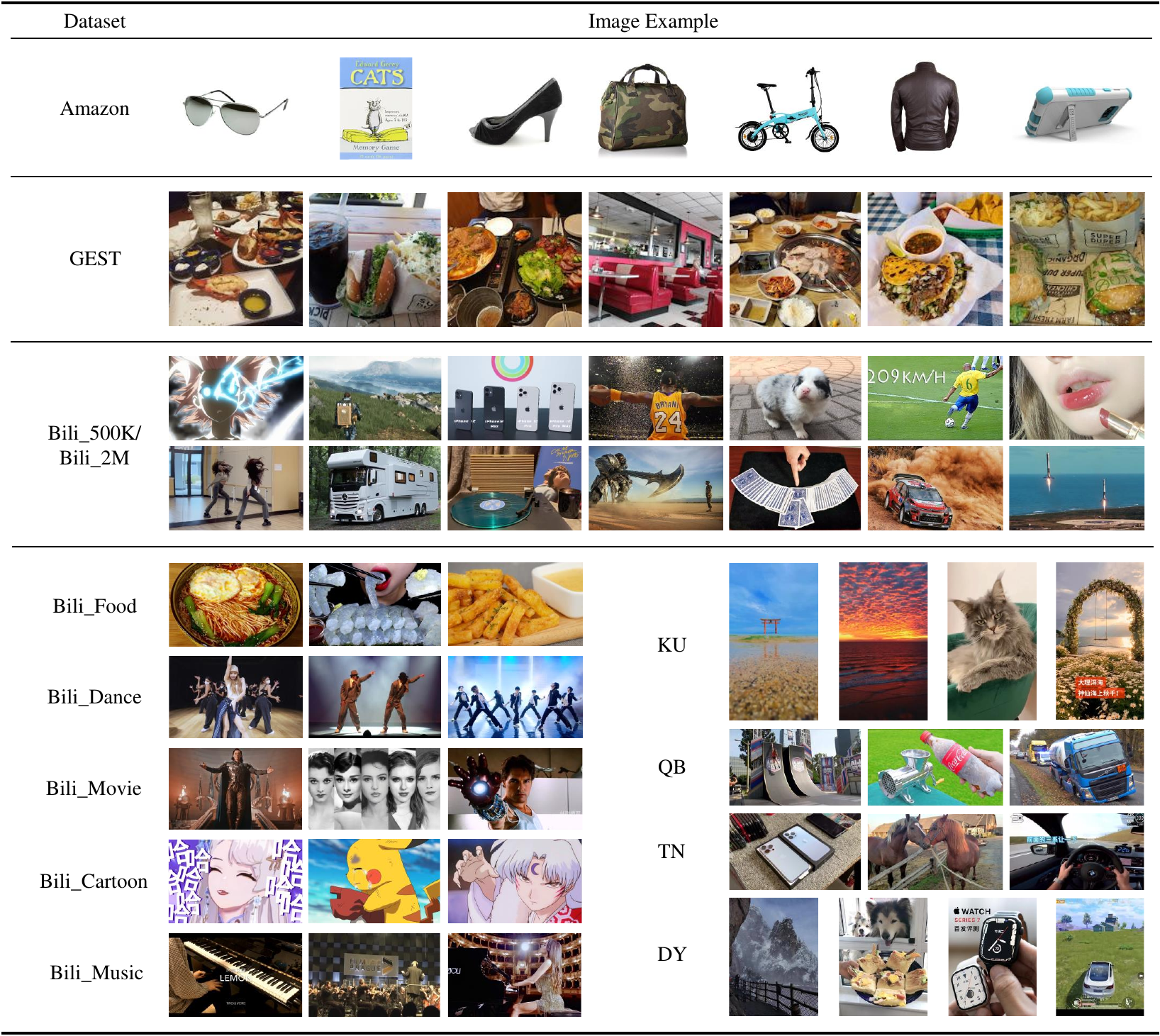}
    \caption{Image Examples of NineRec vs. Amazon vs. GEST. Images in GEST are mainly about food and restaurants.  Images in Amazon are mainly about single products with very low semantics.
    }
    \label{apx:fig:ImageExample}
    \vskip 1.0in
\end{figure*}

\begin{figure*}
	\centering
	\subfloat[Bili\_500K]{
		\begin{minipage}[htbp]{0.46\columnwidth}
			\includegraphics[width=1\columnwidth]{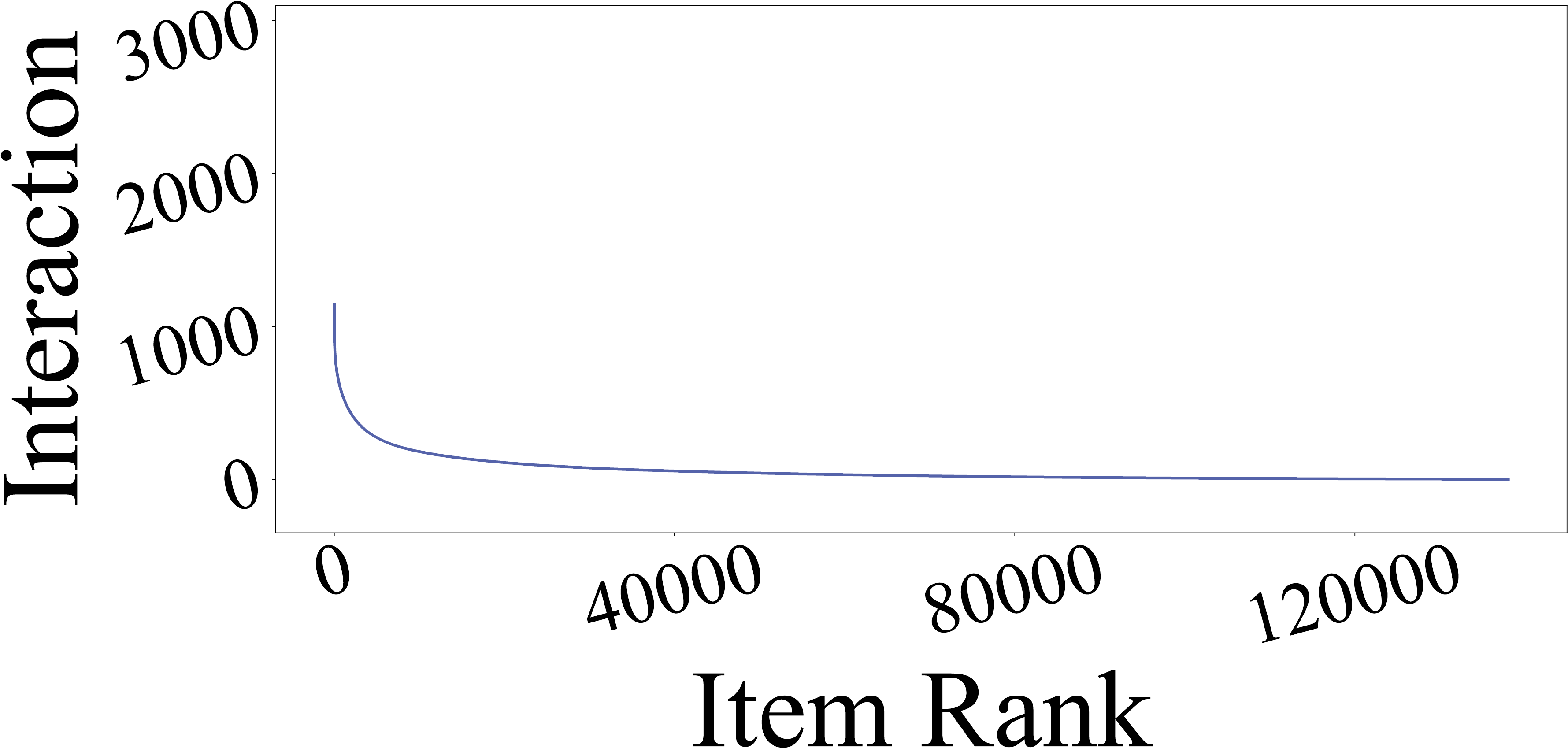}\vspace{1mm} \\
                \includegraphics[width=1\columnwidth]{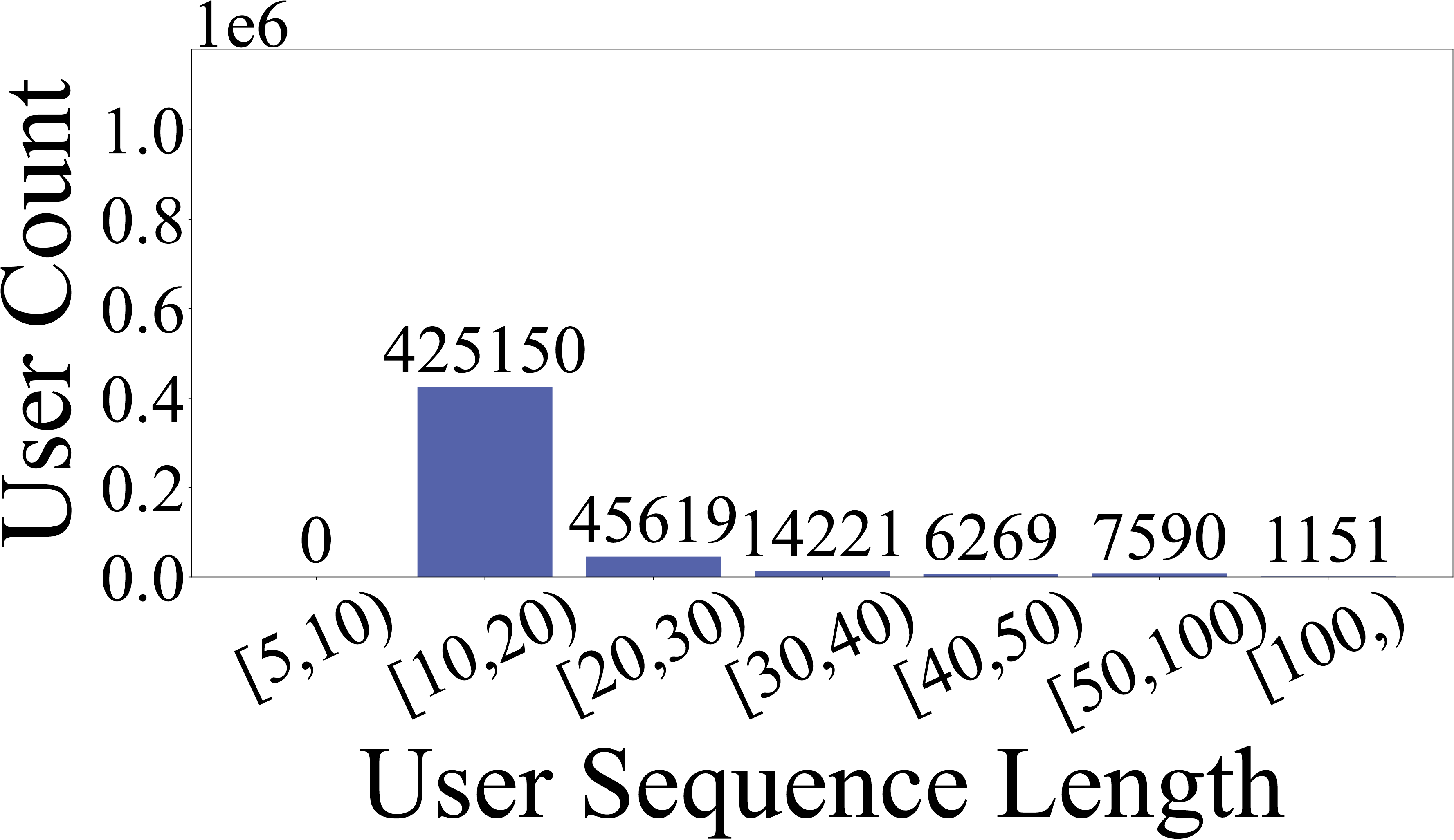}\vspace{1mm} \\ 
                \includegraphics[width=1\columnwidth]{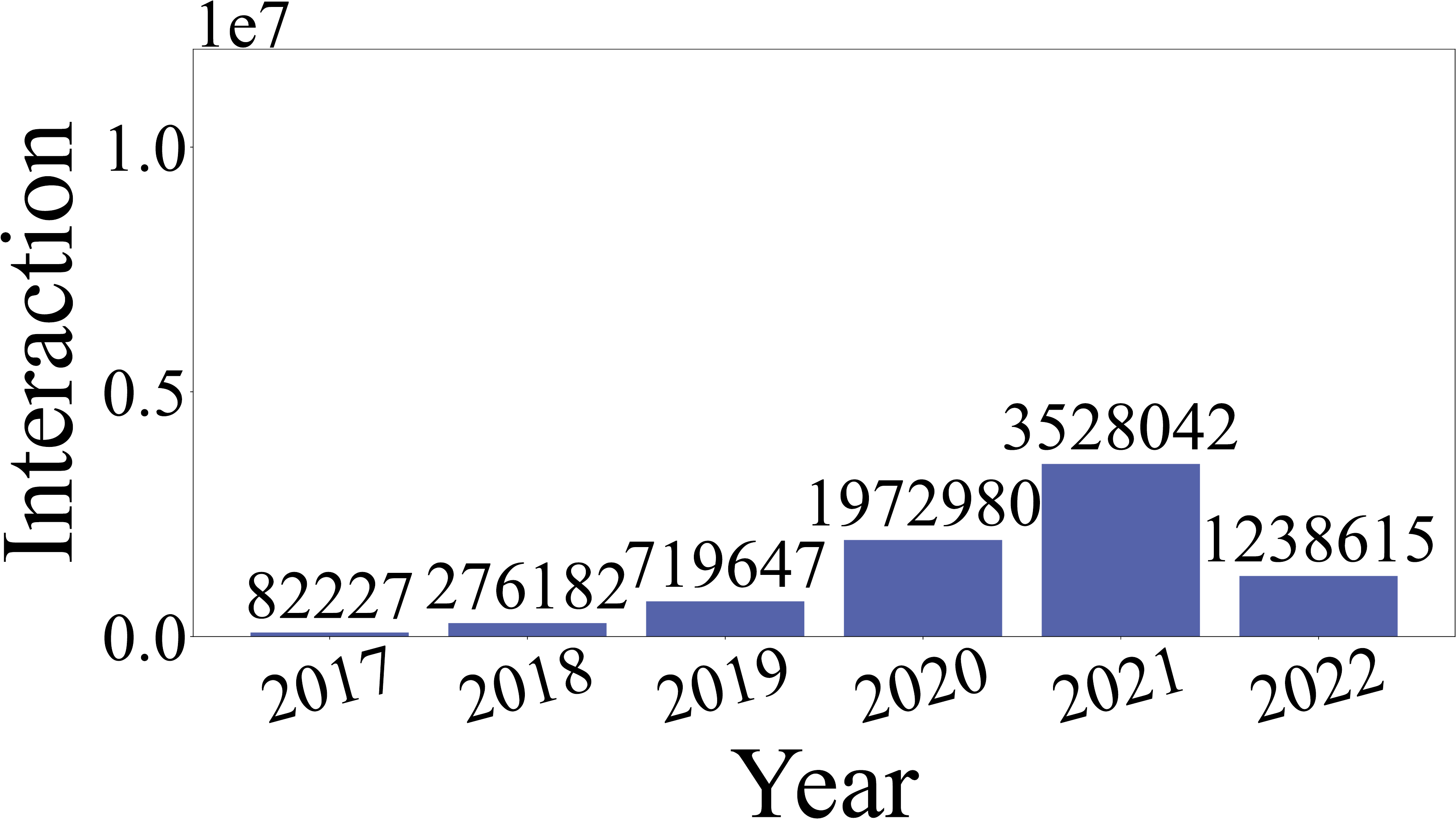}\vspace{1mm}
		\end{minipage}
	}
        \subfloat[Bili\_Food]{
    	\begin{minipage}[htbp]{0.46\columnwidth}
			\includegraphics[width=1\columnwidth]{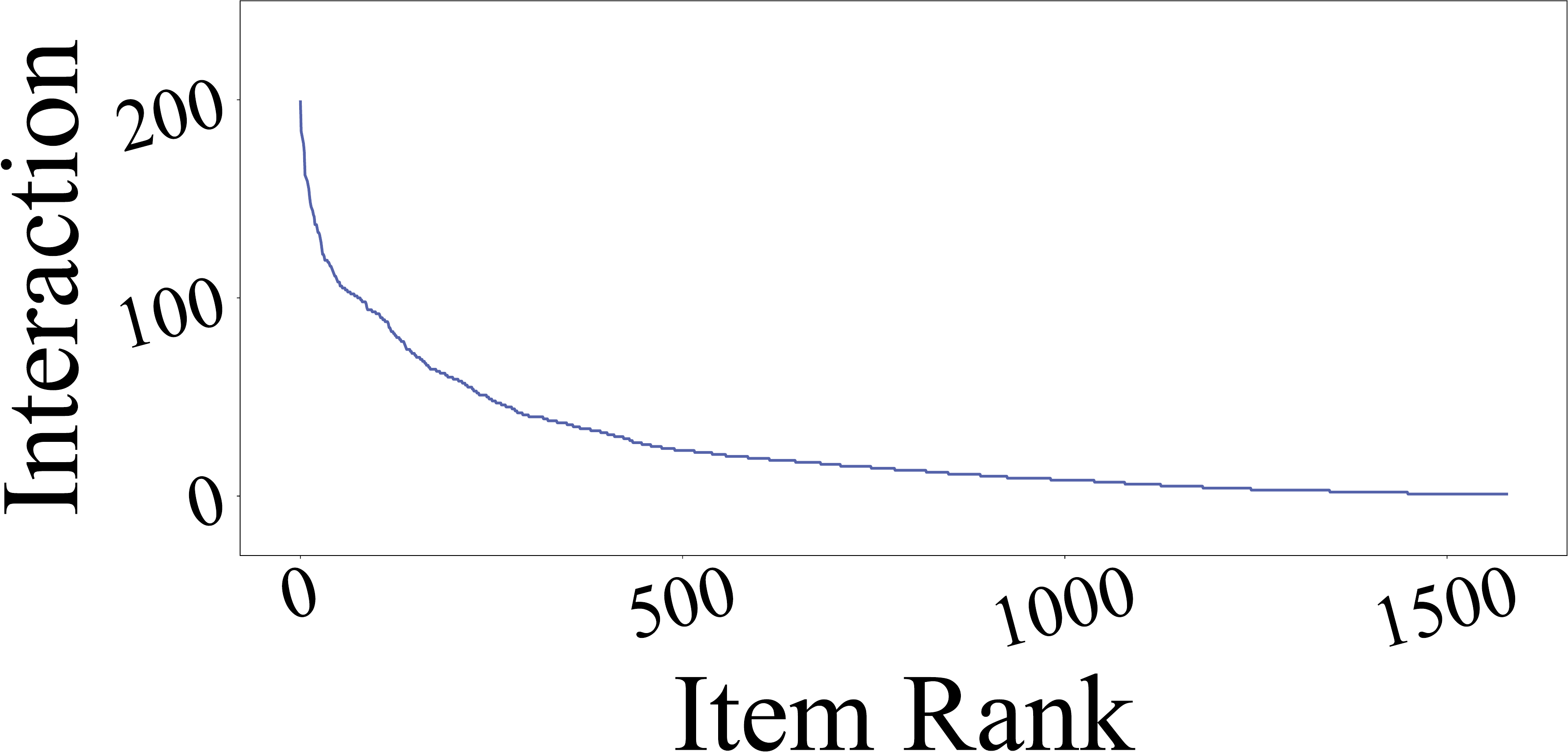}\vspace{1mm} \\
                \includegraphics[width=1\columnwidth]{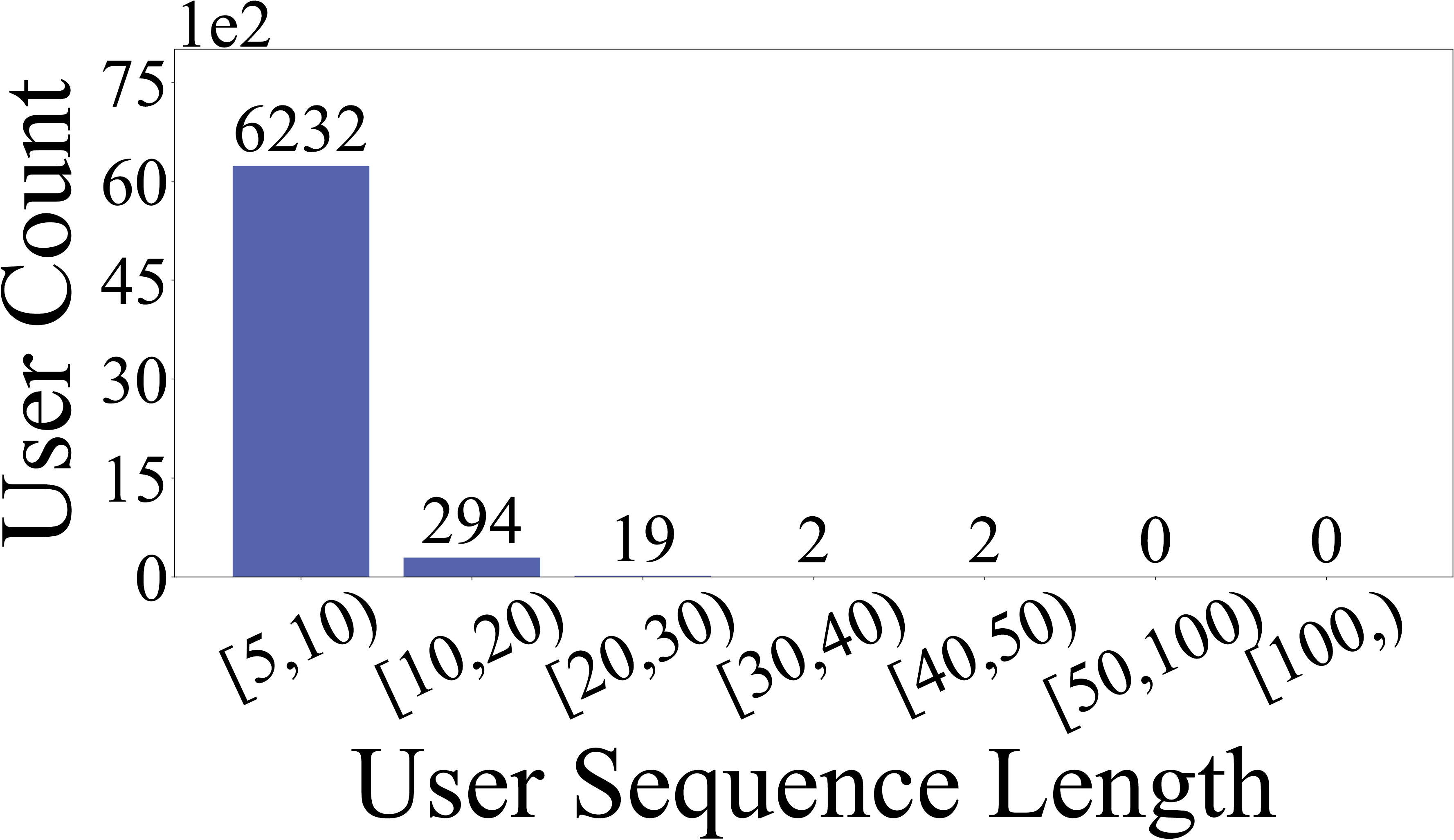}\vspace{1mm} \\ 
                \includegraphics[width=1\columnwidth]{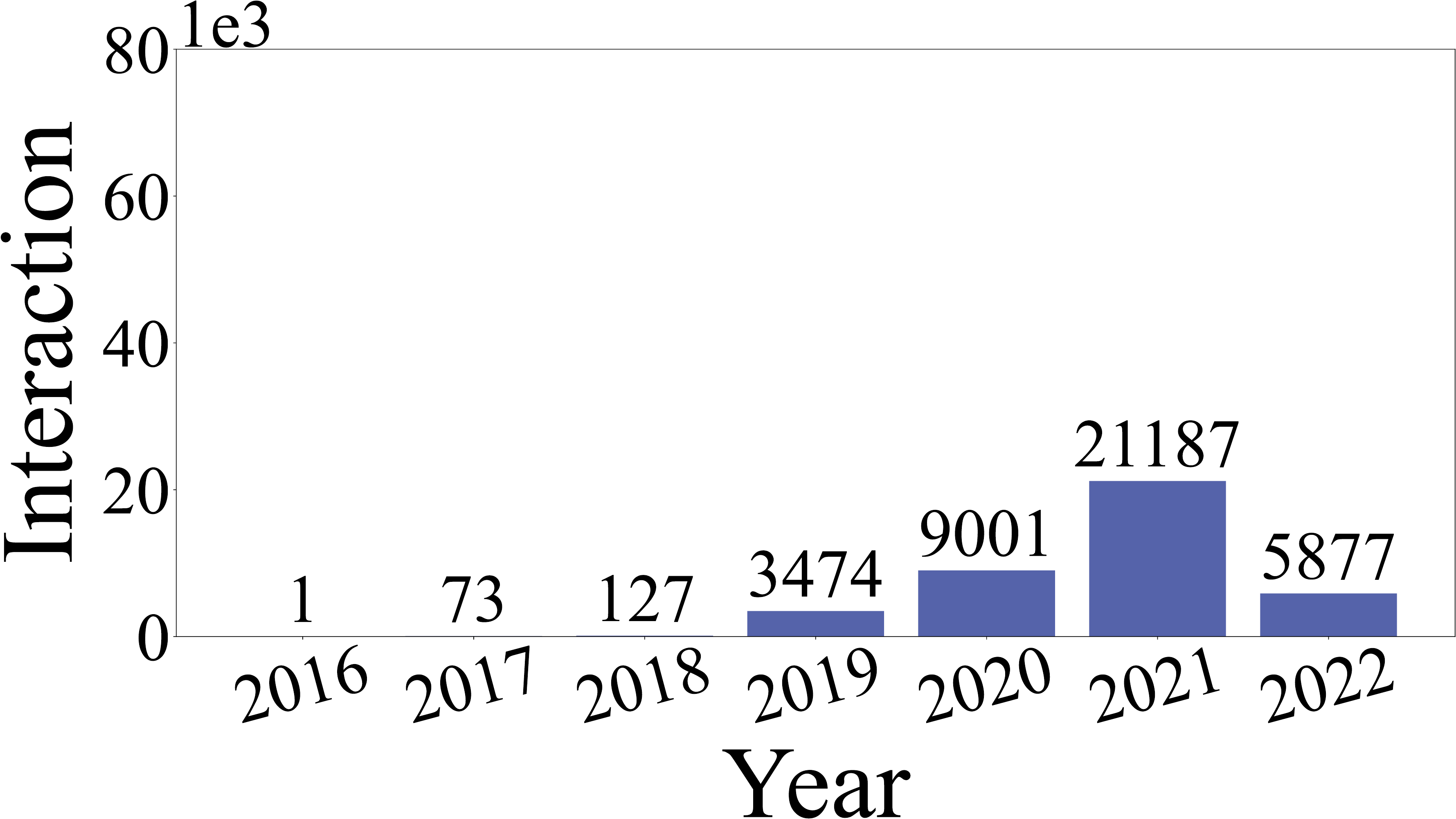}\vspace{1mm}
    	\end{minipage}
        }
        \subfloat[Bili\_Dance]{
    	\begin{minipage}[htbp]{0.46\columnwidth}
			\includegraphics[width=1\columnwidth]{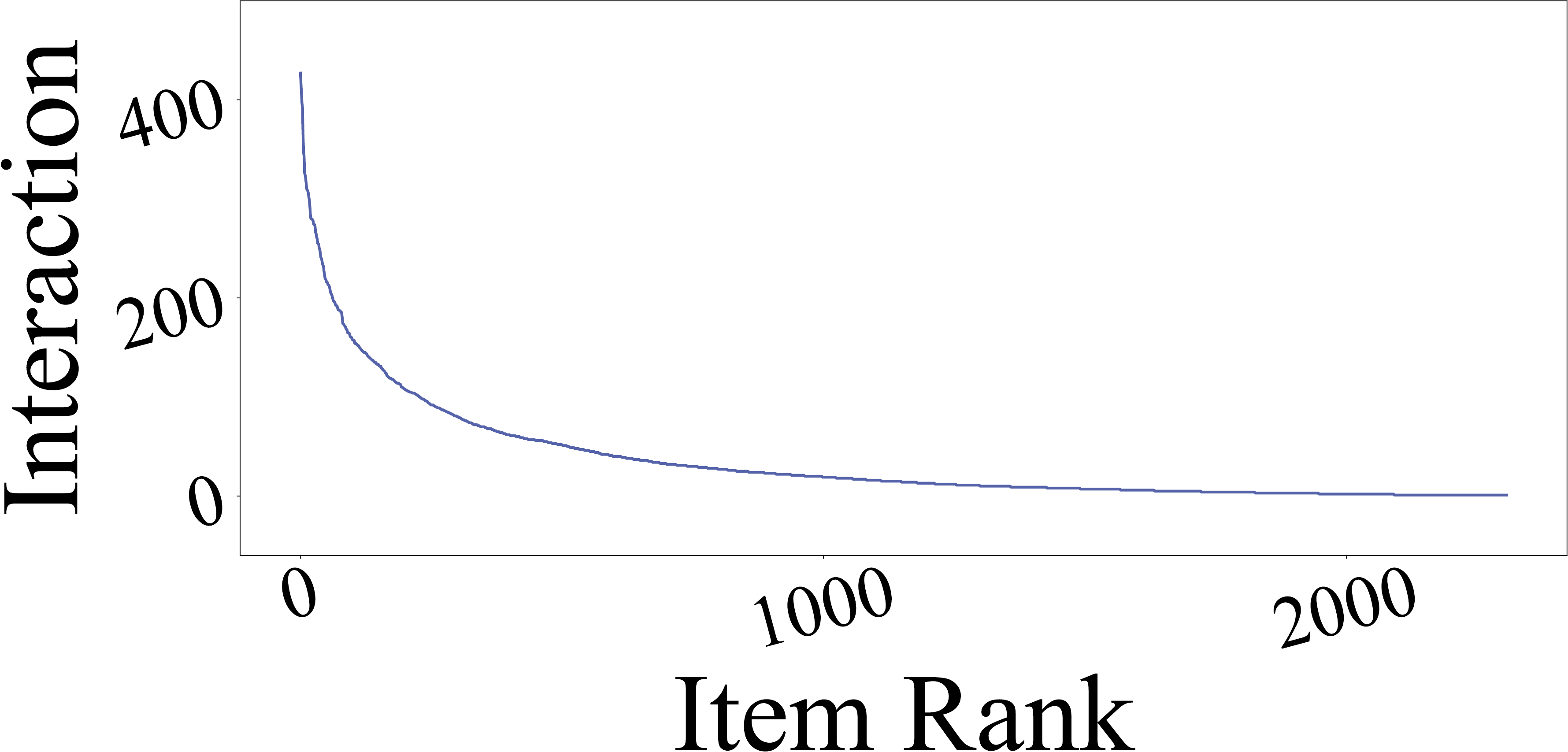}\vspace{1mm} \\
                \includegraphics[width=1\columnwidth]{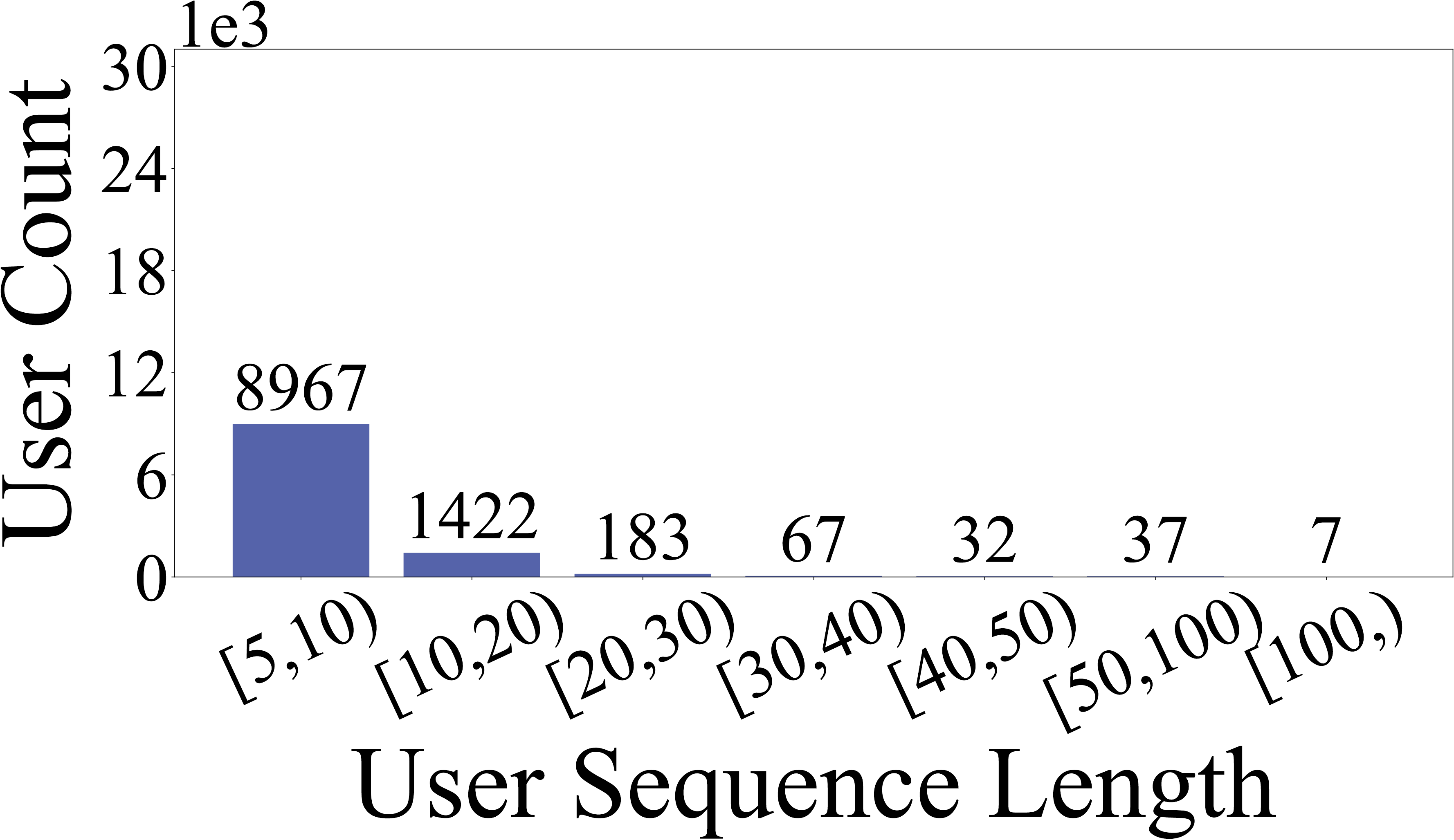}\vspace{1mm} \\ 
                \includegraphics[width=1\columnwidth]{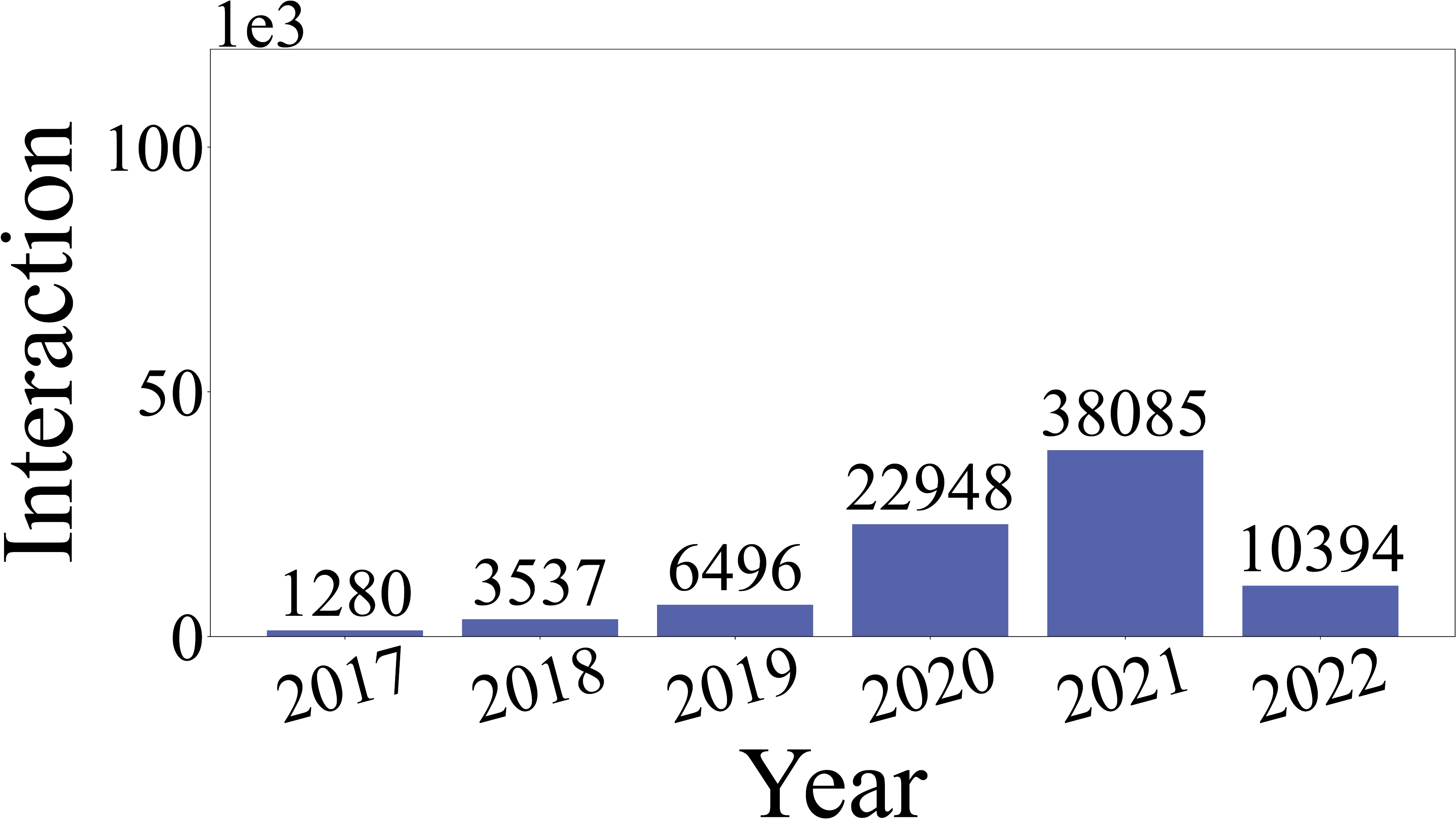}\vspace{1mm}
    	\end{minipage}
        }
        \qquad
        \subfloat[Bili\_Movie]{
    	\begin{minipage}[htbp]{0.46\columnwidth}
			\includegraphics[width=1\columnwidth]{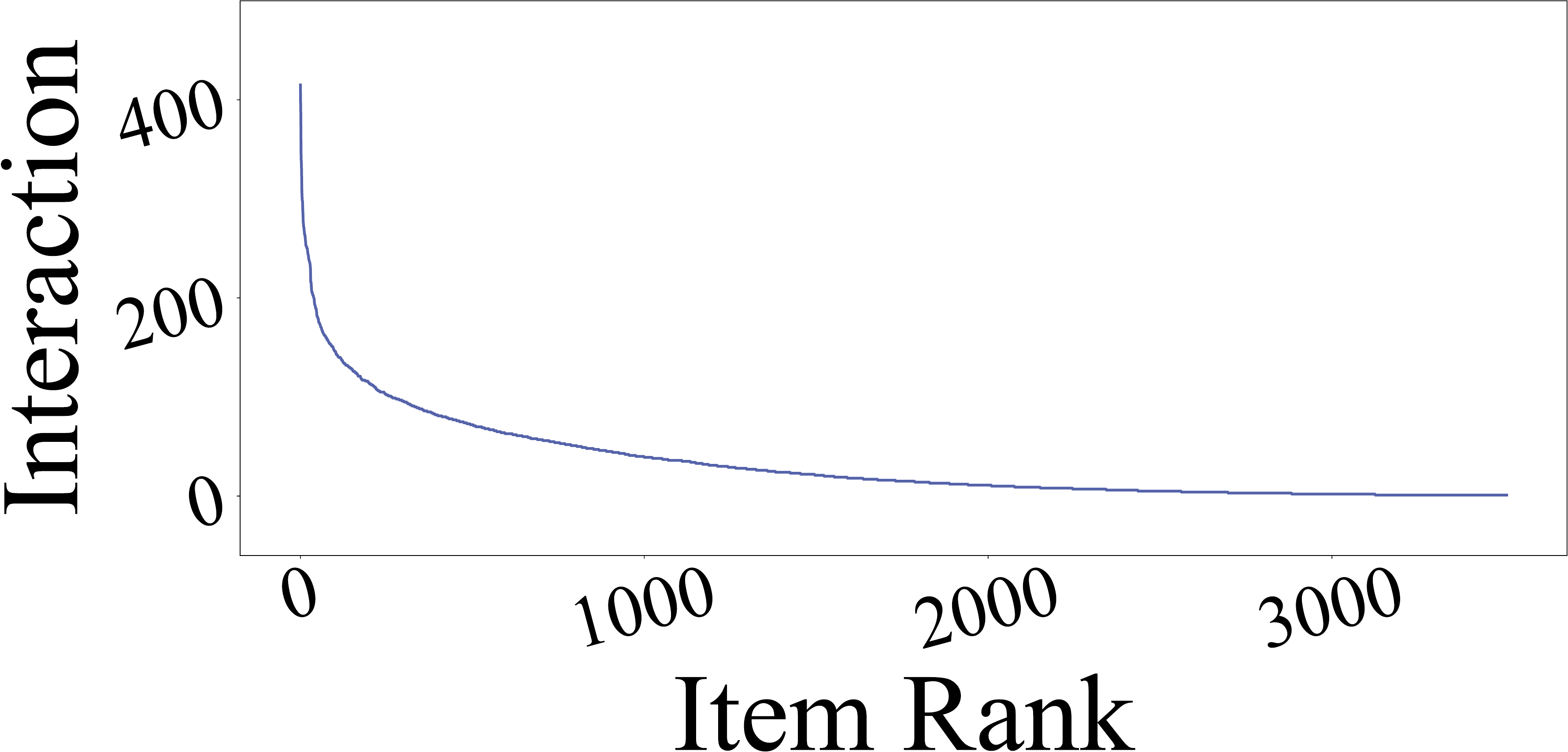}\vspace{1mm} \\
                \includegraphics[width=1\columnwidth]{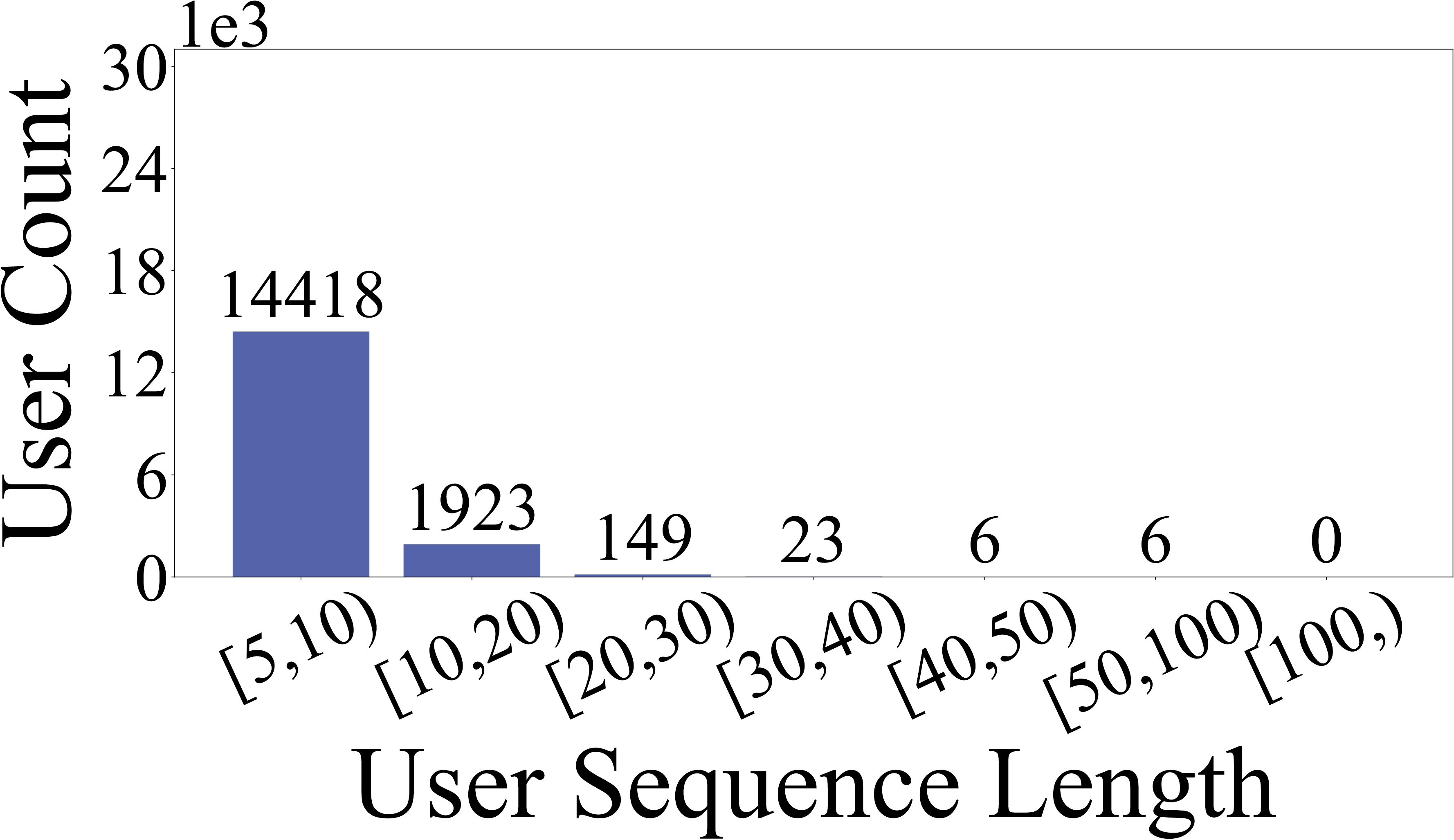}\vspace{1mm} \\ 
                \includegraphics[width=1\columnwidth]{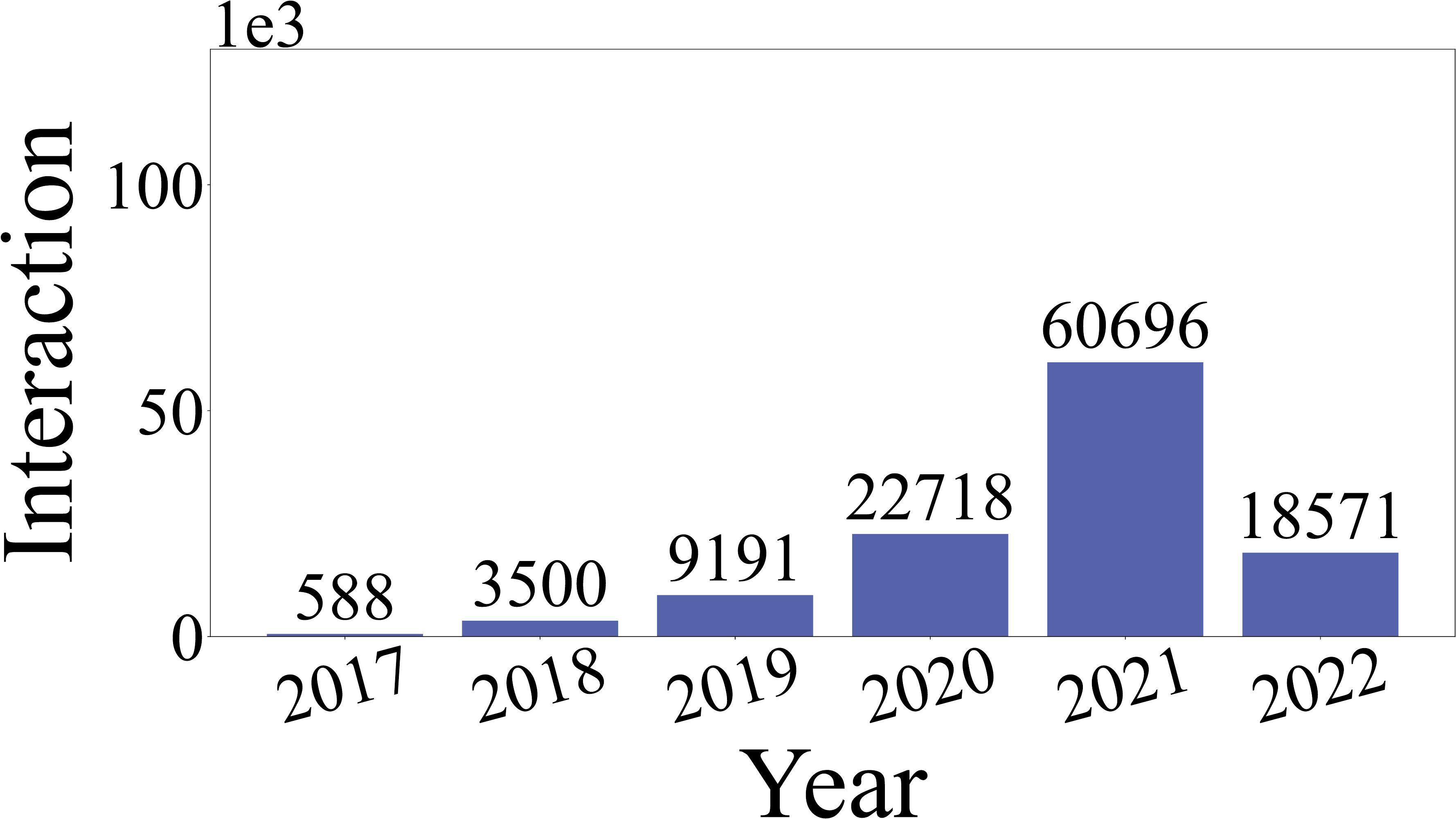}\vspace{1mm}
    	\end{minipage}
        }
        \subfloat[Bili\_Cartoon]{
    	\begin{minipage}[htbp]{0.46\columnwidth}
			\includegraphics[width=1\columnwidth]{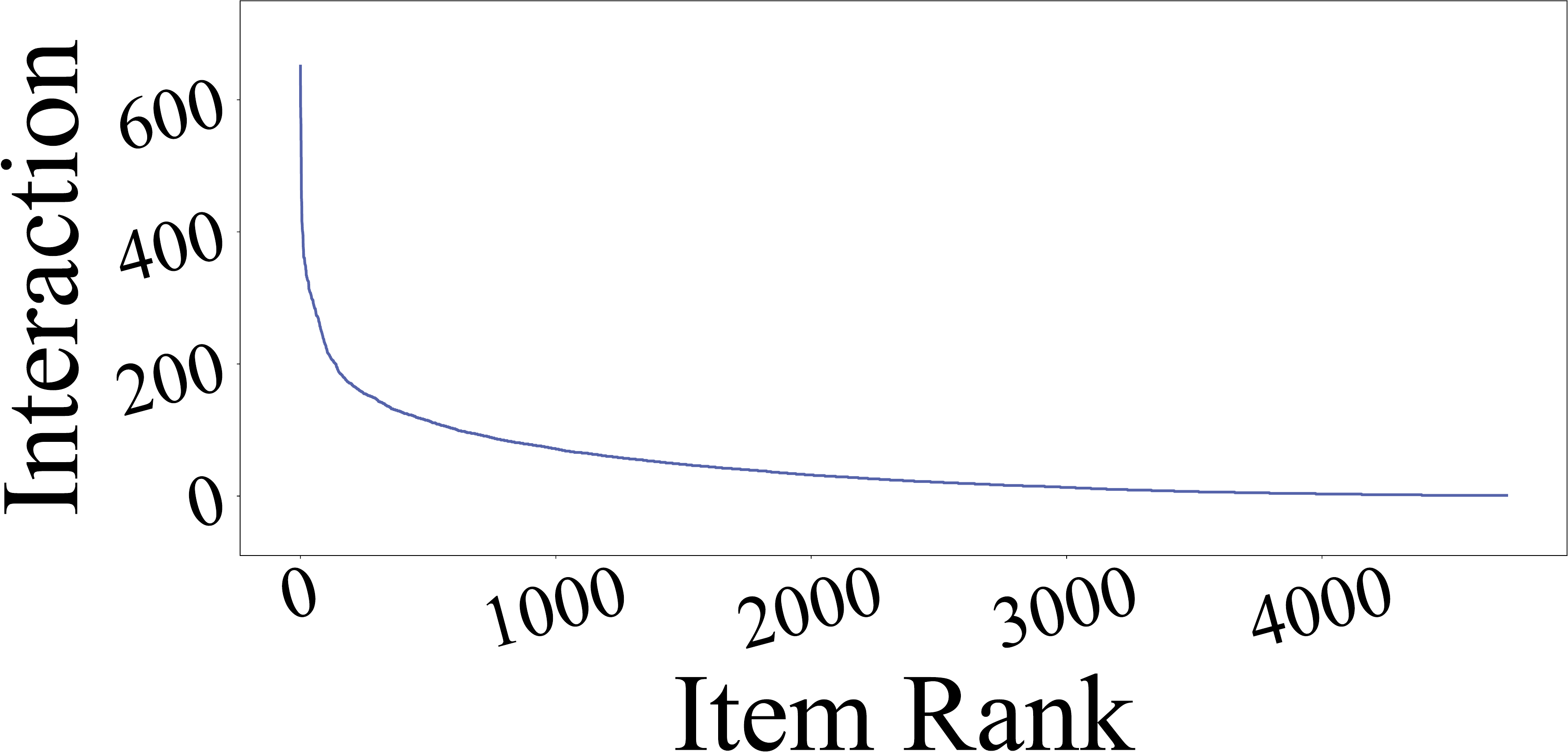}\vspace{1mm} \\
                \includegraphics[width=1\columnwidth]{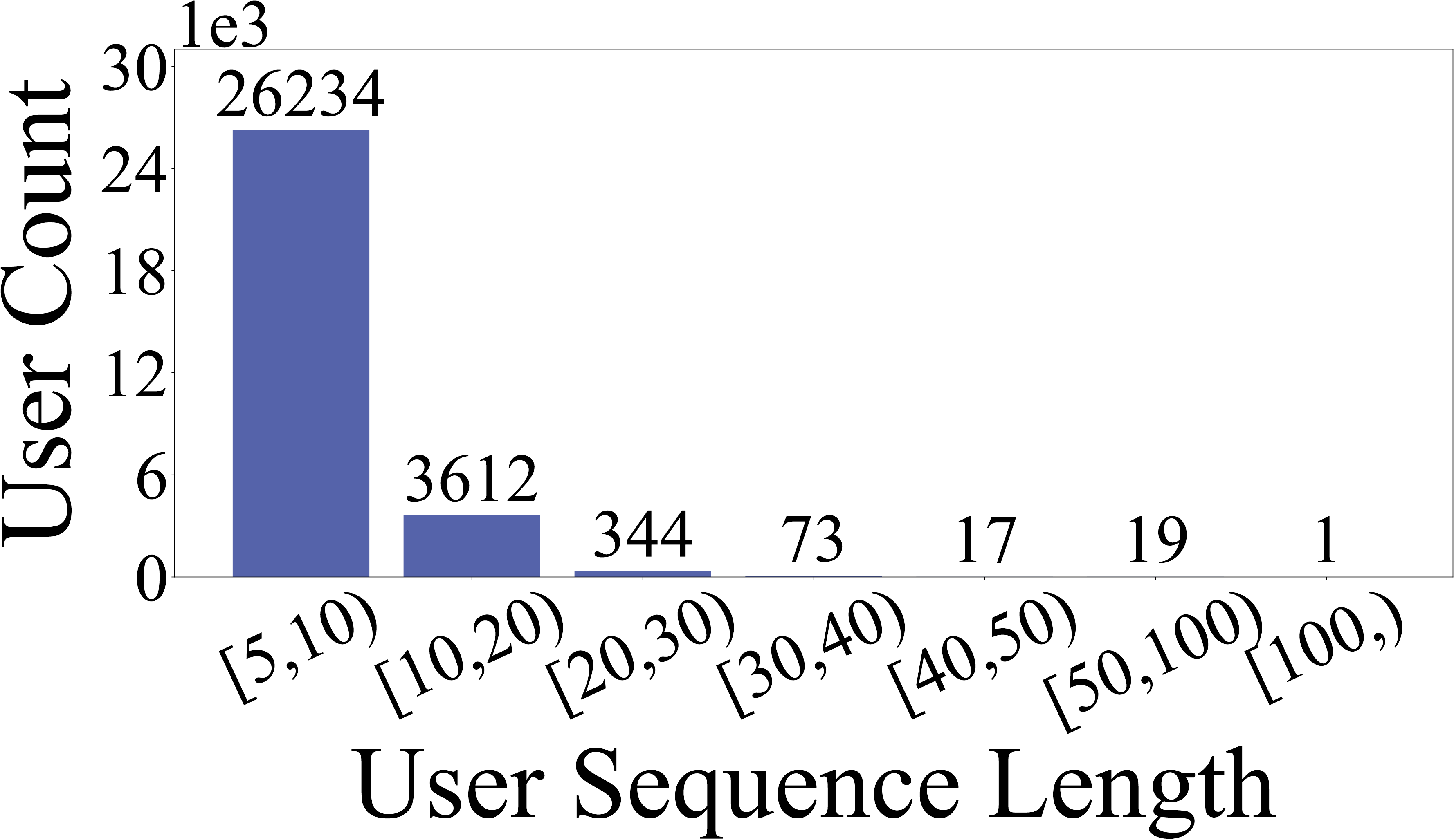}\vspace{1mm} \\ 
                \includegraphics[width=1\columnwidth]{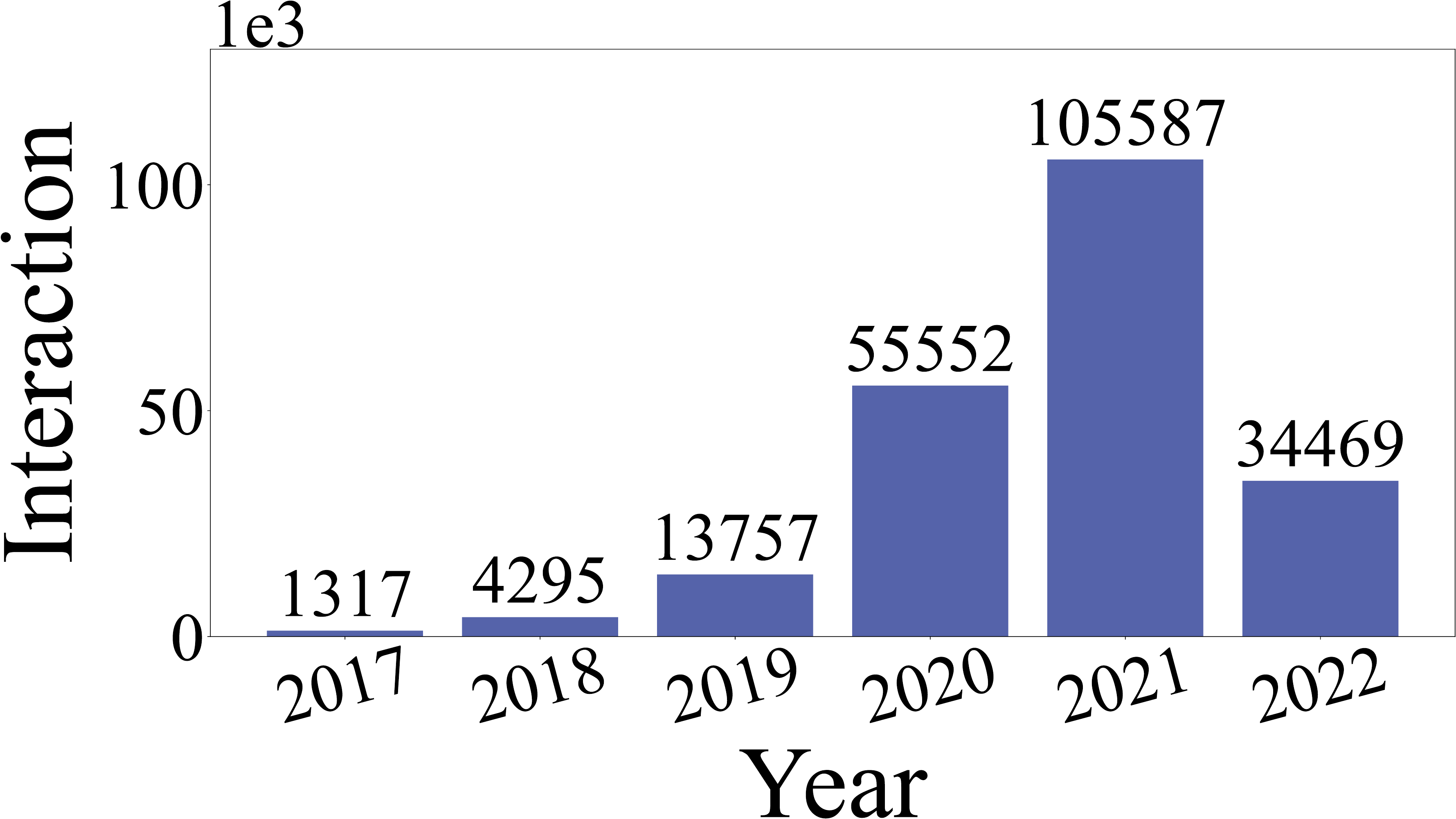}\vspace{1mm}
    	\end{minipage}
        }
        \subfloat[Bili\_Music]{
    	\begin{minipage}[htbp]{0.46\columnwidth}
			\includegraphics[width=1\columnwidth]{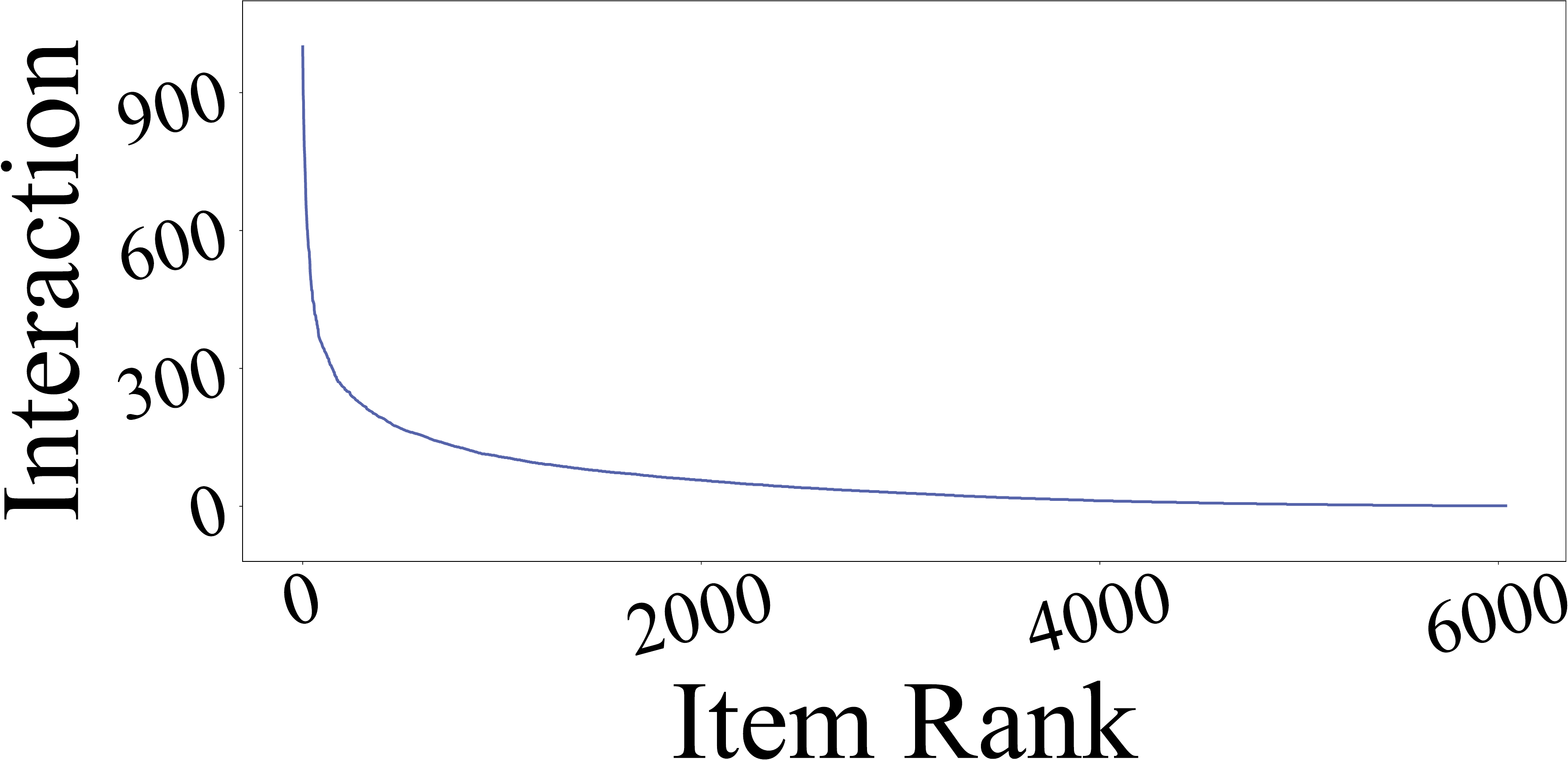}\vspace{1mm} \\
                \includegraphics[width=1\columnwidth]{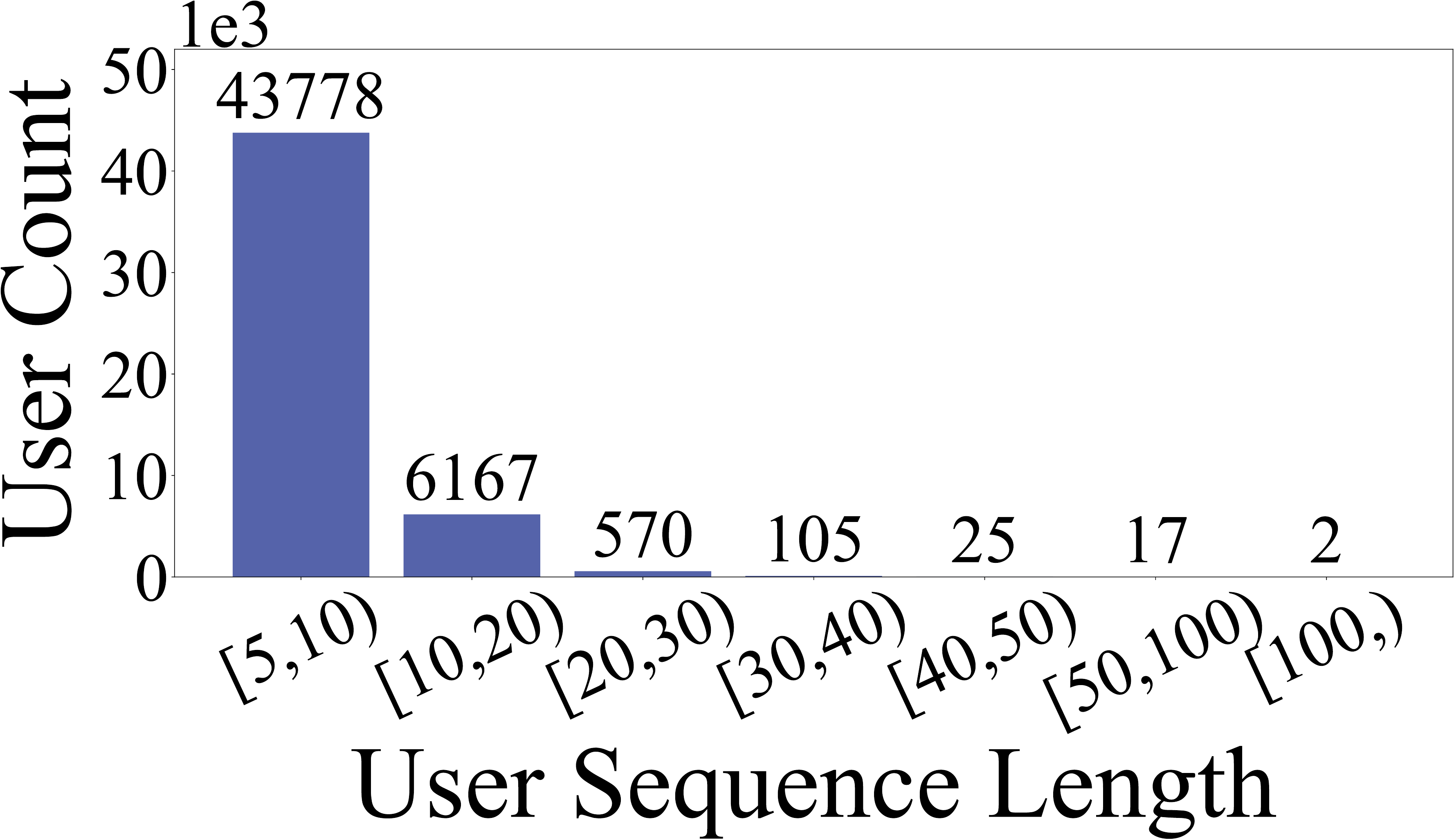}\vspace{1mm} \\ 
                \includegraphics[width=1\columnwidth]{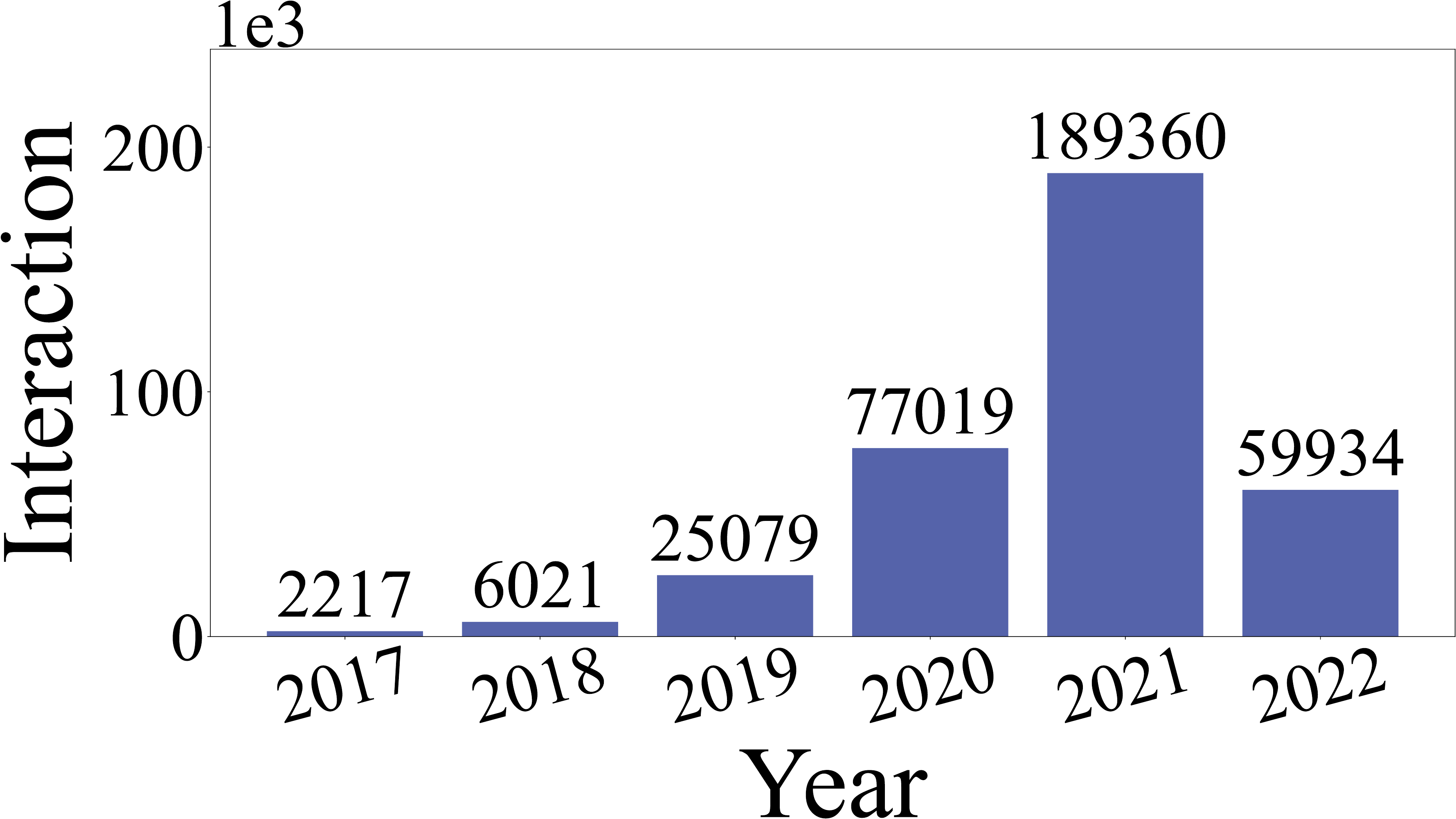}\vspace{1mm}
    	\end{minipage}
        }
	\caption{Dataset details. Top:  item popularity distribution; Middle:  user interaction length distribution;  Bottom: the occurring time of user-item interactions.  Except TN  and QB, every user-item interaction in these datasets has an accurate timestamp. The QB and TN data was collected earlier, we did not keep the timestamps but ranked them by interaction time at that time.}
        \label{apx:fig:ItemDistribution}
\end{figure*}

\begin{figure*}[htbp]
\centering
    \subfloat[DSSM]{
    \includegraphics[width=0.7\columnwidth]{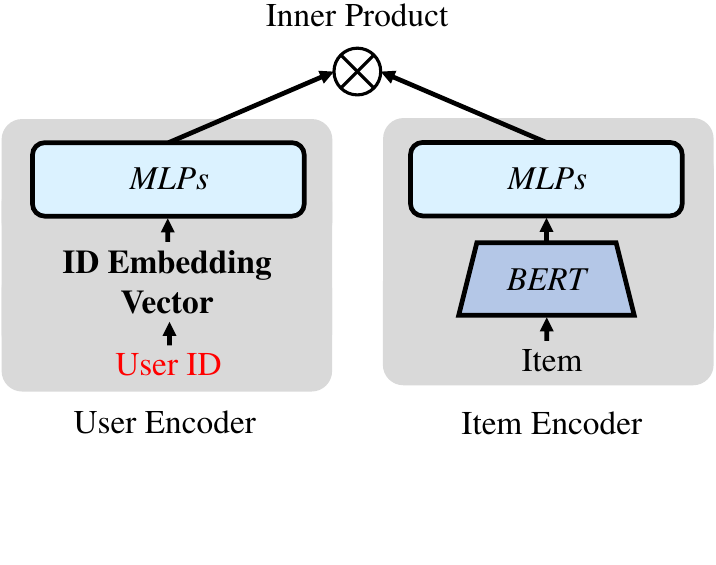}
    }\hspace{1cm}
    \subfloat[DSSM-variant]{
    \includegraphics[width=0.7\columnwidth]{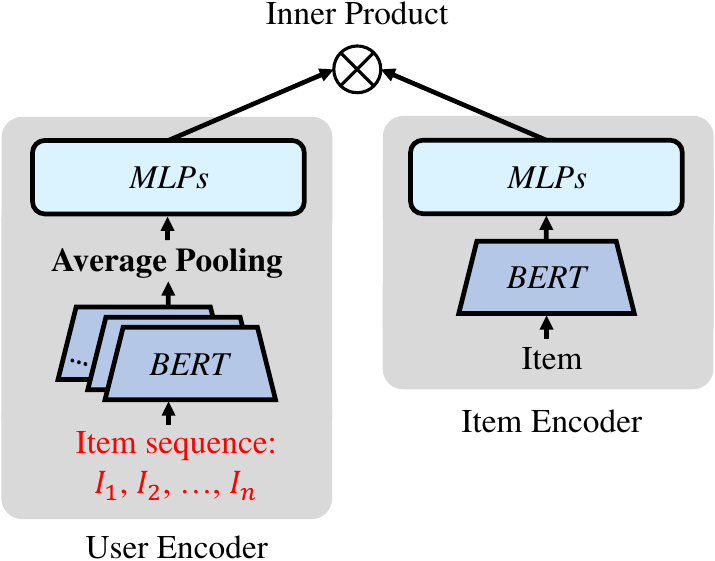}
    }
\caption{Illustration of DSSM and DSSM-variant for text recommendation. DSSM-variant is essentially the same network architecture as the S2O mode, but use DNN/MLP (vs. RNN, CNN, MHSA modules in main paper Figure 3) as the user encoder backbone. Note that parameters of each item encoder of a user sequence are always shared.}
\label{apx:fig:DSSM vs DSSM-variant}
\end{figure*}

\begin{table*}[t]
\caption{NDCG@10 Results of TransRec on the nine target datasets corresponding to main paper Table 2. Results in italics denote model collapse. Results in bold indicate the maximum between NoPT and HasPT. Underlined results are the maximum value among all.}
\label{tab:Comparative_StudyNDCG}
\vskip 0.15in
\begin{center}
\begin{small}
% \begin{sc}
\begin{tabular}{p{1.2cm}<{\centering}  p{1.2cm}<{\centering}  p{0.7cm}<{\centering}  p{0.7cm}<{\centering}  p{0.8cm}<{\centering}  p{0.7cm}<{\centering}  p{0.7cm}<{\centering}  p{0.8cm}<{\centering}  p{0.7cm}<{\centering}  p{0.7cm}<{\centering}  p{0.8cm}<{\centering}  p{0.7cm}<{\centering}  p{0.7cm}<{\centering}  p{0.8cm}<{\centering}}
\toprule
\multirow{2}{*}{Dataset} & \multirow{2}{*}{Metric} &\multicolumn{3}{c}{SASRec} &\multicolumn{3}{c}{BERT4Rec}  &\multicolumn{3}{c}{NextItNet}  &\multicolumn{3}{c}{GRU4Rec} \\
\cmidrule(r){3-5}\cmidrule(r){6-8}\cmidrule(r){9-11}\cmidrule(r){12-14}
&&IDRec &NoPT &HasPT &IDRec &NoPT &HasPT &IDRec &NoPT &HasPT &IDRec &NoPT &HasPT \\
\midrule  
\multicolumn{14}{c}{BERT (base version) for text recommendation} \\
\midrule%      &            SASRec                       &          BERT4Rec                       &          NextItNet          &        GRU4Rec        
Bili\_Food     &N@10 &10.34 &9.95  &\underline{\textbf{10.95}} &9.35  &8.57  &\underline{\textbf{10.75}} &6.85  &8.49  &\underline{\textbf{9.52 }} &6.23  &9.10  &\underline{\textbf{9.25 }} \\
Bili\_Dance    &N@10 &13.77 &13.76 &\underline{\textbf{14.72}} &12.17 &12.27 &\underline{\textbf{15.29}} &9.98  &11.84 &\underline{\textbf{13.30}} &9.23  &11.89 &\underline{\textbf{12.48}} \\
Bili\_Movie    &N@10 &6.32  &6.14  &\underline{\textbf{6.95 }} &5.63  &5.81  &\underline{\textbf{7.10 }} &4.52  &5.50  &\underline{\textbf{6.39 }} &3.49  &4.99  &\underline{\textbf{6.07 }} \\
Bili\_Cartoon  &N@10 &6.52  &6.55  &\underline{\textbf{7.35 }} &6.10  &6.34  &\underline{\textbf{7.66 }} &4.35  &5.97  &\underline{\textbf{6.62 }} &4.55  &5.71  &\underline{\textbf{6.66 }} \\ 
Bili\_Music    &N@10 &11.6  &11.40 &\underline{\textbf{11.91}} &9.91  &9.67  &\underline{\textbf{12.27}} &8.57  &10.27 &\underline{\textbf{11.34}} &8.73  &8.81  &\underline{\textbf{9.86 }} \\
\midrule
KU             &N@10 &24.79 &25.23 &\underline{\textbf{26.19}} &21.74 &12.76 &\underline{\textbf{24.36}} &20.95 &23.70 &\underline{\textbf{23.87}} &10.64 &24.16 &\underline{\textbf{24.39}} \\ 
QB             &N@10 &\underline{28.04} &26.30 &\textbf{27.16} &\underline{24.86} &22.57 &\textbf{24.03} &\underline{25.82} &21.69 &\textbf{23.17} &\underline{25.38} &23.72 &\textbf{24.76} \\
TN             &N@10 &8.74  &8.24  &\underline{\textbf{9.23 }} &8.74  &8.12  &\underline{\textbf{9.55 }} &\underline{7.74 } &\textbf{7.33 } &7.04  &8.36  &\underline{\textbf{8.99 }} &8.00 \\ 
DY             &N@10 &\underline{9.93 } &8.31  &\textbf{8.38 } &7.41  &4.92  &\underline{\textbf{7.56 }} &\underline{7.33 } &\textbf{5.88 } &5.44  &5.86  &8.84  &\underline{\textbf{9.01 }} \\
\midrule 
\multicolumn{14}{c}{Swin Transformer (base version) for image recommendation} \\
\midrule%      &            SASRec                       &                      BERT4Rec                        &          NextItNet                      &        GRU4Rec   
Bili\_Food     &N@10 &\underline{10.34} &9.42  &\textbf{10.15}    &9.35  &\textit{1.19} &\underline{\textbf{10.18}}    &6.85  &7.75  &\underline{\textbf{9.98 }}    &6.23 &9.00  &\underline{\textbf{9.42 }} \\
Bili\_Dance    &N@10 &\underline{13.77} &12.14 &\textbf{12.27}    &\underline{12.17} &8.82                   &\textbf{10.73}    &9.98  &9.84  &\underline{\textbf{12.46}}    &9.23 &10.76 &\underline{\textbf{11.86}} \\
Bili\_Movie    &N@10 &\underline{6.32 } &5.68  &\textbf{6.15 }    &\underline{5.63 } &\textit{0.78} &\textbf{5.15 }    &4.52  &4.43  &\underline{\textbf{5.88 }}    &3.49 &4.81  &\underline{\textbf{5.48 }} \\
Bili\_Cartoon  &N@10 &\underline{6.52 } &6.04  &\textbf{6.48 }    &6.10              &5.34       &\underline{\textbf{6.42 }}    &4.35  &4.77  &\underline{\textbf{5.90 }}    &4.55 &4.68  &\underline{\textbf{5.20 }} \\
Bili\_Music    &N@10 &11.6  &9.64  &\underline{\textbf{9.89 }}    &\underline{9.91 } &6.24                   &\textbf{8.53 }    &8.57  &8.92  &\underline{\textbf{9.60 }}    &8.73 &8.44  &\underline{\textbf{8.93 }} \\
\midrule
KU             &N@10 &24.79 &26.08 &\underline{\textbf{26.41}}    &\underline{21.74} &14.53                  &\textbf{17.94}    &20.95 &21.80 &\underline{\textbf{26.91}}    &10.6 &20.61 &\underline{\textbf{25.00}} \\  
QB             &N@10 &\underline{28.04} &24.98 &\textbf{26.60}    &\underline{24.86} &\textbf{17.04}         &13.86             &25.82 &22.85 &\underline{\textbf{25.85}}    &25.3 &24.37 &\underline{\textbf{25.45}} \\ 
TN             &N@10 &\underline{8.74 } &\textbf{8.30 } &8.00     &\underline{8.74 } &\textbf{7.37 }         &6.85              &\underline{7.74}  &6.89  &\textbf{7.09 }    &\underline{8.36} &\textbf{8.12 } &7.97 \\ 
DY             &N@10 &\underline{9.93 } &8.27  &\textbf{8.50 }    &\underline{7.41 } &\textbf{5.53 }         &4.90              &\underline{7.33}  &6.94  &\textbf{7.26 }    &5.86 &7.65  &\underline{\textbf{8.13 }} \\
\bottomrule
\end{tabular}
% \end{sc}
\end{small}
\end{center}
\vskip 0.1in
\end{table*}

\begin{table*}[!h]
\caption{Results of TransRec (SASRec as UE) on the target domain datasets  pre-trained on Bili\_2M}
\label{tab:bili2mtarget}
% \vskip 0.1in
\begin{center}
\begin{small}
% \begin{sc}
\begin{tabular}{p{1.2cm}<{\centering} r p{1.5cm}<{\centering}  p{1.5cm}<{\centering}  p{1.5cm}<{\centering}   p{1.5cm}<{\centering}   p{1.5cm}<{\centering}}
\toprule
\multirow{2}{*}{Dataset} &&\multirow{2}{*}{Metric} &\multicolumn{2}{c}{BERT} &\multicolumn{2}{c}{Swin-B}\\
\cmidrule(r){4-5}\cmidrule(r){6-7}
&&&NoPT &HasPT &NoPT &HasPT \\
\midrule
\multirow{2}{*}{Bili\_Food}     &&H@10 &18.03 &19.43  &17.20 &18.73 \\
                                &&N@10 &9.95  &10.67  &9.42  &10.23 \\
\multirow{2}{*}{Bili\_Dance}    &&H@10 &23.49 &25.10  &21.63 &22.30 \\
                                &&N@10 &13.76 &14.60  &12.14 &12.86 \\
\multirow{2}{*}{Bili\_Movie}    &&H@10 &11.64 &13.13  &10.30 &11.56 \\
                                &&N@10 &6.14  &7.21   &5.68  &6.49  \\
\multirow{2}{*}{Bili\_Cartoon}  &&H@10 &11.94 &13.75  &11.09 &11.43 \\
                                &&N@10 &6.55  &7.39   &6.04  &6.01 \\
\multirow{2}{*}{Bili\_Music}    &&H@10 &19.42 &21.74  &17.17 &18.05 \\
                                &&N@10 &11.40 &12.37  &9.64  &10.45 \\
\midrule
\multirow{2}{*}{KU}             &&H@10 &30.77 &29.69  &33.08 &31.80 \\
                                &&N@10 &25.23 &25.84  &26.08 &25.69 \\
\multirow{2}{*}{QB}             &&H@10 &34.27 &34.79  &32.39 &33.36 \\
                                &&N@10 &26.30 &26.90  &24.98 &26.23 \\
\multirow{2}{*}{TN}             &&H@10 &15.11 &16.65  &14.12 &15.39 \\
                                &&N@10 &8.24  &9.10   &8.30  &9.03  \\
\multirow{2}{*}{DY}             &&H@10 &14.35 &16.11  &14.08 &15.02 \\
                                &&N@10 &8.31  &9.47   &8.27  &8.63  \\
\bottomrule
\end{tabular}
% \end{sc}
\end{small}
\end{center}
% \vskip -0.1in
\end{table*}

\begin{table*}[htbp]
\caption{HR@10 Result of zero-shot recommendation.  ZeroRec-frozenME denotes parameters of ME is fixed during pre-training on the source dataset. We show ZeroRec vs.Random and ZeroRec vs. HasPT for analysis.
See Figure~\ref{apx:fig:ZeroshotFig} for description.}
\label{apx:tab:Zeroshot}
\vskip 0.15in
\begin{center}
\begin{small}
% \begin{sc}
\begin{tabular}{p{1.2cm}<{\centering} r p{1.5cm}<{\centering} p{1.5cm}<{\centering} p{1.5cm}<{\centering} p{1.5cm}<{\centering} p{2cm}<{\centering}}
\toprule %\multirow{2}{*}{Metric}   \makecell[c]{Cross-\\Domain}
\multirow{2}{*}{Dataset} &&\multirow{2}{*}{Random} &\multirow{2}{*}{ZeroRec} &\multirow{2}{*}{\makecell[c]{ZeroRec-\\frozenME}} &\multirow{2}{*}{vs. Random} &\multirow{2}{*}{\makecell[c]{vs. HasPT}} \\
&&&&&& \\
\midrule    
\multicolumn{7}{c}{BERT} \\
\cmidrule(r){1-7}  
Bili\_Food    &&0.63 &\textbf{4.76} &2.56 &7.55x  &25.42\% \\
Bili\_Dance   &&0.43 &\textbf{8.30} &2.82 &19.30x &32.37\% \\
Bili\_Movie   &&0.28 &\textbf{5.16} &1.33 &18.42x &40.85\% \\
Bili\_Cartoon &&0.21 &\textbf{5.95} &2.50 &28.33x &45.29\% \\
Bili\_Music   &&0.16 &\textbf{11.32}&2.84 &70.75x &54.34\% \\
\midrule             
KU            &&0.49 &\textbf{4.96} &3.53 &10.12x &15.83\% \\
QB            &&0.18 &\textbf{7.58} &3.33 &42.11x &21.81\% \\
TN            &&0.26 &\textbf{1.21} &0.91 &4.65x  &7.82\% \\
DY            &&0.12 &\textbf{0.88} &0.55 &7.33x  &6.90\% \\
\midrule             
\multicolumn{7}{c}{Swin-B} \\
\cmidrule(r){1-7} 
Bili\_Food    &&0.63 &\textbf{4.65} &2.67 &7.38x  &26.06\% \\
Bili\_Dance   &&0.43 &\textbf{3.40} &3.01 &7.90x  &15.36\% \\
Bili\_Movie   &&0.28 &\textbf{3.28} &1.95 &11.71x &28.56\% \\
Bili\_Cartoon &&0.21 &\textbf{3.49} &1.46 &16.61x &29.65\% \\
Bili\_Music   &&0.16 &\textbf{3.71} &2.05 &23.18x &21.15\% \\
\midrule             
KU            &&0.49 &\textbf{7.86} &7.17 &16.04x &23.80\% \\
QB            &&0.18 &\textbf{7.43} &5.77 &41.22x &22.14\% \\
TN            &&0.26 &\textbf{1.38} &1.25 &5.30x  &9.62\% \\
DY            &&0.12 &1.14 &\textbf{1.16} &9.50x  &7.90\% \\
\bottomrule
\end{tabular}
% \end{sc}
\end{small}
\end{center}
\vskip -0.1in
\end{table*}

\begin{figure*}[h]
\centering
\subfloat[Random]{
\includegraphics[width=0.4\columnwidth]{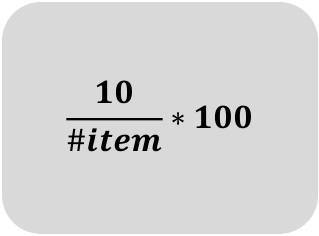}}\hspace{0.5cm}
\subfloat[ZeroRec]{
\includegraphics[width=0.4\columnwidth]{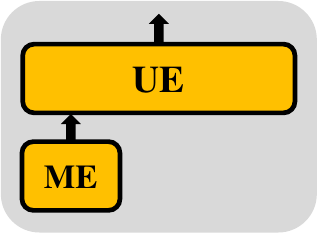}}\hspace{0.5cm}
\subfloat[ZeroRec-frozenME]{
\includegraphics[width=0.4\columnwidth]{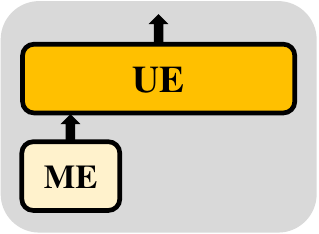}}
\caption{Zero-shot recommendation.  ZeroRec is the TransRec
(SASRec as UE, BERT as ME) pre-trained on the source dataset,
and then predicts (without fine-tuning) on the target datasets. ZeroRec-frozenME means parameters of ME is frozen during pre-training on the source dataset. }
\label{apx:fig:ZeroshotFig}
% \vskip 1.0in
\end{figure*}

\begin{table*}[ht]
% \begin{sc}
% \vskip 0.2in
\caption{Network architecture, parameter size, and download URL of the pre-trained ME we used. L: the number of Transformer blocks, H: the number of multi-head attention, C: the channel number of the hidden layers in the first stage \cite{liu2021swin}, B: the number of layers in each block.
Note that since Roberta, OPT, CLIP and ViLT have no Chinese version, we translated the
Chinese text into English (by DeepL: https://www.deepl.com/translatorand) then performed the evaluation. English translation will be
provided in NineRec.
}
\label{apx:tab:MEDetailsTable}
\vskip 0.15in
\begin{center}
\begin{small}
\begin{tabular}{p{2.8cm}<{\centering}  c  p{0.8cm}<{\centering}  c }%
\toprule
Pre-trained model  &Architecture  &\#Param.  &URL \\
\midrule
chinese-bert-wwm               & L=12, H=768  &102M  &https://huggingface.co/hfl/chinese-bert-wwm \\
$\text{RoBERTa}_{\text{base}}$ & L=12, H=768  &125M  &https://huggingface.co/roberta-base \\
$\text{OPT}_{\text{125M}}$     & L=12, H=768  &125M  &https://huggingface.co/facebook/opt-125M \\
\midrule
ResNet50 &C = 64, B=\{3, 4, 6, 3\} &26M   &https://download.pytorch.org/models/resnet50-19c8e357.pth \\
Swin-T   &C = 96, B=\{2, 2, 6, 2\} &28M   &https://huggingface.co/microsoft/swin-tiny-patch4-window7-224 \\
Swin-S   &C = 96, B=\{2, 2, 18, 2\} &50M   &https://huggingface.co/microsoft/swin-small-patch4-window7-224 \\
Swin-B   &C = 128, B=\{2, 2, 18, 2\} &88M   &https://huggingface.co/microsoft/swin-base-patch4-window7-224 \\
\midrule
CLIP     &L=24, H=768 &144M &https://huggingface.co/openai/clip-vit-base-patch32 \\ 
ViLT     &L=12, H=768 &104M &https://huggingface.co/dandelin/vilt-b32-mlm \\
\bottomrule
\end{tabular}
% \end{sc}
\end{small}
\end{center}
\vskip -0.1in
\end{table*}

\begin{table*}[htbp]
\caption{The best hyper parameters and training cost (SASRec as UE). $\gamma^{\text{UE}}$: learning rate of UE, $\gamma^{\text{ME}}$: learning rate of ME, $\beta^{\text{UE}}$: weight decay of recommendation network, $\beta^{\text{ME}}$: weight decay of modality encoder, $d$: embedding/hidden size, $b$: batch size, $l$: number of transformer layers in UE, FLOPs: computational complexity, 
Time/E: averaged training time per  iteration. As can be seen, MoRec or TransRec (using S2S training mode + SASRec backbone) requires \textbf{100x-1000x} more training computation and time than IDRec (the S2O is 10x-20x faster than S2S mode).   Note that we use \textcolor{black}{1  3090 GPU} for IDRec but \textcolor{black}{4 or 8 most powerful A100 GPUs} for TransRec. The number of GPUs should be considered when comparing their training time. For example, on Bili\_2M, MoRec with Swin-B requires nearly 200x more training than IDRec. In other words, the training of SASRec as UE and Swin-B as ME needs \textbf{nearly 40 days for training 80 iterations using 8 NVIDIA A100}, which  costs about \textbf{11,000 US dollars} (half discount) for one set of hyper-parameters. 
}
\label{apx:tab:hyperparameter-training cost}
\vskip 0.15in
\begin{center}
\scalebox{0.9}{
\begin{tabular}{p{2cm}<{\centering} p{1.5cm}<{\centering} p{1.5cm}<{\centering} | p{0.8cm}<{\centering} p{0.8cm}<{\centering} p{0.8cm}<{\centering} p{0.8cm}<{\centering} p{0.8cm}<{\centering} p{0.8cm}<{\centering} p{0.8cm}<{\centering} | p{1cm}<{\centering} p{0.8cm}<{\centering} p{2cm}<{\centering}}
\toprule
Scenario &Dataset &Method &$\gamma^{\text{UE}}$ &$\gamma^{\text{ME}}$ &$\beta^{\text{UE}}$ &$\beta^{\text{ME}}$ &$d$ &$b$ &$l$ &FLOPs &Time/E &GPU \\
\midrule
\multirow{9}*{Source}
&\multirow{3}*{Bili\_2M}   &IDRec   &5e-5 &-    &0.1 &-   &1024 &64 &2 &0.5G &30m  &\textbf{3090-24G(1)}\\
&                          &BERT    &5e-6 &5e-6 &0.1 &0.1 &1024 &64 &2 &107G &140m &\textbf{A100-80G(4)}\\
&                          &Swin-B  &5e-5 &5e-5 &0.1 &0.0 &1024 &64 &2 &637G &695m &\textbf{A100-80G(8)}\\
\cmidrule(r){2-13}
&\multirow{6}*{Bili\_500K} &IDRec   &5e-5 &-    &0.1 &-   &1024 &64 &2 &0.5G &10m  &3090-24G(1)\\
&                          &BERT    &5e-5 &1e-5 &0.1 &0.1 &1024 &64 &2 &107G &94m  &A100-80G(4)\\
&                          &Res50   &5e-5 &5e-5 &0.1 &0.0 &1024 &64 &2 &174G &88m  &A100-80G(4)\\
&                          &Swin-T  &5e-5 &5e-5 &0.1 &0.0 &1024 &64 &2 &183G &130m &A100-80G(4)\\
&                          &Swin-S  &5e-5 &5e-5 &0.1 &0.0 &1024 &64 &2 &358G &145m &A100-80G(4)\\
&                          &Swin-B  &5e-5 &5e-5 &0.1 &0.0 &1024 &64 &2 &637G &109m &A100-80G(8)\\
\midrule 
\multirow{30}*{\makecell[c]{Cross-\\Domain}}
&\multirow{6}*{Bili\_Food}       &IDRec   &1e-4 &-    &0.1 &-   &256  &64 &2 &0.5G &8s   &3090-24G(1)\\
&                          &BERT    &1e-5 &1e-5 &0.1 &0.1 &1024 &64 &2 &107G &1m   &A100-80G(4)\\
&                          &Res50   &5e-5 &5e-5 &0.1 &0.0 &1024 &64 &2 &174G &1m   &A100-80G(4)\\
&                          &Swin-T  &5e-5 &5e-5 &0.1 &0.0 &1024 &64 &2 &183G &1m   &A100-80G(4)\\
&                          &Swin-S  &5e-5 &5e-5 &0.1 &0.0 &1024 &64 &2 &358G &1m   &A100-80G(4)\\
&                          &Swin-B  &5e-5 &5e-5 &0.1 &0.0 &1024 &64 &2 &637G &2m   &A100-80G(8)\\
\cmidrule(r){2-13}
&\multirow{6}*{Bili\_Dance}      &IDRec   &5e-5 &-    &0.1 &-   &512  &64 &2 &0.5G &12s  &3090-24G(1)\\
&                          &BERT    &1e-5 &1e-5 &0.1 &0.1 &1024 &64 &2 &107G &2m   &A100-80G(4)\\
&                          &Res50   &5e-5 &5e-5 &0.1 &0.0 &1024 &64 &2 &174G &1m   &A100-80G(4)\\
&                          &Swin-T  &5e-5 &5e-5 &0.1 &0.0 &1024 &64 &2 &183G &1m   &A100-80G(4)\\
&                          &Swin-S  &5e-5 &5e-5 &0.1 &0.0 &1024 &64 &2 &358G &2m   &A100-80G(4)\\
&                          &Swin-B  &5e-5 &5e-5 &0.1 &0.0 &1024 &64 &2 &637G &3m   &A100-80G(8)\\
\cmidrule(r){2-13}
&\multirow{6}*{Bili\_Movie}      &IDRec   &5e-5 &-    &0.1 &-   &512  &64 &2 &0.5G &15s  &3090-24G(1)\\
&                          &BERT    &5e-5 &1e-5 &0.1 &0.1 &1024 &64 &2 &107G &2.5m &A100-80G(4)\\
&                          &Res50   &5e-5 &5e-5 &0.1 &0.0 &1024 &64 &2 &174G &2m   &A100-80G(4)\\
&                          &Swin-T  &5e-5 &5e-5 &0.1 &0.0 &1024 &64 &2 &183G &2m   &A100-80G(4)\\
&                          &Swin-S  &5e-5 &5e-5 &0.1 &0.0 &1024 &64 &2 &358G &3m   &A100-80G(4)\\
&                          &Swin-B  &5e-5 &5e-5 &0.1 &0.0 &1024 &64 &2 &637G &4m   &A100-80G(8)\\
\cmidrule(r){2-13}
&\multirow{6}*{Bili\_Cartoon}    &IDRec   &5e-5 &-    &0.1 &-   &512  &128&2 &0.5G &20s  &3090-24G(1)\\
&                          &BERT    &1e-5 &1e-5 &0.1 &0.1 &1024 &64 &2 &107G &4.5m &A100-80G(4)\\
&                          &Res50   &5e-5 &5e-5 &0.1 &0.0 &1024 &64 &2 &174G &3m   &A100-80G(4)\\
&                          &Swin-T  &5e-5 &5e-5 &0.1 &0.0 &1024 &64 &2 &183G &3m   &A100-80G(4)\\
&                          &Swin-S  &5e-5 &5e-5 &0.1 &0.0 &1024 &64 &2 &358G &5m   &A100-80G(4)\\
&                          &Swin-B  &5e-5 &5e-5 &0.1 &0.0 &1024 &64 &2 &637G &7m   &A100-80G(8)\\
\cmidrule(r){2-13}
&\multirow{6}*{Bili\_Music}      &IDRec   &1e-5 &-    &0.1 &-   &1024 &64 &2 &0.5G &30s  &3090-24G(1)\\
&                          &BERT    &1e-5 &1e-5 &0.1 &0.1 &1024 &64 &2 &107G &8m   &A100-80G(4)\\
&                          &Res50   &5e-5 &5e-5 &0.1 &0.0 &1024 &64 &2 &174G &4.5m &A100-80G(4)\\
&                          &Swin-T  &5e-5 &5e-5 &0.1 &0.0 &1024 &64 &2 &183G &4.5m &A100-80G(4)\\
&                          &Swin-S  &5e-5 &5e-5 &0.1 &0.0 &1024 &64 &2 &358G &8.5m &A100-80G(4)\\
&                          &Swin-B  &5e-5 &5e-5 &0.1 &0.0 &1024 &64 &2 &637G &12m  &A100-80G(8)\\
\midrule
\multirow{24}*{\makecell[c]{Cross-\\Platform}}
&\multirow{6}*{KU}         &IDRec   &1e-4 &-    &0.1 &-   &1024 &64 &2 &0.5G &7s   &3090-24G(1)\\
&                          &BERT    &5e-5 &1e-5 &0.1 &0.1 &1024 &64 &2 &107G &0.5m &A100-80G(4)\\
&                          &Res50   &5e-5 &5e-5 &0.1 &0.0 &1024 &64 &2 &174G &0.5m &A100-80G(4)\\
&                          &Swin-T  &5e-5 &5e-5 &0.1 &0.0 &1024 &64 &2 &183G &0.5m &A100-80G(4)\\
&                          &Swin-S  &5e-5 &5e-5 &0.1 &0.0 &1024 &64 &2 &358G &0.5m &A100-80G(4)\\
&                          &Swin-B  &5e-5 &5e-5 &0.1 &0.0 &1024 &64 &2 &637G &1m   &A100-80G(8)\\
\cmidrule(r){2-13}
&\multirow{6}*{QB}         &IDRec   &5e-5 &-    &0.1 &-   &512  &64 &2 &0.5G &15s  &3090-24G(1)\\
&                          &BERT    &1e-5 &5e-5 &0.1 &0.1 &1024 &64 &2 &107G &3m   &A100-80G(4)\\
&                          &Res50   &5e-5 &5e-5 &0.1 &0.0 &1024 &64 &2 &174G &2m   &A100-80G(4)\\
&                          &Swin-T  &5e-5 &5e-5 &0.1 &0.0 &1024 &64 &2 &183G &2m   &A100-80G(4)\\
&                          &Swin-S  &5e-5 &5e-5 &0.1 &0.0 &1024 &64 &2 &358G &3.5m &A100-80G(4)\\
&                          &Swin-B  &5e-5 &5e-5 &0.1 &0.0 &1024 &64 &2 &637G &5m   &A100-80G(8)\\
\cmidrule(r){2-13}
&\multirow{6}*{TN}         &IDRec   &5e-5 &-    &0.1 &-   &512  &128&2 &0.5G &20s  &3090-24G(1)\\
&                          &BERT    &1e-4 &1e-5 &0.1 &0.1 &1024 &64 &2 &107G &3m   &A100-80G(4)\\
&                          &Res50   &5e-5 &5e-5 &0.1 &0.0 &1024 &64 &2 &174G &2m   &A100-80G(4)\\
&                          &Swin-T  &5e-5 &5e-5 &0.1 &0.0 &1024 &64 &2 &183G &2m   &A100-80G(4)\\
&                          &Swin-S  &5e-5 &5e-5 &0.1 &0.0 &1024 &64 &2 &358G &3.5m &A100-80G(4)\\
&                          &Swin-B  &5e-5 &5e-5 &0.1 &0.0 &1024 &64 &2 &637G &5m   &A100-80G(8)\\
\cmidrule(r){2-13}
&\multirow{6}*{DY}         &IDRec   &5e-5 &-    &0.1 &-   &1024 &64 &2 &0.5G &20s  &3090-24G(1)\\
&                          &BERT    &5e-5 &1e-5 &0.1 &0.1 &1024 &64 &2 &107G &3m   &A100-80G(4)\\
&                          &Res50   &5e-5 &5e-5 &0.1 &0.0 &1024 &64 &2 &174G &2m   &A100-80G(4)\\
&                          &Swin-T  &5e-5 &5e-5 &0.1 &0.0 &1024 &64 &2 &183G &2m   &A100-80G(4)\\
&                          &Swin-S  &5e-5 &5e-5 &0.1 &0.0 &1024 &64 &2 &358G &3.5m &A100-80G(4)\\
&                          &Swin-B  &5e-5 &5e-5 &0.1 &0.0 &1024 &64 &2 &637G &5m   &A100-80G(8)\\
\bottomrule
\end{tabular}}
\end{center}
\vskip -0.1in
\end{table*}

\begin{table*}[ht]
\caption{Results of DSSM~\cite{huang2013learning} and DSSM-variant (see Figure~\ref{apx:fig:DSSM vs DSSM-variant}). 
NoPT of DSSM replaces its original itemID tower by ME, and its user encoder is still based on userID tower; NoPT of DSSM-variant  replaces the DSSM userID by her interacted item sequence, where  items are still represented by ME, and item sequence  is modeled by a standard DNN encoder. 
Since userID in DSSM cannot be transferred, we do not show the results of its HasPT version. 
It can be clearly seen that DSSM here performs substantially worse than its IDRec counterpart. 
% \textcolor{black}{This could be due to the ID-based user encoder utilized in DSSM.} 
By comparison, DSSM-variant can perform better than its IDRec variant even without pre-training (i.e. NoPT) on the source dataset.  
Clearly, DSSM and DSSM-variant are not ideal (very weak) baselines (compared with GRU4Rec, NextItNet, SASRec and BERT4Rec by the S2S training mode). 
In other words, the network architecture of recommendation models have very big impact on the performance of MoRec.
 ‘-’ means that there is no pre-training stage
on the source dataset.
The table here reports only text based recommendation. Given its much worse performance compared to main paper Table 2, we did not perform further experiments for
 image recommendation. 
}
\label{apx:tab:DSSMresults}
\vskip 0.15in
\begin{center}
\begin{small}
% \begin{sc}
\begin{tabular}{p{1.2cm}<{\centering}  p{1.5cm}<{\centering}  p{1.2cm}<{\centering} p{1.2cm}<{\centering} p{1.2cm}<{\centering} p{1.2cm}<{\centering}  p{1.2cm}<{\centering}  p{1.6cm}<{\centering}}
\toprule
\multirow{2}{*}{Dataset} &\multirow{2}{*}{Metric} &\multicolumn{2}{c}{DSSM} &\multicolumn{4}{c}{ DSSM-variant}\\
\cmidrule(r){3-4}\cmidrule(r){5-8}
&&IDRec &NoPT &IDRec &NoPT &HasPT  &Improv. \\
\midrule
\multirow{2}{*}{Bili\_500K}   &H@10 &0.97  &0.36 &0.55 &1.29 & - & - \\ 
                              &N@10 &0.47  &0.17 &0.26 &0.63 & - & - \\ 
\midrule
\multirow{2}{*}{Bili\_Food}   &H@10 &6.70  &1.93  &9.50  &9.62  &10.60 &+9.22\% \\ 
                              &N@10 &3.49  &0.89  &4.78  &4.79  &5.25  &+8.80\% \\ 
\multirow{2}{*}{Bili\_Dance}  &H@10 &8.44  &4.05  &10.36 &9.50  &12.24 &+22.40\% \\
                              &N@10 &4.37  &1.94  &5.34  &4.75  &6.62  &+28.25\% \\ 
\multirow{2}{*}{Bili\_Movie}  &H@10 &5.70  &1.88  &6.16  &6.25  &7.50  &+16.77\% \\
                              &N@10 &3.13  &0.94  &3.15  &3.32  &3.85  &+13.67\% \\
\multirow{2}{*}{Bili\_Cartoon}&H@10 &4.91  &1.09  &5.47  &7.01  &8.33  &+15.80\% \\ 
                              &N@10 &2.61  &0.51  &2.70  &3.60  &4.30  &+16.28\% \\
\multirow{2}{*}{Bili\_Music}  &H@10 &8.99  &3.38  &9.22  &10.66 &13.12 &+18.71\% \\
                              &N@10 &4.57  &1.60  &4.93  &5.44  &6.78  &+19.83\% \\ 
\midrule
\multirow{2}{*}{KU}           &H@10 &26.49 &22.02 &20.65 &25.91 &26.70 &+2.94\% \\
                              &N@10 &22.92 &12.34 &20.04 &22.75 &23.30 &+2.37\% \\
\multirow{2}{*}{QB}           &H@10 &24.02 &2.22  &27.10 &27.76 &28.19 &+1.53\% \\
                              &N@10 &17.38 &1.06  &20.32 &19.62 &21.08 &+6.96\% \\ 
\multirow{2}{*}{TN}           &H@10 &4.11  &0.61  &7.21  &7.36  &7.47  &+1.38\% \\
                              &N@10 &2.20  &0.27  &3.89  &3.97  &4.05  &+1.91\% \\ 
\multirow{2}{*}{DY}           &H@10 &6.34  &0.45  &8.24  &7.69  &9.16  &+16.05\% \\ 
                              &N@10 &3.80  &0.21  &4.34  &3.90  &4.82  &+19.04\% \\
\bottomrule
\end{tabular}
% \end{sc}
\end{small}
\end{center}
\vskip -0.1in
\end{table*}

\begin{table*}[t]
\caption{Results on the source dataset Bili\_500K with BERT and Swin-B as ME. TXT  and IMG represent text and image respectively. 
% Results of \textcolor{red}{red} color denote model collapse.
}
\label{apx:tab:source-result-500K}
\vskip 0.15in
\begin{center}
\begin{small}
% \begin{sc}
\begin{tabular}{p{1.5cm}<{\centering}  p{1.2cm}<{\centering}  p{0.7cm}<{\centering}  p{0.7cm}<{\centering}  p{0.7cm}<{\centering}  p{0.7cm}<{\centering}  p{0.7cm}<{\centering}  p{0.7cm}<{\centering}  p{0.7cm}<{\centering}  p{0.7cm}<{\centering}  p{0.7cm}<{\centering}  p{0.7cm}<{\centering}  p{0.7cm}<{\centering}  p{0.7cm}<{\centering}}
\toprule
\multirow{2}{*}{Dataset} & \multirow{2}{*}{Metric} &\multicolumn{3}{c}{SASRec} &\multicolumn{3}{c}{BERT4Rec}  &\multicolumn{3}{c}{NextItNet}  &\multicolumn{3}{c}{GRU4Rec} \\
\cmidrule(r){3-5}\cmidrule(r){6-8}\cmidrule(r){9-11}\cmidrule(r){12-14}
&& IDRec & TXT & IMG & IDRec & TXT & IMG & IDRec & TXT & IMG & IDRec & TXT & IMG \\
\midrule
\multirow{2}{*}{Bili\_500K} &H@10 &3.10 &3.97 &3.01 &2.96 &4.19 &2.19 &2.17 &3.64 &2.74 &2.46 &3.45 &2.34 \\
                            &N@10 &1.66 &2.08 &1.54 &1.54 &2.20 &1.11 &1.11 &1.96 &1.42 &1.24 &1.79 &1.17 \\
\bottomrule
\end{tabular}
% \end{sc}
\end{small}
\end{center}
\vskip -0.1in
\end{table*}

\begin{table*}[ht]
\caption{Results on more source datasets with BERT and Swin-B as ME.
Given that IDRec is generally more powerful in the warm-start recommendation setting, we generate two additional datasets called Bili\_warm20 (\#users: 359.9K, \#items 59.9K, \#interactions: 5.9M) and Bili\_warm50 (\#users: 169.7K, \#items 22.4K, \#interactions: 1.7M) by removing cold users and items. For  Bili\_warm20, we first remove cold items with less than 20 user interactions in Bili\_500K. Then we delete user sequences with less than 10 items. By iterating this many times, we finally ensure that all users had 10+ item interactions, and all items had 20+ user interactions. Similarly, we generate Bili\_warm50 where each item has at least 50 use interactions, which is a very warm dataset.
One can evaluate more cold- or warm-start settings using NineRec.
}
\label{tab:source-result-2M}
\vskip 0.15in
\begin{center}
\begin{small}
% \begin{sc}
\begin{tabular}{p{1.5cm}<{\centering}  p{1.5cm}<{\centering}  p{1.2cm}<{\centering} p{1.2cm}<{\centering}  p{1.2cm}<{\centering}}
\toprule
\multirow{2}{*}{Dataset} & \multirow{2}{*}{Metric} &\multicolumn{3}{c}{SASRec} \\
\cmidrule(r){3-5}
&& IDRec & BERT & Swin-B \\
\midrule
\multirow{2}{*}{Bili\_500K} &H@10 &3.10 &\textbf{3.97} &3.01 \\
                            &N@10 &1.66 &\textbf{2.08} &1.54 \\
\multirow{2}{*}{Bili\_2M}   &H@10 &3.51 &\textbf{4.26} &\textbf{3.71} \\
                            &N@10 &1.87 &\textbf{2.26} &\textbf{1.91} \\
\midrule
\multirow{2}{*}{\textcolor{black}{Bili\_warm20}}   &H@10 &3.79 &\textbf{4.81} &3.87 \\
                                &N@10 &2.05 &\textbf{2.56} &2.00 \\
\multirow{2}{*}{\textcolor{black}{Bili\_warm50}}   &H@10 &4.34 &\textbf{5.35} &4.58 \\
                                &N@10 &2.32 &\textbf{2.81} &2.31 \\
\bottomrule
\end{tabular}
% \end{sc}
\end{small}
\end{center}
\vskip -0.15in
\end{table*}

\begin{table*}[t]
\caption{Results of multimodal recommender system pre-trained on the source dataset. Since there is no suitable Chinese version, CLIP~\cite{radford2021learning} and ViLT~\cite{kim2021vilt} use the English translation as input for the text descriptions. English translation will be provided in the attachment.}
\label{apx:tab:MMRec-source}
\vskip 0.15in
\begin{center}
\begin{small}
% \begin{sc}
\begin{tabular}{p{1.4cm}<{\centering}  p{1.2cm}<{\centering}  p{1cm}<{\centering}  p{1cm}<{\centering}   p{2cm}<{\centering}}
\toprule
\multirow{1}{*}{Dataset} & \multirow{1}{*}{Metric} &\multicolumn{1}{c}{CLIP} &\multicolumn{1}{c}{ViLT} &\multicolumn{1}{c}{BERT+Swin-T} \\
% \cmidrule(r){3-5}\cmidrule(r){6-8}\cmidrule(r){9-11}
% &&TFS &TFS &TFS  \\
\midrule
\multirow{2}{*}{Bili\_500K} &H@10 &2.87	 & 2.96	& 3.67	\\
                            &N@10 &1.47 &1.53 & 1.90 \\
\bottomrule
\end{tabular}
% \end{sc}
\end{small}
\end{center}
% \vskip -0.3in
\end{table*}

\begin{table*}[htbp]
\caption{\textcolor{black}{Comparison of Ninerec with existing datasets. 'r-Image' refers to images with raw image pixels. 
\textcolor{black}{'Semantic.C' refers to the semantic complexity of images. 'Image.D' and 'Scenario.D' refer to the image diversity and scenario diversity.} 
}}
\label{dataset-comparison}
\centering
\begin{small}
\vskip 0.15in
\begin{tabular}{l p{0.4cm}<{\centering} p{0.8cm}<{\centering}  rrr c p{1.4cm}<{\centering} p{1.0cm}<{\centering} p{1.3cm}<{\centering}} 
\toprule
\multirow{3}{*}{Dataset} & \multicolumn{2}{c}{Modality} & \multicolumn{4}{c}{Statistics} & \multicolumn{3}{c}{Complexity/Diversity}  \\ %Complexity\&Scale
\cmidrule(lr){2-3}\cmidrule(lr){4-7}\cmidrule(lr){8-10}
~ & \multirow{1}{*}{Text} & \multirow{1}{*}{r-Image} &\multirow{1}{*}{\#User} &\multirow{1}{*}{\#Item} &\multirow{1}{*}{\#Actions.} &\multirow{1}{*}{Scenario}   &\multirow{1}{*}{\makecell[c]{Semantic.C}} &\multirow{1}{*}{\makecell[c]{Image.D}}  &\multirow{1}{*}{\makecell[c]{Scenario.D}}  \\
 % & & & & & & & \\
\midrule
Amazon           &\usym{2713}  &\usym{2713}  &20.98M  &9.35M   &82.83M  &E-commerce    &Low   &High  &Low  \\
H\&M             &\usym{2713}  &\usym{2713}  &1.37M   &106K    &31.79M  &E-commerce    &Low   &High  &Low  \\
GEST             &\usym{2713}  &\usym{2713}  &1.01M   &4.43M   &1.77M   &E-commerce    &Low   &Low   &Low  \\
\midrule
Reasoner         &\usym{2713}  &\usym{2713}  &3K      &5K      &58K     &Micro-video   &High  &High  &Low  \\
KuaiRec          &\usym{2717}  &\usym{2717}  &7K      &11K     &12.53M  &Micro-video   &Low   &Low   &Low  \\
\midrule
Behance          &\usym{2717}  &\usym{2717}  &63K     &179K    &1.00M   &Social Media  &Low   &Low   &Low  \\
Flickr           &\usym{2717}  &\usym{2717}  &8K      &105K    &5.90M   &Social Media  &Low   &Low   &Low  \\
\midrule
\textbf{NineRec (Source)} &\usym{2713}  &\usym{2713}  &2M      &185.43K &25.75M  &Stream Media  &\multirow{10}{*}{High} &\multirow{10}{*}{High} &\multirow{10}{*}{High} \\
-Bili\_Food      &\usym{2713}  &\usym{2713}  &6.55K   &1.58K   &39.74K  &Short-Video   & & & \\
-Bili\_Dance     &\usym{2713}  &\usym{2713}  &10.72K  &2.31K   &83.39K  &Short-Video   & & & \\
-Bili\_Movie     &\usym{2713}  &\usym{2713}  &16.53K  &3.51K   &115.58K &Short-Video   & & & \\
-Bili\_Cartoon   &\usym{2713}  &\usym{2713}  &30.30K  &4.72K   &215.44K &Short-Video   & & & \\
-Bili\_Music     &\usym{2713}  &\usym{2713}  &50.66K  &6.04K   &360.18K &Short-Video   & & & \\   
-KU              &\usym{2713}  &\usym{2713}  &2.03K   &5.37K   &18.52K  &Micro-Video   & & & \\             
-QB              &\usym{2713}  &\usym{2713}  &17.72K  &6.12K   &133.66K &News \& Videos \& ads & & & \\             
-TN              &\usym{2713}  &\usym{2713}  &20.21K  &3.33K   &122.58K &News \& Videos \& ads          & & & \\             
-DY              &\usym{2713}  &\usym{2713}  &20.40K  &8.30K   &139.83K &Micro-Video   & & & \\             
\bottomrule
\end{tabular}
\end{small}
\end{table*}

\begin{figure*}[htbp]
    \centering
    \includegraphics[width=0.88\textwidth]{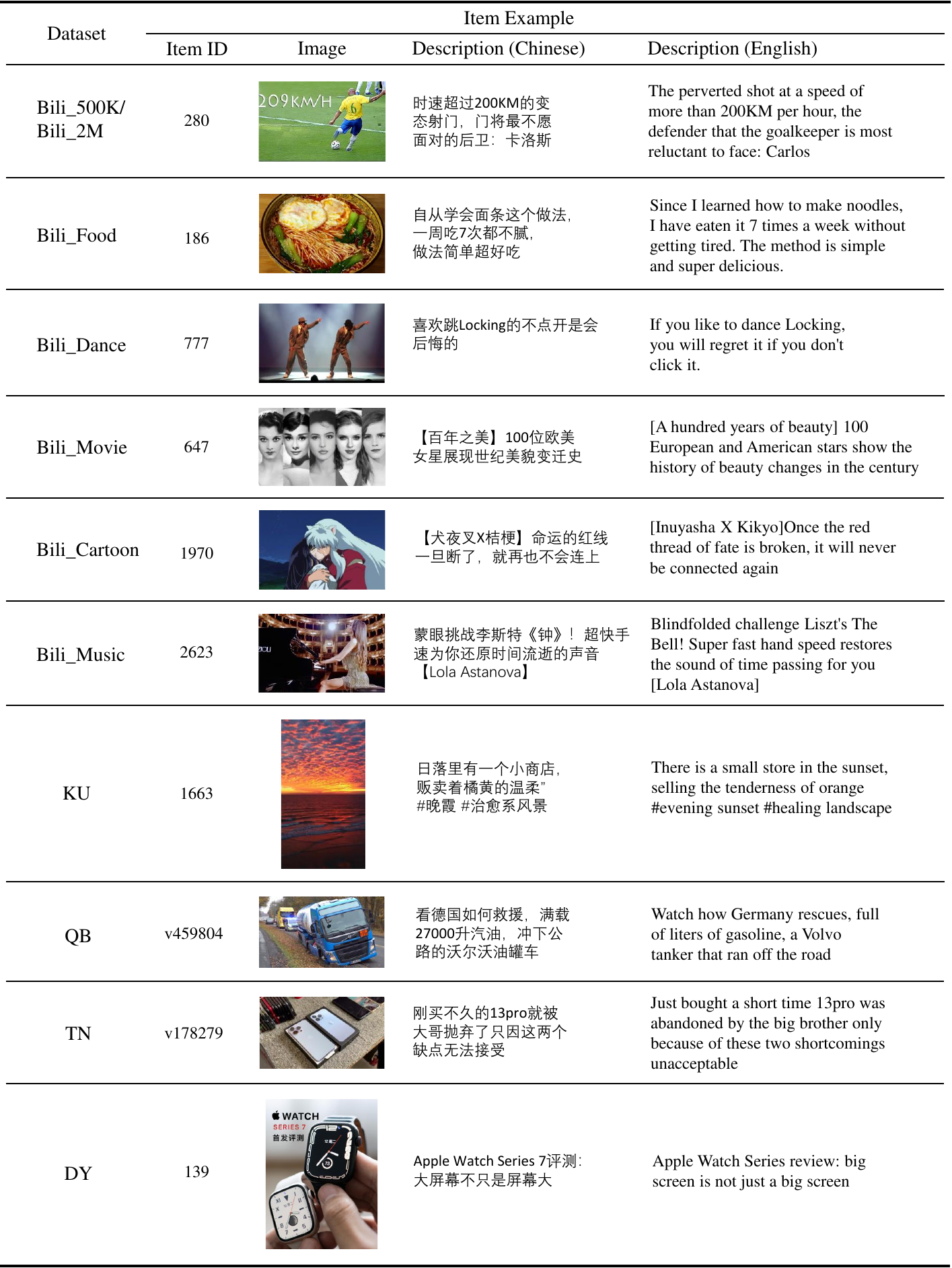}
    \caption{\textcolor{black}{An example of NineRec, including textual descriptions and thumbnails. 
    }}
    \label{apx:fig:entire_raw_dataset}
\end{figure*}

\begin{figure*}
    \centering
    \subfloat[\textcolor{black}{Case study including ground truth.}]{
        \begin{minipage}[htbp]{1.0\textwidth}
            \includegraphics[width=0.99\textwidth]{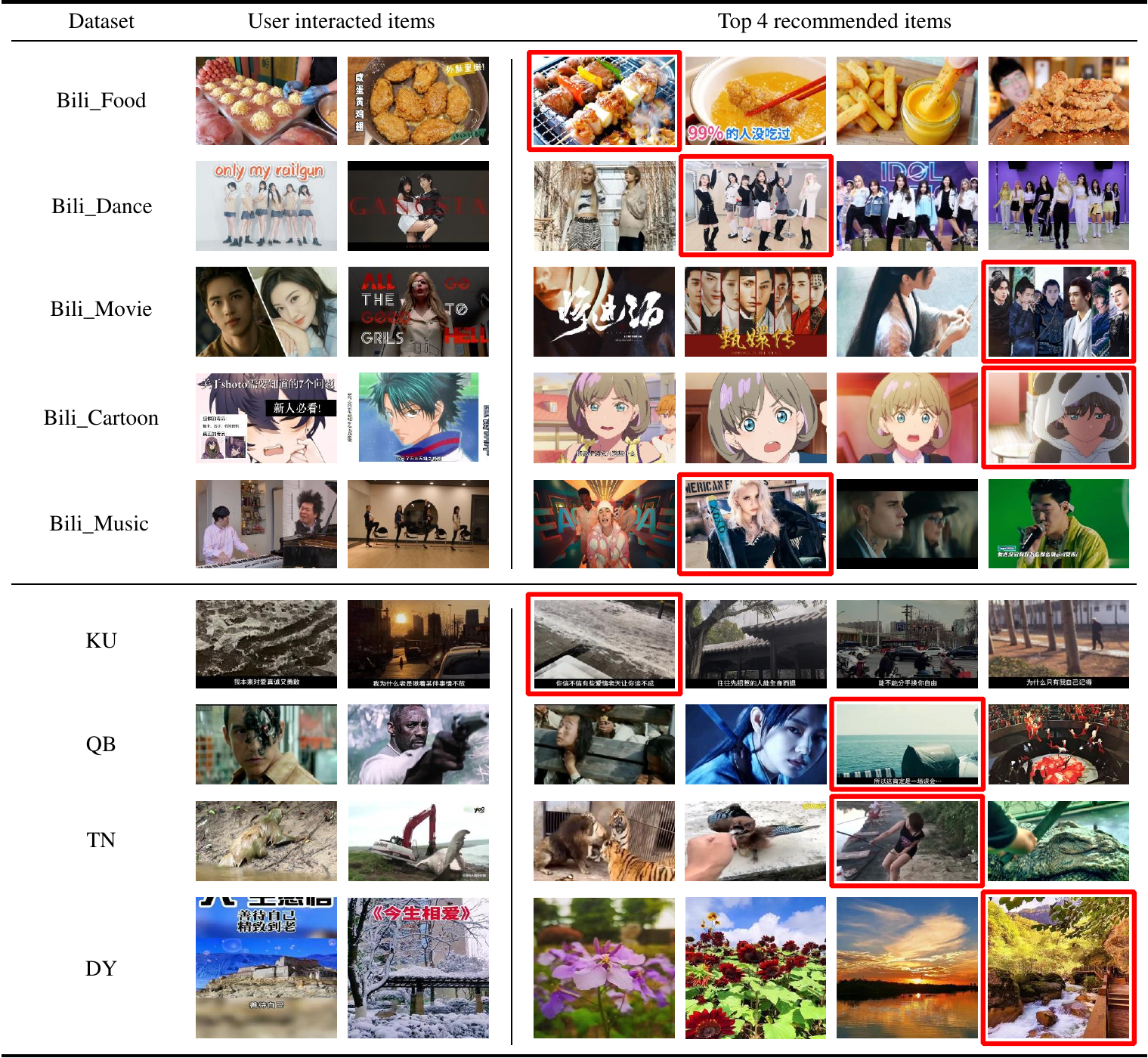}
        \end{minipage}}
\end{figure*}

\begin{figure*} 
    \centering
    \ContinuedFloat
    \subfloat[\textcolor{black}{Case study without ground truth.}]{
        \begin{minipage}[htbp]{1.0\textwidth}
            \includegraphics[width=0.99\textwidth]{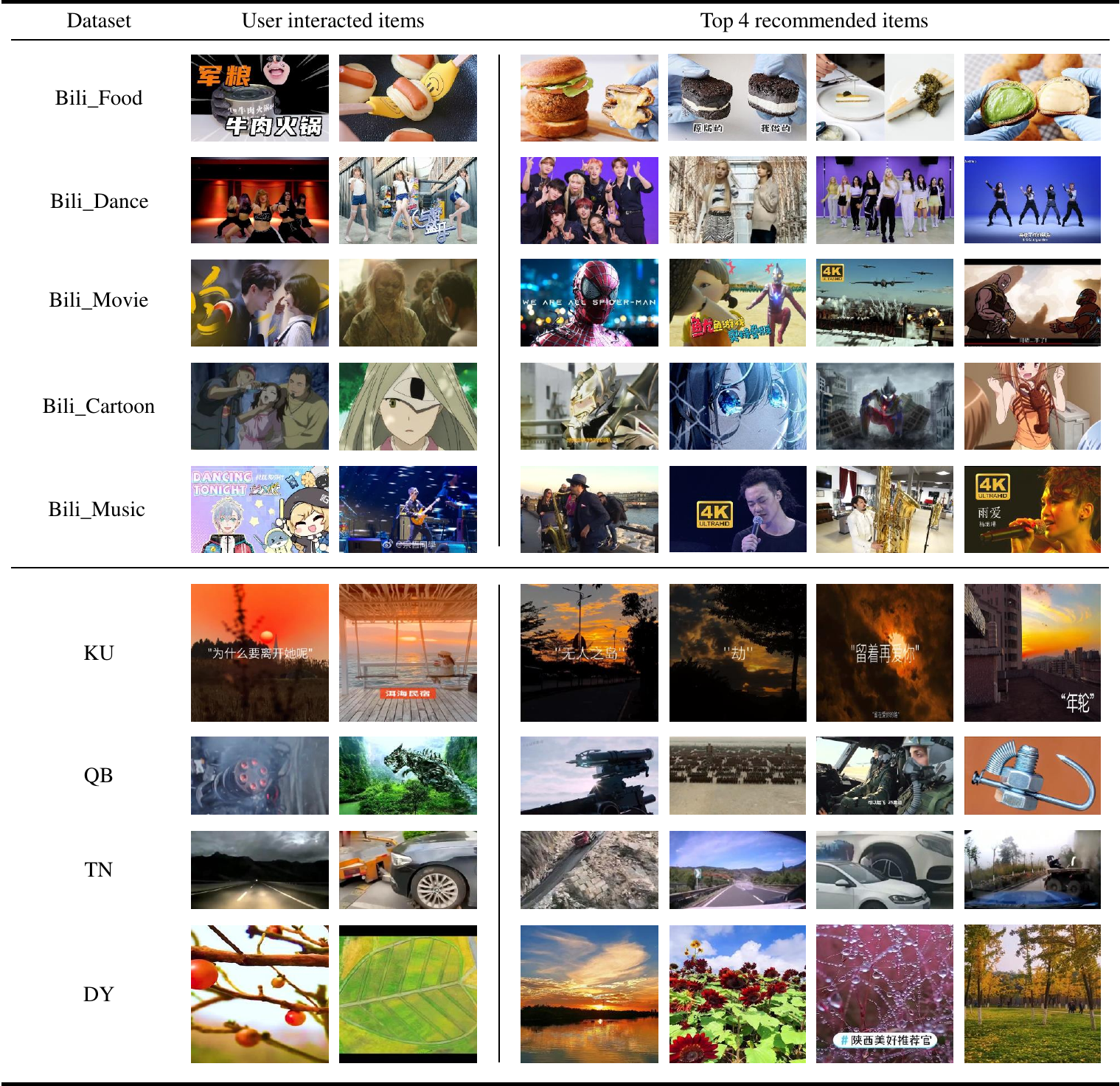}
        \end{minipage}}     
    \caption{\textcolor{black}{A case study of recommendation on NineRec. We show TransRec with SASRec as UE and Swin-B as ME, pre-trained on Bili\_2M. The left column is the user interacted items on nine downstream tasks. The right column shows the top 4 recommended items in the corresponding dataset. The ground truth recommended items have been framed by red lines on (a).  As can be clearly seen, the top-ranked items suggested by TransRec are often  relevant in terms of visual semantics of the input items, indicating a strong level of personalization. However, a weakness of TransRec is that its recommendations may lack diversity, which is a challenge commonly faced by classical recommendation algorithms.  
    }}
    \label{apx:fig:case_study} 
\end{figure*}

% \clearpage
% \bibliography{reference}
% \bibliographystyle{IEEEtran}

\end{document}